\documentclass[11pt]{article}

\usepackage[T1]{fontenc}

\usepackage[usenames,dvipsnames]{xcolor}
\definecolor{Maroon}{rgb}{0.15,0.33,0.52}
\definecolor{Mahogany}{rgb}{0.65,0.0975,0.0845}

\usepackage{amsmath, amssymb,amsthm}

\numberwithin{equation}{section}
\theoremstyle{plain}
\newtheorem{theorem}{Theorem}
\newtheorem{proposition}[theorem]{Proposition}
\newtheorem{corollary}{Corollary}[theorem]
\usepackage{bm,environ,mathrsfs,array,arydshln}
\usepackage[mathcal]{euscript}
\usepackage{mathabx}

\usepackage{graphicx,epsfig,epic}
\usepackage{subcaption,wrapfig}
\usepackage{tikz}
\usetikzlibrary{arrows.meta,calc,intersections}
\usepackage{tikz-feynman} 

\allowdisplaybreaks

\usepackage{framed}						
\definecolor{shadecolor}{rgb}{0.9996078, 0.984314, 0.960784}
\definecolor{framecolor}{rgb}{0,0,0}
\definecolor{TFTitleColor}{RGB}{1,1,1}
\definecolor{TFFrameColor}{RGB}{249	218	181}		
\definecolor{TFFrameColor}{RGB}{230 230 230 }

\newenvironment{frshaded}{%
    \MakeFramed {\FrameRestore}}%
    {\endMakeFramed}

\definecolor{newgreen}{RGB}{0,110,60}

\usepackage{mdframed}

\definecolor{lightpeach}{RGB}{255, 247, 235}
\definecolor{mediumpeach}{RGB}{166, 83,1}

\newmdenv[
  backgroundcolor=lightpeach,
  linecolor=mediumpeach,
  roundcorner=5pt,
  innerleftmargin=8pt,
  innerrightmargin=8pt,
  innertopmargin=6pt,
  innerbottommargin=6pt,
]{remarkbox}

\usepackage{jheppub}
\hypersetup{colorlinks=true,linkcolor=Maroon,citecolor=Maroon,urlcolor=Maroon}

\newcommand{\bs}{\begin{frshaded}}			
\newcommand{\es}{\end{frshaded}\noindent}

\def\ba#1\ea{\begin{align}#1\end{align}}		         
\newcommand{\be}{\begin{equation}}
\newcommand{\ee}{\end{equation}}
\newcommand{\bea}{\begin{equation} \begin{aligned}} 
\newcommand{\eea}{\end{aligned} \end{equation}}
\newcommand{\mc}{\mathcal }
\newcommand{\wh}{\widehat}
\newcommand{\wt}{\widetilde}

\newcommand{\mk}{\mathfrak}
\newcommand{\la}{\label}

\DeclareMathOperator{\Tr}{\text{Tr}}
\DeclareMathOperator{\tr}{\text{tr}}

\newcommand{\cf}{\textit{cf.} }
\newcommand{\ie}{\textit{i.e.} }
\newcommand{\eg}{\textit{e.g.} }

\renewcommand{\l}{\lambda}

\newcommand{\ket}[1]{|#1\rangle}
\newcommand{\bra}[1]{\langle #1|}
\newcommand{\braket}[2]{\langle #1|#2\rangle}
\newcommand{\mmm}[3]{\langle #1|#2|#3\rangle}

\definecolor{diagramInk}{HTML}{24313A}
\definecolor{diagramTeal}{HTML}{007B83}
\definecolor{diagramMatter}{HTML}{B65625}
\definecolor{diagramWash}{HTML}{F1F4F5}
\definecolor{diagramLoop}{HTML}{6A4C93}

\pgfmathsetmacro{\rneck}{0.38}   
\pgfmathsetmacro{\Lhalf}{1.25}   
\pgfmathsetmacro{\kflare}{0.50}  
\pgfmathsetmacro{\ecc}{0.28}     
\pgfmathsetmacro{\phiL}{-55}     
\pgfmathsetmacro{\phiR}{-25}     
\pgfmathsetmacro{\phiN}{-40}     
\pgfmathdeclarefunction{prof}{1}{\pgfmathparse{sqrt(\rneck*\rneck+\kflare*#1*#1)}}

\colorlet{glueSurf}{gray!10}
\colorlet{glueHole}{gray!28}
\colorlet{glueMatter}{diagramMatter}
\tikzset{
  outline/.style={line width=0.7pt},
  neckcut/.style={diagramTeal,line width=0.7pt},
  neckhidden/.style={diagramTeal,line width=0.5pt,dashed},
  matterline/.style={glueMatter,line width=1.1pt},
  threaded/.style={draw=glueMatter,line width=1.1pt},
  ins/.style={circle,fill=glueMatter,inner sep=1.3pt},       
}

\newcommand{\surfpt}[2]{({#1+\ecc*prof(#1)*cos(#2)},{prof(#1)*sin(#2)})}

\newcommand{\tube}[2]{%
  \fill[glueSurf] plot[domain=#1:#2,samples=60,smooth] (\x,{prof(\x)})
     -- plot[domain=#2:#1,samples=60,smooth] (\x,{-prof(\x)}) -- cycle;
  \draw[outline] plot[domain=#1:#2,samples=60,smooth] (\x,{prof(\x)});
  \draw[outline] plot[domain=#1:#2,samples=60,smooth] (\x,{-prof(\x)});
}
\newcommand{\openend}[2]{%
  \pgfmathsetmacro{\rr}{prof(#1)}%
  \filldraw[#2,fill=glueHole] (#1,0) ellipse[x radius={\ecc*\rr},y radius=\rr];
}
\newcommand{\threadedneck}{%
  \filldraw[fill=glueMatter!18,threaded] (0,0) ellipse[x radius={\ecc*\rneck},y radius=\rneck];
}
\newcommand{\matterpath}[4]{%
  \draw[matterline] plot[domain=0:1,samples=40,smooth,variable=\s]
    ({(#1)+\s*((#3)-(#1)) + \ecc*prof((#1)+\s*((#3)-(#1)))*cos((#2)+\s*((#4)-(#2)))},
     {prof((#1)+\s*((#3)-(#1)))*sin((#2)+\s*((#4)-(#2)))});
}

\newcommand{\glueWormhole}{%
\begin{tikzpicture}[baseline={(0,-0.08)}]
  \tube{-\Lhalf}{\Lhalf}
  \draw[neckhidden] (0,\rneck) arc[start angle=90,end angle=270,x radius={\ecc*\rneck},y radius=\rneck];
  \draw[neckcut]    (0,-\rneck) arc[start angle=-90,end angle=90,x radius={\ecc*\rneck},y radius=\rneck];
  \openend{-\Lhalf}{outline}
  \openend{\Lhalf}{outline}
  \matterpath{-\Lhalf}{\phiL}{\Lhalf}{\phiR}
  \node[ins] at \surfpt{-\Lhalf}{\phiL} {};
  \node[ins] at \surfpt{\Lhalf}{\phiR} {};
  \node[below=3pt] at (-\Lhalf,-0.95) {$\beta_1$};
  \node[below=3pt] at ( \Lhalf,-0.95) {$\beta_2$};
\end{tikzpicture}}

\newcommand{\glueTrumpetL}[1]{%
\begin{tikzpicture}[baseline={(0,-0.08)}]
  \tube{-\Lhalf}{0}
  \openend{-\Lhalf}{outline}
  \threadedneck
  \matterpath{-\Lhalf}{\phiL}{0}{\phiN}
  \node[ins] at \surfpt{-\Lhalf}{\phiL} {};
  \node[below=3pt] at (-\Lhalf,-0.95) {$#1$};
  \node[right=1pt] at (0.08,0.25) {$\scriptstyle m$};
\end{tikzpicture}}

\newcommand{\glueTrumpetR}[1]{%
\begin{tikzpicture}[baseline={(0,-0.08)}]
  \tube{0}{\Lhalf}
  \threadedneck
  \openend{\Lhalf}{outline}
  \matterpath{0}{\phiN}{\Lhalf}{\phiR}
  \node[ins] at \surfpt{\Lhalf}{\phiR} {};
  \node[below=3pt] at (\Lhalf,-0.95) {$#1$};
  \node[left=1pt] at (-0.08,0.25) {$\scriptstyle m$};
\end{tikzpicture}}

\begin{document}

\title{Matter-decorated trumpets in double-scaled SYK}
\author[a,b]{M. Beccaria}
\affiliation[a]{Universit\`a del Salento,
Dipartimento di Matematica e Fisica
\textit{Ennio De Giorgi},
Via Arnesano, I-73100 Lecce, Italy}
\affiliation[b]{INFN, Sezione di Lecce,
Via Arnesano, I-73100 Lecce, Italy}
\emailAdd{matteo.beccaria@le.infn.it}
\abstract{

In sine-dilaton gravity, the double trumpet, the wormhole amplitude without matter, is obtained by gluing two trumpets along a closed geodesic of discrete length and summing over this length. When the wormhole carries the propagator of a matter field of dimension $\Delta$, joining the two asymptotic boundaries and crossing the neck, such a gluing is known only in the Jackiw--Teitelboim (JT) limit. We construct the gluing beyond JT in double-scaled SYK (DSSYK), the proposed boundary dual of sine-dilaton gravity, whose disk amplitudes it reproduces. In this context, we adopt the definition of the matter-decorated wormhole in terms of the ETH two-matrix model with Gaussian matter. We build matter-decorated trumpets from an end-of-the-world brane trace with a matter insertion, and prove that the wormhole is glued from them exactly, sector by sector in the number of chords through the neck, with a positive weight that accounts for the twists and windings of these chords and for the closed loops around the neck. Trumpets and weight are obtained in closed form, the former as transforms of Rogers polynomials, and we discuss the uniqueness of the construction. In the JT limit, both reproduce the known gluing separately, and not only in combination. In the classical limit, each trumpet takes the form expected in the bulk: a pure-gravity sine-dilaton trumpet dressed by the factor that a probe matter geodesic contributes at its boundary.
}


\maketitle


\setcounter{footnote}{0}

\section{Introduction and summary}

Double-scaled Sachdev--Ye--Kitaev (DSSYK) quantum mechanics is exactly solvable in terms of chord diagrams: its disorder-averaged partition function is a sum over diagrams of chords joining pairs of
Hamiltonian insertions on a circle, weighted by $q=e^{-\lambda}$ for each chord crossing
\cite{Berkooz:2018qkz,Berkooz:2018jqr,Berkooz:2024lgq}. We study the
two-boundary amplitude defined by these rules on an annulus, with
boundaries of lengths $\beta_1,\beta_2$ and one insertion of a
matter operator of dimension $\Delta$ on each boundary. The two insertions are joined by a
matter chord, whose crossings with Hamiltonian chords are weighted by $q^\Delta$. The resulting amplitude is a trace over
the chord Hilbert space of DSSYK \cite{Okuyama:2023yat}, with a spectral
representation that depends only on $\beta_1+\beta_2$,
\be
\la{1.1}
\mc W_\Delta(\beta_1+\beta_2)=\int dE\;\rho_\Delta(E)\,e^{-(\beta_1+\beta_2)E},
\qquad \rho_\Delta(E)>0,
\ee
where $\rho_\Delta$ is the DSSYK density of states dressed by the matter
two-point kernel. This amplitude is the genus-zero two-boundary amplitude of the ETH two-matrix model of
DSSYK \cite{Jafferis:2022uhu,Okuyama:2023yat} with Gaussian matter, that is, with a single matter propagator and no matter loops. We call it the \emph{matter-decorated wormhole}.

Our reason to reconsider this amplitude is the following. DSSYK has been proposed to be dual to sine-dilaton gravity, a duality established at the level of the disk
\cite{Blommaert:2023opb,Blommaert:2024ymv,Bossi:2024ffa}.
Beyond the disk, only some amplitudes of the matrix model have a bulk derivation, among them the one studied here. Without matter, the two-boundary amplitude has been derived in sine-dilaton gravity
\cite{Blommaert:2025avl} by cutting the wormhole along the closed geodesic at its neck, of length $\lambda b$ with integer $b$. Each half is an ordinary trumpet, a geometry with one asymptotic boundary and the geodesic as its neck, whose amplitude is the Bessel function $I_b$, and the wormhole is the sum over $b$ of products of two trumpets glued with the measure $b$.

With matter, the amplitude \eqref{1.1} has been obtained in sine-dilaton gravity in \cite{Cui:2025sgy} by cutting the wormhole \emph{along} the matter line, on slices joining the two boundaries. A gluing \emph{across} the matter line, into matter-decorated trumpets whose neck is threaded by it, is known only in the JT limit \cite{Saad:2019pqd,Stanford:2020wkf,Jafferis:2022wez}. In sine-dilaton gravity, \ie at generic finite $q$, no counterpart is available for the trumpet states labelled by the closed geodesic around the neck, or for their gluing measure.%
\footnote{Closed slices of
sine-dilaton gravity with a pointlike observer of dynamical mass are quantized in
\cite{Blommaert:2025rgw}, in a basis labelled by
the length of the slice and the value of the dilaton on it; the neck states for a matter line of fixed dimension and the
decomposition of the matter-decorated wormhole into one-boundary factors are not addressed there.}
Finding a suitable gluing into matter-decorated trumpets at finite $q$ is the subject of this paper. In the absence of a fully-fledged closed-channel quantization of sine-dilaton gravity with matter, we address it on the DSSYK side, where the chord diagrams of the wormhole are explicit and can be cut at the neck diagram by diagram.

We remark that, by \eqref{1.1}, the amplitude may be written as a sum of products of one-boundary factors in many obvious ways: every orthonormal basis of
functions of the energy for the weight $\rho_\Delta$ gives one. Since $\rho_\Delta$ has compact support, its orthonormal polynomials are such a basis, and the resulting gluing identity, with positive gluing weights, is just Parseval's identity for them. For $\rho_\Delta$ these are the continuous $q$-ultraspherical (Rogers) polynomials. They give this decomposition in closed form, but only up to the normalization of its one-boundary factors, our candidate matter-decorated trumpets. In this paper we address a finer question: whether each term of the decomposition is a sector of chord diagrams of the wormhole, with a fixed number of chords crossing a cut \emph{across} the matter line, and whether, suitably normalized, the candidate trumpets are sums of chord diagrams with a single asymptotic boundary, as in gravity.

To answer, we define one-boundary amplitudes directly by chord diagrams, but guided by the JT gravity construction, which extends the matter-free gluing of trumpets at a neck \cite{Saad:2018bqo,Saad:2019JTMatrixIntegral,Stanford:2019vob} to a matter line. The
JT limit $\mc W^{\rm JT}_\Delta(\tau_1+\tau_2)$ of \eqref{1.1}, with
$\tau_{1,2}=\lambda^{3/2}\beta_{1,2}$ the Schwarzian boundary lengths held
fixed as $\lambda\to0$, factorizes at fixed neck circumference
$\mathfrak b$: it is the integral over $\mathfrak b$, with the flat
measure $\frac{\Gamma(\Delta)^2}{\Gamma(2\Delta)}\,d\mathfrak b$, of the
product of two \emph{fixed-neck JT amplitudes}
$Z^{\rm fix}_\Delta(\mathfrak b,\tau_i)$.

 The fixed-neck amplitudes are obtained as Schwarzian path integrals with a boundary-profile constraint that has no direct counterpart at finite $q$. To go beyond JT, we start from their Laplace transform in the neck circumference, taken with the inverse of the Schwarzian orbit factor $2\sinh(\mathfrak b/2)$, the Jacobian between the Schwarzian measure and the standard path-integral measure of the boundary profile. This transform replaces the fixed neck by an end-of-the-world (EOW) brane of tension $\kappa$. The result is a thermal trace in a one-dimensional Morse quantum mechanics, whose coordinate $\varphi$ is the logarithm of the local boundary stretching at the matter endpoint,
\be
\la{1.2}
\int_0^\infty d\mathfrak b\,
\frac{e^{-\kappa\mathfrak b}}{2\sinh(\mathfrak b/2)}\,
Z^{\rm fix}_\Delta(\mathfrak b,\tau)
=\Tr\!\left(e^{-\tau H_\kappa}e^{\Delta\varphi}\right),
\qquad
H_\kappa=-\partial_\varphi^2+\tfrac14e^{2\varphi}+\kappa e^\varphi.
\ee
Without the matter insertion $e^{\Delta\varphi}$, the right-hand side of \eqref{1.2} is the Morse description of the JT brane \cite{Gao:2021uro,Okuyama:2023byh},
with the tension giving the term $\kappa e^\varphi$.

The advantage of this reformulation is that the EOW brane admits a chord realization in DSSYK at any $q$, as a
boundary state on which Hamiltonian chords can end
\cite{Okuyama:2023byh}. The DSSYK counterpart of $H_\kappa$ is the brane transfer operator $T_a$, acting on states with $\widehat N$ open Hamiltonian chords between the boundary and the brane, on which a chord ends with weight $a=q^{\kappa+\frac12}$. The matter line crosses these chords, which gives the insertion
$q^{\Delta\widehat N}$.  On the chord Hilbert space
$\mathcal H$,
we define the \emph{matter-decorated trumpet amplitudes}
$Z_\Delta(m,\beta)$ by the expansion in the brane parameter, the finite-$q$ counterpart of the Laplace transform in \eqref{1.2}
\be
\la{1.3}
\Tr_{\mc H}\!\left[q^{\Delta\wh N}e^{-\beta T_a}\right]
=\sum_{m\geq0}\frac{a^m}{1-q^{m+\Delta}}\,
Z_\Delta(m,\beta).
\ee
The integer $m\geq0$ labels the amplitudes. The factors $1/(1-q^{m+\Delta})$ are the finite-$q$ form of the inverse orbit factor $1/[2\sinh(\mathfrak b/2)]$ of \eqref{1.2} at $\mathfrak b=\lambda(m+\Delta)$, and they resum the closed loops, Hamiltonian chords that close on themselves around the neck, each crossing the $m$ chords ending on the brane and the matter line once. We prove that $Z_\Delta(m,\beta)$ is the sum of the chord diagrams of
this trace with $m$ chords ending on the brane and no closed loop.%
\footnote{Without matter, the same expansion of the brane amplitude gives the ordinary trumpets, in the chord description \cite{Okuyama:2023byh} and in sine-dilaton gravity \cite{Blommaert:2025avl}.}
\begin{figure}[t]
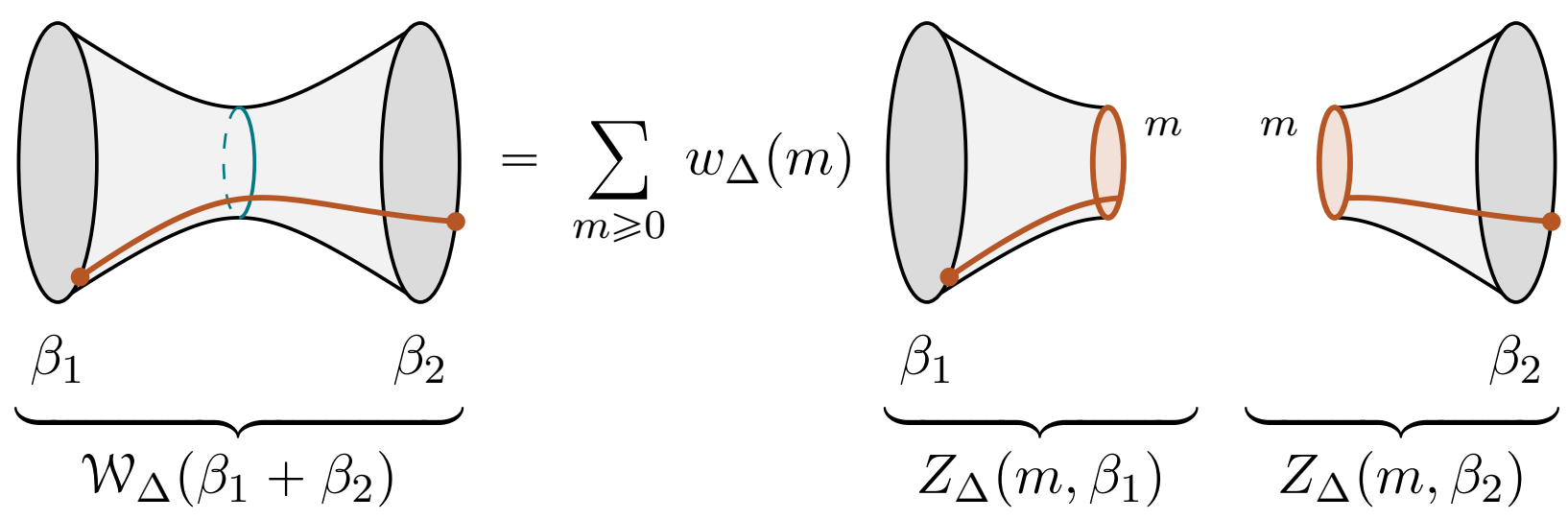

\centering
\[
\underbrace{\glueWormhole}_{\displaystyle\mathcal W_\Delta(\beta_1+\beta_2)}
\;=\;\sum_{m\ge0}\,w_\Delta(m)\;
\underbrace{\glueTrumpetL{\beta_1}}_{\displaystyle Z_\Delta(m,\beta_1)}
\;\;
\underbrace{\glueTrumpetR{\beta_2}}_{\displaystyle Z_\Delta(m,\beta_2)}
\]
\caption{Sector decomposition \eqref{1.4}, \eqref{1.6} of the matter-decorated
wormhole, drawn as a gluing of matter-decorated trumpets. Left: the
wormhole $\mathcal W_\Delta(\beta_1+\beta_2)$, with boundaries
of lengths $\beta_1,\beta_2$ and a matter line of dimension $\Delta$
(orange) joining two operator insertions (dots), is cut along its neck
(teal). Right: each half is a one-boundary amplitude
$Z_\Delta(m,\beta_i)$, a trumpet whose neck (drawn in the matter
color) is threaded by the matter line. The halves are glued with weight $w_\Delta(m)$ and summed
over the integer neck label $m$. By the sector factorization \eqref{1.4}, the term with label $m$ is the sum of the chord diagrams of the wormhole with $m$ Hamiltonian chords joining its two boundaries.}
\label{fig:gluing}
\end{figure}
In detail, our results are the following (see Fig.~\ref{fig:gluing}).
\begin{enumerate}
\item \textbf{Sector factorization.} The chord-diagram expansion of the
wormhole factorizes \emph{sector by sector} (Theorem~\ref{prop:sectors}): if
$S_m(\beta_1,\beta_2)$ denotes the sum of the chord diagrams of the
wormhole with exactly $m$ Hamiltonian chords joining the two
boundaries, so that the wormhole amplitude is $\mc W_\Delta(\beta_1+\beta_2)=\sum_{m\geq0}S_m$,
then
\be
\la{1.4}
S_m(\beta_1,\beta_2)=w_\Delta(m)\,
Z_\Delta(m,\beta_1)\,Z_\Delta(m,\beta_2),
\qquad m\geq0.
\ee
 The proof rests on splitting each diagram into a part near each boundary and a band around the neck. The crossings near the boundary of length $\beta_i$ involve only the chords with both endpoints on that boundary, the endpoints of the $m$ chords on it and the matter endpoint, and they sum to $Z_\Delta(m,\beta_i)$; the crossings in the band depend only on how the $m$ chords are permuted and wound around the neck and on the closed loops, and they give $w_\Delta(m)$.
 This positive weight is thus a sum over twists and windings, dressed by the closed chord loops around the neck,
\be
\la{1.5}
w_\Delta(m)=\frac{(1-q)^m}{1-q^{m+\Delta}}
\sum_{\sigma\in\mathsf S_m}\sum_{\bm k\in\mathbb Z^m}
q^{N(\sigma,\bm k)}\,q^{\Delta\sum_i|k_i|}
=\frac{(q;q)_m\,(q^{2\Delta};q)_m}{(1-q^{m+\Delta})\,(q^\Delta;q)_m^2}.
\ee
Here the relative twist $\sigma\in\mathsf S_m$ of the $m$ chords through the neck is the permutation relating the order of their endpoints on the two boundaries, and $\bm k\in\mathbb Z^m$ are their windings around the neck relative to the matter line. The exponent $N(\sigma,\bm k)$ is the number of crossings of the $m$ chords among themselves, which depends only on the twist and the windings, and the factor $q^{\Delta\sum_i|k_i|}$ counts their crossings with the matter line. The denominator resums the closed loops, each of which crosses the $m$ chords and, once, the matter line. The second equality is a summation identity, the twist--winding identity of Proposition~\ref{prop:shi}.

That the weight does not depend on the two boundaries is the
finite-$q$ form of the decoupling of the twist integral in JT gravity.

Summing \eqref{1.4} over $m$ gives the gluing identity
\be
\la{1.6}
\mc W_\Delta(\beta_1+\beta_2)
=\sum_{m\geq0}w_\Delta(m)\,
Z_\Delta(m,\beta_1)Z_\Delta(m,\beta_2),
\ee
for all $0<q<1$, $\Delta>0$ and $\beta_1,\beta_2>0$.

\item \textbf{Threshold property.} The amplitude with label $m$ satisfies $Z_\Delta(m,\beta)=O(\beta^m)$ at short boundary length, since each chord ending on the brane costs a power of the boundary length, and similarly $S_m=O(\beta_1^m\beta_2^m)$.
\item \textbf{Uniqueness and normalization.} Whatever their origin, one-boundary factors that glue into the wormhole and have the threshold property are fixed by it up to an $m$-dependent
normalization (Sec.~\ref{sec:general}). Since $Z_\Delta(m,\beta)$ glue into the wormhole by \eqref{1.4} and have this property, the uniqueness statement also gives them in closed form, as transforms of the Rogers polynomials. In our case the normalization is fixed by the \emph{loop criterion}, the exclusion of the closed loops stated below \eqref{1.3}, which assigns them to the weight.
\item \textbf{JT and matter-free limits.} For every $\Delta>0$ the
amplitudes reproduce the fixed-neck JT amplitudes in the JT limit, with
the weights reproducing the flat neck measure of \eqref{4.13} (Proposition~\ref{prop:JT}). With $\wt\beta=\beta E_0$ and $E_0$ the DSSYK spectral edge, and for
$\lambda m\to\mathfrak b>0$,
\be
\la{1.7}
\lambda^{-(\Delta+1)}e^{-\wt\beta}\,Z_\Delta(m,\beta)\ \longrightarrow\ Z^{\rm fix}_\Delta(\mathfrak b,\tau),
\qquad
\lambda\,w_\Delta(m)\ \longrightarrow\ \frac{\Gamma(\Delta)^2}{\Gamma(2\Delta)}.
\ee
 As $\Delta\to0$ the amplitudes reduce to the ordinary trumpets $I_m(\wt\beta)$. For $m\geq1$ the weights reduce to the matter-free measure $b=m$ once the common winding of the chords through the neck is quotiented: without the matter line to measure it against, this winding is an unphysical redundancy. The sector $m=0$, which diverges through its closed loops and has no matter-free counterpart, is omitted.\footnote{A similar quotient is suggested in \cite[Sec.~5.2]{Cui:2025sgy} for the matter-free double trumpet of \cite{Blommaert:2025avl}.}
\item \textbf{Classical limit.} In the limit $\l\to0$ at fixed $\l m$ and $\l\wt\beta$, which corresponds to the classical regime of sine-dilaton gravity under the proposed duality, each amplitude becomes a classical pure-gravity sine-dilaton trumpet dressed by the factor that a probe matter geodesic contributes at its boundary. This suggests its interpretation as a matter-decorated trumpet in the bulk.

\end{enumerate}

These are exact properties of the matrix model amplitude. Whether sine-dilaton gravity is described by a matrix model beyond the disk is open already without matter, where the agreement of the wormhole with a finite-cut matrix integral, the matter-free sector of the ETH matrix model, makes the identification possible, but does not guarantee it \cite[Sec.~6.1]{Blommaert:2025avl}. With matter, our results make this question sharper: since the gluing is fixed by the threshold property, a closed-channel quantization of sine-dilaton gravity with matter can be compared with the ETH matrix model not only in the wormhole, but term by term, in the neck states, their measure $w_\Delta(m)$ and the trumpets $Z_\Delta(m,\beta)$. The comparison is made in a definite basis of neck states: the one labelled by the number of chords through the neck, which reduces to the geodesic length $b$ without matter and is singled out by the threshold property. In any other basis, the decomposition differs from ours by the change of basis.

\paragraph{Plan of the paper.}
Sec.~\ref{sec:review} defines the matter-decorated wormhole and states its chord rules, and Sec.~\ref{sec:general} collects what follows from its spectral representation alone. Sec.~\ref{sec:neck-fixed} recalls the fixed-neck JT
amplitudes and Sec.~\ref{sec:fq-candidate} relates them to the Morse trace \eqref{1.2}, defines the amplitudes and fixes
their normalization. Sec.~\ref{sec:fq-chords} states the sector
factorization and Sec.~\ref{sec:fq-spectral} collects further exact properties.
Sec.~\ref{sec:fq-JT-fixed} is devoted to the JT limit and
Sec.~\ref{sec:sinedilaton} to the classical limit and its interpretation from the sine-dilaton perspective. The appendices
contain the technical derivations. Different boundary lengths and neck labels are used, each natural in its own regime (finite $q$, the JT limit, the classical limit); to keep them apart, Table~\ref{tab:notation} collects them, with the matter crossing weight, and indicates where each is defined.

\begin{table}[!htb]
\centering
\renewcommand{\arraystretch}{1.25}
\begin{tabular}{>{\raggedright\arraybackslash}p{0.17\textwidth}>{\raggedright\arraybackslash}p{0.75\textwidth}}
\hline
\textbf{Symbol} & \textbf{Meaning}\\
\hline
$\beta$ & boundary length in DSSYK units (inverse temperature)\\
$\wt\beta=\beta E_0$ & boundary length in units of the inverse spectral edge\\
$\tau=\l^{3/2}\beta$ & Schwarzian boundary length, held fixed in the JT limit\\
$b$ & integer label of the ordinary trumpets $I_b(\wt\beta)$; in sine-dilaton gravity $\l b$ is the length of the neck in the rescaled metric; continued to real $b>0$ in \eqref{9.5}\\
$m$ & label of the matter-decorated trumpet amplitudes, the number of Hamiltonian chords through the neck\\
$\mathfrak b$ & neck circumference of JT gravity, limit of $\l m$\\
$\mathfrak b_m=\l(m+\Delta)$ & shifted label\\
$t=q^\Delta$ & weight of a crossing of a Hamiltonian chord with the matter chord\\
$\beta_s=\l\wt\beta$ & renormalized boundary length in gravity units, held fixed in the classical limit\\
$\eta$ & parameter of the classical trumpet, $\sinh\eta=b/\wt\beta$\\
\hline
\end{tabular}
\caption{Lengths, neck labels and the matter crossing weight.}
\label{tab:notation}
\end{table}

\section{DSSYK chord rules and the decorated wormhole}
\la{sec:review}

\subsection{DSSYK and the chord Hilbert space}
\la{sec:dssyk}

The boundary theory is constructed from the SYK model \cite{Sachdev:1992fk,Kitaev:2015Talks,Maldacena:2016hyu,Berkooz:2018jqr}
with $M$ Majorana fermions $\{\psi_i\}_{i=1,\dots, M}$, with
$\{\psi_i,\psi_j\}=2\delta_{ij}$, and a random even-$p$-body Hamiltonian
\be
H=i^{p/2}\sum_{i_1<\cdots<i_p}J_{i_1\cdots i_p}
\psi_{i_1}\cdots\psi_{i_p},
\qquad
\langle J_{i_1\cdots i_p}^{\,2}\rangle_J=\binom Mp^{-1}.
\ee
The couplings are independent real Gaussian variables with zero mean,
and $\langle\cdots\rangle_J$ denotes the average over them.
The double-scaled limit takes
$M,p\to\infty$ at fixed $\lambda=2p^2/M$
\cite{Erdos:2014zgc,Cotler:2016fpe}. In the Gaussian average of a moment $\langle\tr H^k\rangle_J$, Wick contractions pair the Hamiltonian insertions into chords,
and in this limit each crossing of two chords contributes a factor \cite{Berkooz:2018qkz,Berkooz:2018jqr}
\be
\la{2.2}
q=e^{-2p^2/M}=e^{-\lambda},\qquad 0<q<1.
\ee

The chord Hilbert space $\mathcal H$ is spanned by orthonormal states
$\{\ket n\}_{n=0,1,2,\dots}$, labeled by the number of open Hamiltonian chords at a given boundary time, with
$\widehat N\ket n=n\ket n$.\footnote{For quantum-group formulations of the chord algebra see
\cite{Berkooz:2022mfk,Blommaert:2023opb,Schouten:2025tvn,Belaey:2025ijg},
and \cite{Blommaert:2023wad} for the relation to the $q$-Schwarzian.}
The operators $A_+$ and $A_-$, which open and close a
Hamiltonian chord, are defined by
\be
A_+\ket n=\sqrt{1-q^{n+1}}\ket{n+1},\qquad
A_-\ket n=\sqrt{1-q^n}\ket{n-1},\qquad A_-\ket0=0,
\ee
and satisfy the $q$-deformed commutation relation $A_-A_+-qA_+A_-=1-q$.
A Hamiltonian insertion is represented on $\mathcal H$ by the transfer operator $T$:
in the double-scaled limit, the moments of $H$ are
\cite{Berkooz:2018qkz,Berkooz:2018jqr,Berkooz:2024lgq}
\be
\la{2.4}
\big\langle \tr H^{k}\big\rangle_J
=
\mmm{0}{T^{k}}{0},
\qquad
\widehat D\equiv A_++A_-,
\qquad
T=-E_0\,\frac{\widehat D}{2},
\qquad
E_0=\frac{2}{\sqrt{1-q}},
\ee
where $\tr$ is the microscopic trace, normalized to $\tr\mathbb I=1$, $\ket0$ is the
state with no open Hamiltonian chords, and $E_0$ sets the DSSYK spectral edges $\pm E_0$.
The transfer operator has a continuous spectrum, with
energy eigenstates $|\theta\rangle$ satisfying
\be
\la{2.5}
T\ket\theta
=
E(\theta)\ket\theta,
\qquad
E(\theta)=-E_0\cos\theta,
\qquad
0\le\theta\le\pi,
\ee
and wavefunctions\footnote{We use standard notation: $(a;q)_n=\prod_{j=0}^{n-1}(1-aq^j)$ for integers $n\geq0$,
with $(a;q)_0=1$, and
$(a;q)_\infty=\prod_{j\geq0}(1-aq^j)$. Multiple arguments denote a product, and independent $\pm$ signs
denote a product over all sign choices.}
\be
\la{2.6}
\braket{\theta}{n}
=
\frac{H_n(\cos\theta|q)}{\sqrt{(q;q)_n}},
\ee
where $H_n(x|q)$ are the continuous $q$-Hermite polynomials
\cite{Berkooz:2018jqr,gasper2011basic}.
The spectral measure, the DSSYK density of states in $\theta$, and resolution of the identity read
\be
\la{2.7}
\mu(\theta)
=
(q,e^{\pm2i\theta};q)_\infty,
\qquad
\mathbb I
=
\int_0^\pi\frac{d\theta}{2\pi}\,
\mu(\theta)\,
\ket\theta\bra\theta
=
\sum_{n\ge0}\ket n\bra n.
\ee
For a boundary of length $\beta$ in DSSYK units (the inverse temperature) we set
$\wt\beta=\beta E_0$, so that $e^{-\beta T}=e^{\wt\beta\wh D/2}$.

We also need the two-boundary amplitude without matter, which is not a single trace on $\mathcal H$.
In terms of the modified Bessel functions of the first kind,
\be
\la{2.8}
I_b(\wt\beta)=\int_0^\pi\frac{d\theta}{2\pi}\,2\cos(b\theta)\,e^{\wt\beta\cos\theta},\qquad b=0,1,2,\ldots,
\ee
the genus-zero two-boundary amplitude of the matrix model of DSSYK
\cite{Jafferis:2022uhu,Okuyama:2023byh,Okuyama:2023kdo} is
\be
\la{2.9}
Z_{\rm wormhole}(\beta_1,\beta_2)=\sum_{b\geq1}b\,I_b(\wt\beta_1)\,I_b(\wt\beta_2).
\ee
  We call the functions
$I_b(\wt\beta)$ \emph{ordinary trumpets} and $b$ their label.

\subsection{The matter-decorated wormhole}
\la{sec:decorated}

In DSSYK, a matter operator of even size $p'$ is \cite{Berkooz:2018jqr}
\be
\mathcal O=i^{p'/2}\sum_{i_1<\cdots<i_{p'}}
\widetilde J_{i_1\cdots i_{p'}}\psi_{i_1}\cdots\psi_{i_{p'}},
\qquad
\big\langle\widetilde J_{i_1\cdots i_{p'}}^{\,2}\big\rangle_{\widetilde J}
=\binom M{p'}^{-1},
\ee
where the couplings $\widetilde J$ are independent real Gaussian variables with zero
mean, also independent of the Hamiltonian couplings. Their variance gives $\langle\tr\mathcal O^2\rangle_{\widetilde J}=1$.
A matter chord represents the Gaussian contraction of the random
couplings between two insertions of $\mathcal O$.
In the double-scaled limit at fixed $\Delta=p'/p>0$, which in the JT limit is the conformal
dimension of the corresponding boundary operator
\cite{Berkooz:2018jqr,Berkooz:2024lgq}, each crossing
of a matter chord with a Hamiltonian chord contributes
$e^{-2pp'/M}=q^\Delta$. We denote this matter crossing weight by $t\equiv q^\Delta\in(0,1)$ throughout. A matter chord crossing $n$ open Hamiltonian chords thus acquires
the weight $q^{\Delta n}$, and its insertion is represented
on $\mc H$ by $q^{\Delta\wh N}$, whose energy-basis kernel is the matter two-point kernel
\cite{Berkooz:2018jqr}
\be
\la{2.11}
G_\Delta(\theta_1,\theta_2)
\equiv
\mmm{\theta_1}{q^{\Delta\wh N}}{\theta_2}
=
\frac{(q^{2\Delta};q)_\infty}
{(q^\Delta e^{i(\pm\theta_1\pm\theta_2)};q)_\infty}.
\ee

The matter-decorated wormhole amplitude is the genus-zero amplitude of the ETH two-matrix model of DSSYK \cite{Jafferis:2022uhu,Jafferis:2022wez} with one matter operator on each of two boundaries, of lengths $\beta_1$ and $\beta_2$, and the potential of the matter matrix kept to quadratic order.
In this model the eigenvalues of the Hamiltonian have density $\mu(\theta)$ and the matter operator has, in the energy
eigenbasis, the Gaussian propagator $\langle\mc O_{ab}\,\mc O_{cd}\rangle\propto G_\Delta(\theta_a,\theta_b)\,\delta_{ad}\delta_{bc}$.
With the full potential of \cite{Jafferis:2022wez}, which is not Gaussian in the matter matrix, the same amplitude receives further contributions from matter loops around the wormhole.
The amplitude with Gaussian matter is the trace \cite{Okuyama:2023yat}
\be
\la{2.12}
\mc W_\Delta(\beta_1,\beta_2)
=\Tr_{\mc H}\!\left[q^{\Delta\wh N}e^{-\beta_1T}e^{-\beta_2T}\right]
=\Tr_{\mc H}\!\left[e^{-\beta T}q^{\Delta\wh N}\right]
=\sum_{n\ge0}q^{\Delta n}\,\mmm{n}{e^{-\beta T}}{n}.
\ee
Cyclicity brings the two boundary evolutions together, so that the
amplitude depends only on $\beta=\beta_1+\beta_2$, and we write
$\mc W_\Delta(\beta)\equiv\mc W_\Delta(\beta_1,\beta_2)$.
Inserting the spectral resolution \eqref{2.7} and using
the kernel definition \eqref{2.11}, the trace is expressed through the \emph{dressed} spectral weight
\be
\la{2.13}
\mc K_\Delta(\theta)=\mu(\theta)G_\Delta(\theta,\theta),
\ee
as
\be
\la{2.14}
\mc W_\Delta(\beta)
=
\int_0^\pi\frac{d\theta}{2\pi}\,
\mc K_\Delta(\theta)\,
e^{-\beta E(\theta)},
\ee
which is \eqref{1.1} with $\rho_\Delta(E)\,dE=\mc K_\Delta(\theta)\,d\theta/(2\pi)$.
Explicitly, we can write
\be
\la{2.15}
\mc K_\Delta(\theta)
=
\frac{(q;q)_\infty(q^{2\Delta};q)_\infty}
{(q^\Delta;q)_\infty^2}
\frac{(z;q)_\infty(z^{-1};q)_\infty}
{(q^\Delta z;q)_\infty(q^\Delta z^{-1};q)_\infty},
\qquad
z=e^{2i\theta}.
\ee
It is, up to its overall normalization, the orthogonality weight of
the continuous $q$-ultraspherical (Rogers) polynomials $C_m(\cos\theta;t|q)$, given in
\eqref{6.6} \cite{AskeyIsmail:1983Ultraspherical,Ismail:2009OrthogonalPolynomials}. In the matrix model \cite{Okuyama:2023yat}, the representation \eqref{2.14} arises directly: with one matter operator in each of two traces, the propagator sets the two energies equal, leaving the integral over the remaining one.

The trace \eqref{2.12} has an expansion in chord diagrams with two boundaries, whose rules we now state.
 For a single trace, as in
\eqref{2.4}, the insertions lie on one circle and the chords are drawn in the
disk it bounds; here each boundary is a circle, of length $\beta_1$ or
$\beta_2$, and the chords are drawn on the annulus between the two circles.
The circles carry the Hamiltonian insertions of
$e^{-\beta_1 H}$ and $e^{-\beta_2 H}$, while
one matter chord joins the two boundaries. Each Hamiltonian chord
pairs within a boundary, joins the two, or closes on itself around
the annulus (Fig.~\ref{fig:annulus-chords}a).
A chord of the last kind, a closed loop, is attached to no insertion and is not a Wick contraction:
it stands for a chord that remains open all the way around in the trace \eqref{2.12}.
Crossings are weighted as for a single trace, by $q$ for two Hamiltonian chords and by $q^\Delta$ for
a Hamiltonian chord and the matter chord, and this includes the closed loops.\footnote{Other DSSYK wormhole observables with matter are studied in \cite{Aguilar-Gutierrez:2025pqp,Aguilar-Gutierrez:2025hty,Aguilar-Gutierrez:2026ogo}.}

On the disk the pairing of the insertion points determines the number of crossings, counted for straight chords.
On the annulus two things change. First, chords can no longer be drawn straight
(crossings that a deformation can remove are not counted).
Second, the pairing no longer fixes the diagram, since chords with the same endpoints can wind differently around the annulus (Fig.~\ref{fig:annulus-chords}b).
A diagram consists of the matter chord, of Hamiltonian chords joining pairs of insertion points, and of any number of
closed loops, each winding once around the annulus, all considered up to a common twist around the annulus.
Chords with the same endpoints that wind differently are different diagrams, and all are summed over; since the common twist
is not counted, only windings relative to the matter chord are meaningful. Each
closed loop crosses once each chord joining the two boundaries and the
matter chord, and nothing else.
\begin{figure}[t]
\centering
\begin{tikzpicture}[x=1cm,y=1cm,line cap=round,line join=round,font=\small,text=diagramInk]
\tikzset{annHd/.style={circle,fill=diagramInk,inner sep=1.1pt},
  annMd/.style={circle,fill=diagramMatter,inner sep=1.3pt},
  annLd/.style={circle,fill=diagramLoop,inner sep=1.3pt},
  annPt/.style={circle,fill=diagramInk,inner sep=0.9pt}}
\def\annR{1.55}\def\annr{0.50}
\newcommand{\annulusBase}{%
  \fill[diagramWash,even odd rule] (0,0) circle[radius=\annR] (0,0) circle[radius=\annr];
  \draw[very thick,diagramInk] (0,0) circle[radius=\annR];
  \draw[very thick,diagramInk] (0,0) circle[radius=\annr];
  \draw[thick,dashed,diagramMatter,name path=M] (90:\annR)--(90:\annr);
  \node[annMd] at (90:\annR) {}; \node[annMd] at (90:\annr) {};}
\begin{scope}
\annulusBase
\draw[thick,diagramInk] (200:\annR) .. controls (208:1.08) and (242:1.08) .. (250:\annR);
\node[annPt] at (200:\annR) {}; \node[annPt] at (250:\annR) {};
\draw[thick,diagramTeal,name path=T] (328:\annR) .. controls (322:1.15) and (316:0.85) .. (312:\annr);
\node[annPt] at (328:\annR) {}; \node[annPt] at (312:\annr) {};
\draw[line width=1.1pt,diagramLoop,name path=L] (0,0) circle[radius=0.93];
\fill[name intersections={of=L and M,name=a}] (a-1) circle[radius=0] node[annLd]{};
\fill[name intersections={of=L and T,name=b}] (b-1) circle[radius=0] node[annLd]{};
\node at (-1.85,1.45) {\textbf{(a)}};
\node at (150:\annR+0.17) {\scriptsize 1}; \node at (150:\annr-0.17) {\scriptsize 2};
\end{scope}
\begin{scope}[shift={(4.0,0)}]
\annulusBase
\draw[thick,diagramInk,name path=W]
  plot[domain=250:560,samples=160,smooth] (\x:{1.05+(\annR-1.05)*(exp(-(\x-250)/13)+exp(-(560-\x)/13))});
\node[annPt] at (200:\annR) {}; \node[annPt] at (250:\annR) {};
\fill[name intersections={of=W and M,name=c}] (c-1) circle[radius=0] node[annMd]{};
\node at (-1.85,1.45) {\textbf{(b)}};
\end{scope}
\begin{scope}[shift={(6.3,-1.1)}]
\def\annL{4.4}\def\annH{2.2}
\fill[diagramWash] (0,0) rectangle (\annL,\annH);
\draw[very thick,diagramInk] (0,\annH)--(\annL,\annH);
\draw[very thick,diagramInk] (0,0)--(\annL,0);
\draw[thick,dashed,diagramMatter] (0,0)--(0,\annH);
\draw[thick,dashed,diagramMatter] (\annL,0)--(\annL,\annH);
\foreach \x in {0,\annL}{\node[annMd] at (\x,0) {}; \node[annMd] at (\x,\annH) {};}
\draw[line width=1.1pt,diagramLoop,name path=cL] (0,0.9)--(\annL,0.9);
\node[annLd] at (0,0.9) {}; \node[annLd] at (\annL,0.9) {};
\draw[thick,diagramInk,name path=cA] (1.35,\annH) .. controls (1.45,1.35) and (1.85,1.35) .. (1.95,\annH);
\node[annPt] at (1.35,\annH) {}; \node[annPt] at (1.95,\annH) {};
\draw[thick,diagramTeal,name path=cT] (2.91,\annH) .. controls (2.85,1.4) and (2.77,0.8) .. (2.71,0);
\node[annPt] at (2.91,\annH) {}; \node[annPt] at (2.71,0) {};
\fill[name intersections={of=cL and cT,name=g}] (g-1) circle[radius=0] node[annLd]{};
\draw[densely dotted,line width=.8pt,diagramInk] (1.52,0)--(1.52,\annH);
\node[anchor=north] at (1.52,-0.03) {\scriptsize slice};
\node[anchor=west] at (\annL+0.08,\annH) {\scriptsize 1};
\node[anchor=west] at (\annL+0.08,0) {\scriptsize 2};
\node at (-0.42,2.55) {\textbf{(c)}};
\end{scope}
\end{tikzpicture}
\caption{
Chord diagrams on the annulus. The two circles are the boundaries 1 and 2, and the dashed line is the matter chord.
(a)~A Hamiltonian chord pairing two points of one boundary, a chord joining the two boundaries, and a closed loop. The loop
crosses the matter chord (weight $q^\Delta$) and the chord joining the boundaries (weight $q$).
(b)~The same two points paired by a chord that winds around the annulus, and therefore crosses the matter chord.
(c)~Diagram (a) opened along the matter chord into a strip, whose two dashed edges are identified. A slice joining the
two boundaries crosses the open chords, here two, one of which is the closed loop.}
\la{fig:annulus-chords}
\end{figure}
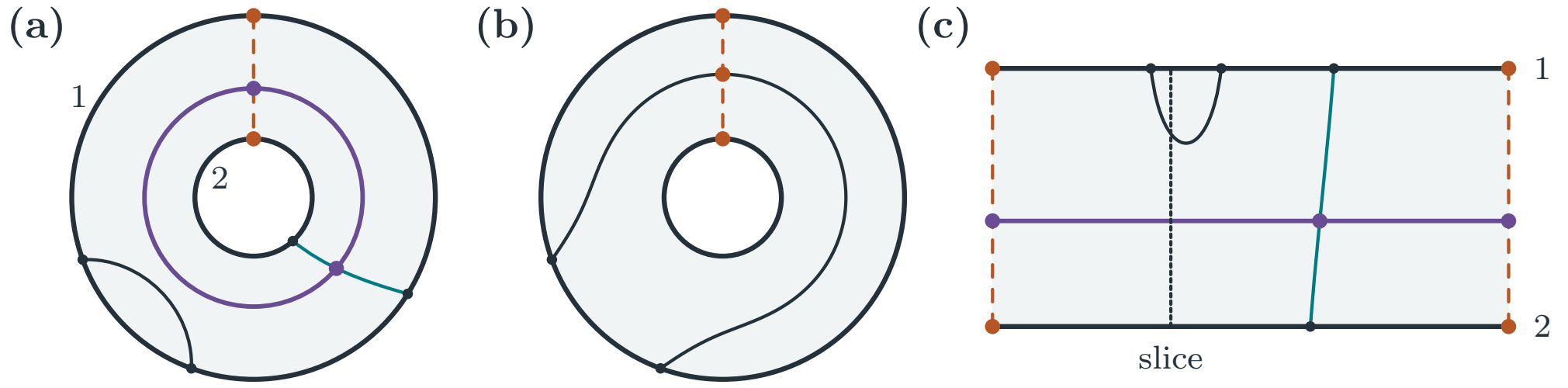

These are the rules generated by the transfer matrix (App.~\ref{app:sectors-histories}). Opening the
annulus along the matter chord gives a strip, whose two edges the trace identifies, and whose slices, joining the
two boundaries, cross a stack of open Hamiltonian chords (Fig.~\ref{fig:annulus-chords}c). The insertions
of boundary~1 open and close chords at one end of the stack, those of
boundary~2 at the other end, and the closed loops are chords that remain open all around the annulus.
In the rescaled basis $\ket n\to\sqrt{(q;q)_n}\,\ket n$, the opening
weight $1$ and the closing weights $(1-q)q^c$, summed over the number $c$
of chords crossed, reproduce at either end of the stack the matrix elements $1$ and $1-q^n$ of $\wh D$, hence the action of $T$. The matter chord, along which the annulus is opened, crosses the whole stack, which gives $q^{\Delta\wh N}$.
 The sum over
diagrams is therefore the trace \eqref{2.12}.

 Unlike the disk rules, the annulus chord rules are not derived here from microscopic SYK:
for light operators with random couplings (fixed $\Delta$), the coupling average of the
product of two SYK traces with matter insertions is not given by these
rules alone, since it also receives non-geometric contributions
\cite{Berkooz:2020fvm} (see also \cite{Bettaque:2026syk} and
Sec.~\ref{sec:sd-bettaque}).

At zero boundary length only closed loops remain, each crossing the
matter chord once, and
$\mathcal W_\Delta(0)=\sum_{n\geq0}q^{\Delta n}=(1-q^\Delta)^{-1}$.
The limit $\Delta\to0$ is singular and does not yield
the undecorated wormhole \eqref{2.9} \cite{Cui:2025sgy}. At $\Delta=0$ the insertion $q^{\Delta\wh N}$ is the identity and \eqref{2.12} becomes the trace of
$e^{-\beta T}$ over $\mc H$, which diverges: for large $n$ the operators $A_\pm$ act as shifts, and one finds
$\mmm{n}{e^{-\beta T}}{n}\to I_0(\wt\beta)$, so that the terms of the sum do not decrease.
In terms of diagrams, closed loops are no longer suppressed.
This divergence, its relation to the JT winding divergence, and how
the undecorated measure $b$ of \eqref{2.9} is recovered once the common winding is quotiented
are discussed in Sec.~\ref{sec:fq-zero-label}.

\section{General spectral properties of gluing identities}
\la{sec:general}

The amplitude \eqref{2.14} depends on $\beta_1+\beta_2$ only and has a positive spectral weight. We collect what follows from
these facts for its decompositions into one-boundary factors. Nothing in this section uses the explicit form
\eqref{2.15} of the weight: the statements hold for any positive integrable weight on a bounded range of energies that is even in the energy, as \eqref{2.15} is. Since the weight vanishes outside a bounded interval of energies, its orthonormal polynomials are complete, which their gluing identity requires; evenness is used only for the absence of a diagonal term in the recurrence \eqref{3.8} and of the order $\wt\beta^{\,m+1}$ in \eqref{3.6}, not for uniqueness.

\paragraph{Gluing identities.}
Consider the inner product
\be
\la{3.1}
\langle f,g\rangle=\int_0^\pi\frac{d\theta}{2\pi}\,\mc K_\Delta(\theta)\,f(\cos\theta)\,g(\cos\theta),
\ee
and let $g_m$, $m\geq0$, be an orthonormal basis of the corresponding real $L^2$ space. For any positive numbers $w(m)$, the functions
$Z(m,\beta)$ defined by
\be
\la{3.2}
\sqrt{w(m)}\,Z(m,\beta)=\int_0^\pi\frac{d\theta}{2\pi}\,\mc K_\Delta(\theta)\,g_m(\cos\theta)\,e^{\wt\beta\cos\theta}
\ee
satisfy, for independent $\beta_1,\beta_2>0$,
\be
\la{3.3}
\mc W_\Delta(\beta_1+\beta_2)=\sum_{m\geq0}w(m)\,Z(m,\beta_1)\,Z(m,\beta_2),
\ee
which is Parseval's identity for the basis. We call \eqref{3.3} a gluing identity, with one-boundary factors $Z(m,\beta)$ and weights $w(m)$. Gluing identities are thus abundant: every orthonormal basis gives one, and so does, more generally, any family with $\sum_m|g_m\rangle\langle g_m|=1$. Clearly, in each of them only the products $\sqrt{w(m)}\,Z(m,\beta)$ are determined, since a
rescaling $Z(m,\beta)\mapsto s_mZ(m,\beta)$ by a $\beta$-independent factor, compensated by $w(m)\mapsto s_m^{-2}w(m)$, leaves
\eqref{3.3} unchanged: the normalization of the factors is invisible to gluing.

\paragraph{The threshold property.}
One-boundary factors $Z(m,\beta)$ have the \emph{threshold property} if $Z(m,\beta)=O(\beta^m)$ as $\beta\to0$ for every $m\geq0$. Among all decompositions of the form \eqref{3.3}, only one has factors with this property, up to the rescaling just described, as we now show. Let $p_m$, $m\geq0$, be the orthonormal polynomials of the weight, of degree $m$ in
$x=\cos\theta$ and with leading coefficient $\zeta_m>0$. Let $Z(m,\beta)$, $m\geq0$, be real functions, entire in $\wt\beta$ as the factors \eqref{3.2} are, with
the threshold property, and $w(m)$ be independent of $\beta$, such that \eqref{3.3} holds term by term in the Taylor
expansion at $\wt\beta_1=\wt\beta_2=0$. This is the case, for instance, when \eqref{3.3} holds for $\beta_1,\beta_2>0$ and its terms have non-negative Taylor coefficients, so that the sum over $m$ and the Taylor expansion can be exchanged, as for the chord sectors of Sec.~\ref{sec:fq-chords}. Then $w(m)>0$ and, for every $m\geq0$,
\be
\la{3.4}
\sqrt{w(m)}\,Z(m,\beta)=\pm\int_0^\pi\frac{d\theta}{2\pi}\,\mc K_\Delta(\theta)\,p_m(\cos\theta)\,e^{\wt\beta\cos\theta},
\ee
where the sign is that of $Z(m,\beta)$ at small $\beta>0$.

The argument is simply Gram--Schmidt orthogonalization in matrix form. Write $Z(m,\beta)=\sum_jz_{mj}\,\wt\beta^{\,j}/j!$, so that $z_{mj}=0$ for $j<m$, and expand
$\mc W_\Delta(\beta_1+\beta_2)=\sum_{i,j}\langle x^i,x^j\rangle\,\wt\beta_1^{\,i}\wt\beta_2^{\,j}/(i!\,j!)$. Term by term, \eqref{3.3} reads
\be
\la{3.5}
\sum_{m\leq\min(i,j)}w(m)\,z_{mi}\,z_{mj}=\langle x^i,x^j\rangle ,
\ee
that is, $H=LDL^{\rm T}$ for the matrix of moments $H_{ij}=\langle x^i,x^j\rangle$, $i,j<N$, with $L_{jm}=z_{mj}$ lower triangular and $D={\rm diag}\,(w(0),\ldots,w(N-1))$. Expanding $x^j$ in the orthonormal polynomials gives a second factorization, $H=AA^{\rm T}$ with $A_{jm}=\langle x^j,p_m\rangle$, also lower triangular. Since the weight is positive, no nonzero polynomial has zero norm, so $H$ is positive definite and $L$ and $A$ are invertible. Then $M=A^{-1}L$ satisfies $DM^{\rm T}=M^{-1}$, where the left-hand side is upper and the right-hand side lower triangular: both are diagonal, and $D=M^{-2}>0$. Hence $w(m)>0$ and $\sqrt{w(m)}\,z_{mj}=\pm\langle x^j,p_m\rangle$, which is \eqref{3.4} order by order in $\wt\beta$. The sign follows from $\langle p_m,x^m\rangle=1/\zeta_m>0$.

\paragraph{Normalization and weights.}
The normalization of the factors $Z(m,\beta)$ in the products $\sqrt{w(m)}\,Z(m,\beta)$ is fixed by a condition on the threshold coefficient $z_{mm}$, the coefficient of $\wt\beta^{\,m}/m!$ in $Z(m,\beta)$. We use the condition
\be
\la{3.6}
Z(m,\beta)=\frac{1}{m!}\Big(\frac{\wt\beta}{2}\Big)^m+O(\wt\beta^{\,m+2}),
\ee
whose leading term is that of the ordinary trumpet $I_m(\wt\beta)$, independently of $\Delta$; it is the normalization of the amplitudes defined in Sec.~\ref{sec:fq-candidate}, by \eqref{5.15}. The right-hand side of \eqref{3.4} is
$\pm\wt\beta^{\,m}/(\zeta_m\,m!)$ up to higher orders, since $\langle p_m,x^m\rangle=1/\zeta_m$, so that \eqref{3.6} gives the plus sign and determines the weights,
\be
\la{3.7}
w(m)=\frac{4^m}{\zeta_m^2}.
\ee
In particular $\zeta_0^{-2}=\mc W_\Delta(0)$, so that $w(0)=\mc W_\Delta(0)$.

\paragraph{Local evolution and positivity.}
The orthonormal polynomials obey the three-term recurrence $2x\,p_m=B_{m+1}p_{m+1}+B_mp_{m-1}$, with $B_m=2\zeta_{m-1}/\zeta_m>0$ for
$m\geq1$ and $B_0=0$. Used under the integral in \eqref{3.4}, it gives for $\widetilde Z_m(\beta)=\sqrt{w(m)}\,Z(m,\beta)$,
with the plus sign,
\be
\la{3.8}
2\,\partial_{\wt\beta}\widetilde Z_m=B_{m+1}\widetilde Z_{m+1}+B_m\widetilde Z_{m-1},
\qquad
\widetilde Z_m\big|_{\wt\beta=0}=\delta_{m,0}\,\sqrt{\mc W_\Delta(0)},
\ee
an evolution in the boundary length that connects only neighboring labels. In terms of the Jacobi matrix $\mathsf J_\Delta$, defined by $(\mathsf J_\Delta\widetilde Z)_m=B_{m+1}\widetilde Z_{m+1}+B_m\widetilde Z_{m-1}$, it reads $2\,\partial_{\wt\beta}\widetilde Z=\mathsf J_\Delta\widetilde Z$, and its solution is $e^{\wt\beta\mathsf J_\Delta/2}$ applied to the initial condition. The matrix has non-negative entries and connects the
label $m=0$, the only one occupied at $\wt\beta=0$, to every other label, so that $\widetilde Z_m(\beta)>0$ for all $m\geq0$ and
$\beta>0$. In the normalization \eqref{3.6}, by \eqref{3.7}, the weights are $w(m)=\mc W_\Delta(0)\prod_{j=1}^mB_j^2$. In the
language of Krylov complexity \cite{Parker:2018yvk,Balasubramanian:2022tpr,Muck:2022xfc,Rabinovici:2025otw}, $m$ is the Krylov
index of this evolution and the $B_m$ are its Lanczos coefficients.

\paragraph{What gluing identities do not determine.}

In this section the label $m$ is a basis index, defined by orthogonalization. Whether it counts anything in the
chord diagrams of Sec.~\ref{sec:decorated}, which normalization the diagrams select, and whether the weights have a combinatorial
meaning are questions that the spectral representation leaves to the diagrams. So is the factorization itself. Sorting the chord diagrams by the number $m$ of Hamiltonian chords that join the two boundaries gives terms that sum to the wormhole and are of order $\beta_1^m\beta_2^m$ at short boundary lengths, since $m$ insertions are needed on each boundary. This does not make each term a product of two one-boundary factors: moving a fraction of the term $m=1$ of the decomposition \eqref{3.4} into the term $m=0$ preserves both properties, and the positivity of each term, but turns the term $m=0$ into a sum of two products.

\section{Gluing the decorated wormhole in JT gravity}
\la{sec:neck-fixed}

In the JT limit of DSSYK, the decorated wormhole decomposes into amplitudes at fixed neck circumference $\mathfrak b$, glued with a flat measure \cite{Saad:2019pqd,Stanford:2020wkf,Jafferis:2022wez}. The two-boundary correlator with the matter line through the neck was computed in \cite[App.~E]{Jafferis:2022wez}, where the integral over the relative twist and the sum over the windings of the matter line produce this measure. The matter is a probe, with the measure of pure JT gravity and no matter determinant.\footnote{Closed-worldline observables winding around the neck, as in \cite{Freivogel:2026loz}, keep the usual $\mathfrak b\,d\mathfrak b$ gluing.}  Here we recast the result of \cite{Jafferis:2022wez} as the gluing of two one-boundary amplitudes at fixed neck, each a Schwarzian path integral with a local insertion at the endpoint of the matter line, and derive the properties of these amplitudes used below; this formulation guides the finite-$q$ construction of Sec.~\ref{sec:fq-candidate}.\footnote{Notice that we use the geometric language of the JT double trumpet, although only the Schwarzian modes of the two boundaries enter, each constrained by $\mathfrak b$.}

\subsection{JT limit of the decorated wormhole}
\la{sec:spectral-ingredients}

The JT limit \cite{Berkooz:2024lgq,Maldacena:2016upp,Engelsoy:2016xyb,Jensen:2016pah} holds the Schwarzian boundary length $\tau$ fixed as $\l\to0$,
\be
\la{4.1}
\text{\sc JT limit:}\qquad
\l\to0,
\qquad
\tau=\beta\l^{3/2}>0
\quad\text{fixed},\qquad \Delta\ \text{fixed}.
\ee
In this regime, we have
\be
\la{4.2}
\wt\beta
=
\beta E_0=\frac{2\beta}{\sqrt{1-e^{-\l}}}
\sim
\frac{2\tau}{\l^2}.
\ee
Following \cite{Okuyama:2023yat,Cui:2025sgy}, we define the JT wormhole
amplitude by\footnote{\la{f:1}%
The factor $e^{-\wt\beta}$ implements the energy shift $E\mapsto E+E_0$, with $\wt\beta=\beta E_0$ kept exact. With $\theta=\l k$ in \eqref{2.14}, $E(\l k)+E_0\sim\l^{3/2}k^2$ at fixed $k$, the Schwarzian momentum, so the shifted Boltzmann factor tends to $e^{-\tau k^2}$. For the weight, with $(q^a;q)_\infty=(q;q)_\infty(1-q)^{1-a}/\Gamma_q(a)$ and $\Gamma_q\to\Gamma$ at complex arguments \cite[App.~B, Thm.~B.2]{Koornwinder:1990Jacobi}, the three factors $(q;q)_\infty$ of $\mu$ cancel against those of $G_\Delta$, the powers of $1-q\sim\l$ combine to $\l^{2\Delta-1}$, and the Jacobian $d\theta=\l\,dk$ supplies one more power of $\l$: this is the origin of the normalization $\l^{-2\Delta}$ in \eqref{4.3}.}
\be
\la{4.3}
\mc W^{\rm JT}_\Delta(\tau)
=
\lim_{\substack{\l\to0^+\\ \beta\l^{3/2}=\tau}}
\l^{-2\Delta}e^{-\wt\beta}\mc W_\Delta(\beta).
\ee
Introducing $\theta=\l k$ and taking the limit under the $k$-integral, we obtain from \eqref{4.3}
\be
\la{4.4}
\mc W^{\rm JT}_\Delta(\tau)
=\int_0^\infty dk\,\rho_{\rm Sch}(k)\,e^{-\tau k^2}\,\mc V_\Delta(k,k),
\ee
with
\be
\la{4.5}
\rho_{\rm Sch}(k)
=
\frac{k\sinh(2\pi k)}{\pi^2},
\qquad
\mc V_\Delta(k,k)
=
\frac{
\Gamma(\Delta)^2
\Gamma(\Delta+2ik)\Gamma(\Delta-2ik)
}{
\Gamma(2\Delta)
}.
\ee
Here $\rho_{\rm Sch}$ is the Schwarzian spectral density
in momentum space, and $\mathcal V_\Delta(k,k)$ is the diagonal
value of the two-energy JT matter kernel $\mathcal V_\Delta(k,k')$ of
\cite{Jafferis:2022wez}.
The amplitude \eqref{4.4} is the JT wormhole with a matter line of \cite{Stanford:2020wkf,Jafferis:2022wez}, and the JT limit of the trace \eqref{2.12} was
taken, up to normalization, in \cite{Okuyama:2023yat}.
Using $|\Gamma(2ik)|^{-2}=2k\sinh(2\pi k)/\pi$,
we can write $\mc W^{\rm JT}_\Delta(\tau)$
in a form useful for later discussion of neck gluing:
\bea
\la{4.6}
\mc W^{\rm JT}_\Delta(\tau)
&=C_\Delta^{\rm Sch}\int_0^\infty dk\,
\big|h_\Delta^{\rm Sch}(k)\big|^2
e^{-\tau k^2},
\\
h_\Delta^{\rm Sch}(k)&\equiv\frac{\Gamma(\Delta+2ik)}{\Gamma(2ik)},\qquad
C_\Delta^{\rm Sch}
=\frac{\Gamma(\Delta)^2}{2\pi\,\Gamma(2\Delta)}.
\eea
Note that, by \eqref{B.7}, the integrand of \eqref{4.6} is the pointwise limit of the integrand of \eqref{2.14}, rescaled as in \eqref{4.3}.

\subsection{The geometric neck cut}
\la{sec:neck-cut}

In the JT limit \eqref{4.1} the boundaries have Schwarzian lengths $\tau_i=\lambda^{3/2}\beta_i>0$, and we work at a fixed geodesic neck circumference $\mathfrak b$.
When discussing a single boundary, we suppress the index $i$
and denote its length simply by $\tau$.
The Euclidean double trumpet (the JT wormhole of Sec.~\ref{sec:spectral-ingredients}) is locally AdS$_2$. With the AdS$_2$ radius set to one, its metric is $ds^2=d\sigma^2+\cosh^2\sigma\,dx^2$,
with $x\sim x+\mathfrak b$, the closed geodesic neck at $\sigma=0$, and asymptotic boundaries at $\sigma\to\pm\infty$.
Each asymptotic boundary, with renormalized time $u$ of period $\tau_i$ and
induced metric $du^2/\epsilon^2$ as $\epsilon\to0$, winds once around the
neck, $x(u+\tau_i)=x(u)+\mathfrak b$.
Up to a constant shift of $x$, the classical boundary profile,
which extremizes the Schwarzian boundary action, is
$x=\mathfrak b u/\tau_i$.

\paragraph{Boundary fluctuations and the cut.}

Without matter, the integral over the fluctuations of an asymptotic boundary at fixed
neck circumference gives the JT trumpet amplitude \cite{Saad:2019JTMatrixIntegral},
\be
\la{4.7}
Z_{\rm trumpet}^{\rm JT}(\mathfrak b,\tau)
=\int_{-\infty}^{\infty}\frac{dk}{2\pi}\,e^{-ik\mathfrak b-\tau k^2}
=\frac{e^{-\mathfrak b^2/(4\tau)}}{\sqrt{4\pi\tau}}.
\ee
The matter propagator depends on the boundary profile at the endpoints of the matter line, so we make the Schwarzian path
integral explicit: the boundary profile $x(u)$ introduced above is
now integrated over, with the same winding condition
\cite{Mertens:2019JTDefects} and modulo rigid translations
$x\to x+{\rm const}$, which change only the origin of $x$.
The allowed trajectories are monotonic, $x'(u)\equiv dx/du>0$, and we introduce
\be
\varphi(u)\equiv\log x'(u).
\ee
Since $dx=x'(u)\,du$, a piece $du$ of the boundary covers the range $e^{\varphi}du$ of the neck coordinate.
We call $e^{\varphi}$ the local stretching of the boundary; by the winding
condition its average over the boundary is $\mathfrak b/\tau$.
The Schwarzian boundary dynamics then becomes Liouville quantum mechanics \cite{Mertens:2018fds},\footnote{We use the Schwarzian derivative
$\{f,u\}=f'''/f'-\frac32(f''/f')^2$, with primes denoting
$u$ derivatives. In our conventions the hyperbolic Schwarzian action is
$-\frac12\int du\,\{e^x,u\}$. With $x'=e^\varphi$, the identity $\{e^x,u\}=\varphi''-\frac12\varphi'^2-\frac12e^{2\varphi}$ and periodicity of $\varphi$ give the action in \eqref{4.9} after dropping the total derivative.} subject to the fixed-neck constraint
\be
\la{4.9}
I=\frac14\int_0^{\tau}du\,\big(\varphi'^2+e^{2\varphi}\big),
\qquad
\int_0^{\tau}du\,e^{\varphi}=\mathfrak b.
\ee
The configuration space of these profiles is the hyperbolic coadjoint
orbit of the Virasoro group associated with the fixed circumference
$\mathfrak b$, and the Schwarzian measure $\mc D\mu$ is the invariant measure induced
by its symplectic form
\cite{Stanford:2017thb,Mertens:2019JTDefects}.
In the variable $\varphi$, it differs from the constrained
measure $\mc D\varphi$ by a neck-dependent factor,
\be
\la{4.10}
\mc D\mu=\mathcal J(\mathfrak b)\,\mc D\varphi\;
\delta\!\left(\int_0^\tau du\,e^\varphi-\mathfrak b\right),
\ee
where $\mc D\varphi$ has the standard quantum-mechanical normalization
for the kinetic term $\frac14\int du\,\varphi'^2$.
The Schwarzian orbit factor $\mathcal J(\mathfrak b)$ is known, see \eg \cite{Jafferis:2022wez},
\be
\la{4.11}
\mathcal J(\mathfrak b)=2\sinh\frac{\mathfrak b}{2}.
\ee
We normalize the fixed-neck expectation value of a boundary
functional $\mathcal O_{\rm bdy}[\varphi]$ by
\be
\la{4.12}
\big\langle\mc O_{\rm bdy}\big\rangle_{\mathfrak b,\tau}
\equiv
\frac{1}{Z_{\rm trumpet}^{\rm JT}(\mathfrak b,\tau)}
\int_{\text{fixed }\mathfrak b}\mc D\mu\;
\mc O_{\rm bdy}\,e^{-I},
\qquad
\int_{\text{fixed }\mathfrak b}\mc D\mu\;e^{-I}
=Z_{\rm trumpet}^{\rm JT}(\mathfrak b,\tau).
\ee
At fixed $\mathfrak b$, the factor $\mathcal J(\mathfrak b)$
cancels between the numerator and denominator of
\eqref{4.12}. Normalized expectation values can therefore be
computed
using just $\mathcal D\varphi$ with the
fixed-neck constraint. The factor $\mathcal J(\mathfrak b)$ remains in
$Z^{\rm JT}_{\rm trumpet}$, hence in $Z^{\rm fix}_\Delta$, and is not
part of the gluing measure.

The matter line contributes the boundary-to-boundary propagator $e^{-\Delta L}$, with $L$ the renormalized length of the geodesic
joining its endpoints, summed over its winding sectors around the neck. Together with this sum, the integral over the
relative twist of the two boundaries unfolds the relative position of the endpoints over the real line, and leaves the factor
$\Gamma(\Delta)^2/\Gamma(2\Delta)$, independent of the boundary profiles and of the neck circumference, times an endpoint factor $[x_i'(u_i)]^\Delta$ on each boundary, with $u_i$ the endpoint of the matter line on boundary $i$ (App.~\ref{app:neck-unfold}). The two boundary path integrals therefore decouple at fixed
neck.
When discussing either trumpet separately, we denote
its marked endpoint by $u_0$.\footnote{The insertion point $u_0$ is held fixed, but its value is irrelevant:
the fixed-neck ensemble is translation invariant along the boundary
circle.}
 The remaining insertion
is then $[x'(u_0)]^\Delta=e^{\Delta\varphi(u_0)}$.
The resulting geometric gluing formula \cite{Saad:2019pqd,Stanford:2020wkf,Jafferis:2022wez} is finally
\bea
\la{4.13}
\mc W^{\rm JT}_\Delta(\tau_1+\tau_2)
&=
\frac{\Gamma(\Delta)^2}{\Gamma(2\Delta)}
\int_0^{\infty}d\mathfrak b\;
Z^{\rm fix}_\Delta(\mathfrak b,\tau_1)\,
Z^{\rm fix}_\Delta(\mathfrak b,\tau_2),
\\
Z^{\rm fix}_\Delta(\mathfrak b,\tau)
&=
Z_{\rm trumpet}^{\rm JT}(\mathfrak b,\tau)\,
\big\langle e^{\Delta\varphi(u_0)}\big\rangle_{\mathfrak b,\tau}.
\eea
We call $Z^{\rm fix}_\Delta$ the fixed-neck JT amplitude; it is positive, since the fixed-neck measure is. The left-hand side of \eqref{4.13} is defined by the geometric path integral, with the propagator normalized as $e^{-\Delta L}$. By \eqref{A.12}, with the exact $Z^{\rm fix}_\Delta$ of Sec.~\ref{sec:neck-generic} it coincides with \eqref{4.6}, normalization included, which justifies the notation.

\paragraph{The case $\Delta=1$.}
At $\Delta=1$, translation invariance and the fixed-neck constraint determine the one-point function exactly,
\be
\la{4.14}
\big\langle e^{\varphi(u_0)}\big\rangle_{\mathfrak b,\tau}
=
\frac1\tau\int_0^{\tau}du\,
\big\langle e^{\varphi(u)}\big\rangle_{\mathfrak b,\tau}
=
\frac{\mathfrak b}{\tau}.
\ee
Hence
$Z^{\rm fix}_1=(\mathfrak b/\tau)\,Z_{\rm trumpet}^{\rm JT}$ for every
$\mathfrak b$ and $\tau>0$, and \eqref{4.13} evaluates to
$\mc W^{\rm JT}_1=(\tau_1+\tau_2)^{-3/2}/(2\sqrt\pi)$, in agreement
with the spectral representation \eqref{4.6}, where $|h^{\rm Sch}_1(k)|^2=4k^2$. The classical value $\mathfrak b/\tau$ of the insertion is
therefore exact at $\Delta=1$, including all Schwarzian boundary fluctuations, as a consequence of the constraint and translation invariance alone, not of the dynamics (see also Sec.~\ref{sec:fq-D1}).

\paragraph{Small-$\tau$ limit and probe-geodesic value.}
At fixed $\mathfrak b$ and small $\tau$ the boundary is classical,
$x=\mathfrak b u/\tau$, and the insertion takes its \emph{probe-geodesic
value}, $\big\langle e^{\Delta\varphi(u_0)}\big\rangle_{\mathfrak b,\tau}\simeq(\mathfrak b/\tau)^\Delta$, the endpoint factor of a probe geodesic
on this background (App.~\ref{app:JT-classical}), with corrections that vanish as $\tau\to0$.

\subsection{Exact expression for \texorpdfstring{$Z^{\rm fix}_\Delta$}{Zfix}}
\la{sec:neck-generic}

The fixed-neck JT amplitude follows from the computation of \cite{Jafferis:2022wez} at fixed boundary energy $k^2$, for general $\Delta>0$, with the boundary-particle propagator \cite{Yang:2018gdb,Kitaev:2018wpr}. In terms of the ratio $h_\Delta^{\rm Sch}$ of \eqref{4.6}, it is
\be
\la{4.15}
Z^{\rm fix}_\Delta(\mathfrak b,\tau)
=\frac1\pi\int_0^\infty dk\,
e^{-\tau k^2}|h_\Delta^{\rm Sch}(k)|\chi_{\Delta,k}(\mathfrak b),
\qquad \tau>0,
\ee
where the neck profiles $\chi_{\Delta,k}(\mathfrak b)$, given in closed
hypergeometric form in \eqref{A.6}, carry the neck dependence at fixed
boundary energy $k^2$, have unit oscillatory amplitude at large
$\mathfrak b$, and are orthogonal in $\mathfrak b$ \eqref{A.11}.

\paragraph{P\"oschl--Teller equation and small-neck behavior.}
The hypergeometric expression \eqref{A.6} implies that $\chi_{\Delta,k}$ is an
eigenfunction of a P\"oschl--Teller operator,
\be
\la{4.16}
H^{\rm PT}_\Delta\ \chi_{\Delta,k}
=k^2\chi_{\Delta,k},\qquad
H^{\rm PT}_\Delta
=-\partial_{\mathfrak b}^2
+\frac{\Delta(\Delta-1)}{4\sinh^2(\mathfrak b/2)}.
\ee
This operator acts on the neck circumference only. Its spectral parameter is the same $k^2$ that appears in the
thermal factor, so \eqref{4.15} also implies the neck
heat equation
$\partial_\tau Z^{\rm fix}_\Delta
=-H^{\rm PT}_\Delta Z^{\rm fix}_\Delta$.
Of the two solutions of \eqref{4.16} near $\mathfrak b=0$, behaving as $\mathfrak b^\Delta$
and $\mathfrak b^{1-\Delta}$, the expression \eqref{A.6} selects the first
(App.~\ref{app:neck-bp}), as does, independently, the JT limit of the
finite-$q$ amplitudes (App.~\ref{app:fq-JT}). Expanding $\chi_{\Delta,k}$ at small
$\mathfrak b$ under the integral in \eqref{4.15}, and using \eqref{4.6}, gives the small-neck expansion
\be
\la{4.17}
Z^{\rm fix}_\Delta(\mathfrak b,\tau)
=\frac{\mathfrak b^\Delta}{\Gamma(\Delta)}
\mc W^{\rm JT}_\Delta(\tau)+O(\mathfrak b^{\Delta+2}).
\ee

\section{Decorated trumpet amplitudes from the inserted brane trace}
\la{sec:fq-candidate}

The next step is to extend the JT gluing construction to finite $q$
by finding one-boundary amplitudes that reconstruct the
matter-decorated wormhole and have the correct fixed-neck
JT limit.

\subsection{From the neck constraint to the Morse thermal trace}
\la{sec:neck-Morse-bridge}

Still within JT gravity, the fixed-neck amplitudes of Sec.~\ref{sec:neck-fixed} have a second representation, which is the starting point of our construction.

By using \eqref{4.10}, \eqref{4.12} and \eqref{4.13}, we get
\be
\la{5.1}
Z^{\rm fix}_\Delta(\mathfrak b,\tau)
=\mathcal J(\mathfrak b)
\int_{\varphi(\tau)=\varphi(0)}\mathcal D\varphi\;
\delta\!\left(\int_0^\tau du\,e^\varphi-\mathfrak b\right)
e^{\Delta\varphi(u_0)}\,e^{-I}.
\ee
In the Laplace transform of $Z^{\rm fix}_\Delta/\mathcal J$, with conjugate variable $\kappa>0$, the division by $\mathcal J$
cancels the prefactor in \eqref{5.1}, and the delta function does
the $\mathfrak b$ integral:
\be
\la{5.2}
\begin{split}
&\int_0^\infty d\mathfrak b\,
\frac{e^{-\kappa\mathfrak b}}{2\sinh(\mathfrak b/2)}
Z^{\rm fix}_\Delta(\mathfrak b,\tau)
=\int_{\varphi(\tau)=\varphi(0)}\mathcal D\varphi\,
e^{\Delta\varphi(u_0)}
\exp\!\left[-I-\kappa\int_0^\tau du\,e^\varphi\right],
\end{split}
\ee
where the constraint has been replaced by the additional potential term
$\kappa e^\varphi$.
The periodic path integral is a thermal trace in quantum mechanics with Morse potential and the insertion $e^{\Delta\varphi}$ at the time $u_0$ of the matter endpoint:
\bea
\la{5.3}
\int_0^\infty d\mathfrak b\,
\frac{e^{-\kappa\mathfrak b}}{2\sinh(\mathfrak b/2)}
Z^{\rm fix}_\Delta(\mathfrak b,\tau)
&=\Tr\!\left(e^{-\tau H_\kappa}e^{\Delta\varphi}\right),
\qquad
H_\kappa =-\partial_\varphi^2+\frac14e^{2\varphi}+\kappa e^\varphi.
\eea

The Laplace variable $\kappa$ can be read as the
JT EOW-brane tension: for a geodesic brane of length
$\mathfrak b$, the worldline action
$\kappa\int ds=\kappa\mathfrak b$ reproduces the Laplace weight $e^{-\kappa\mathfrak b}$ in \eqref{5.3}. Multiplied by $d\mathfrak b$, the neck kernel $e^{-\kappa\mathfrak b}/\mathcal J(\mathfrak b)$ is the closed-brane measure of \cite{Gao:2021uro,Okuyama:2023byh,Belaey:2023jtr}.%
\footnote{The same function of $\mathfrak b$, with $\kappa=\Delta'-\frac12$, appears in the
one-loop determinant of a matter field on the double trumpet, in the term of a particle of dimension $\Delta'$ circling the neck
\cite{Jafferis:2022wez}. There it is integrated against the product of the two one-boundary factors, and both
boundaries remain; here it acts on one factor and trades the neck for a brane.}
Eq.~\eqref{5.3}
is the relation \eqref{1.2} of the Introduction. Besides the path-integral derivation, we prove it directly in App.~\ref{app:neck-bp}, as an identity between spectral densities valid for $\kappa>-\frac12$, which confirms the neck kernel and its normalization.

\subsection{The brane expansion as a discrete Laplace inversion}
\la{sec:fq-motivation}

To arrive at a finite-$q$ definition of the one-boundary
amplitudes, we start from the observation that the DSSYK EOW-brane
problem provides a finite-$q$ counterpart of
the Morse Hamiltonian, and that, without matter, its brane trace, expanded in the brane parameter, gives the ordinary trumpets
\cite{Okuyama:2023byh}.

In the DSSYK description, the EOW brane is represented by a
boundary state that allows Hamiltonian chords to end on it,
with dimensionless weight $a$ per chord \cite{Okuyama:2023byh}, the brane parameter.
We organize the resulting chord diagrams by a slice at an
intermediate step of the transfer-matrix construction.
The chord states $|n\rangle$ of Sec.~\ref{sec:dssyk} now count the
open Hamiltonian chords across this slice.
In this basis, the EOW brane modifies the transfer matrix by
adding to the usual chord-opening and chord-closing terms a diagonal
term $aq^n$, from a Hamiltonian chord that crosses the $n$ open chords
and ends on the brane.

The brane transfer operator $T_a$ that generates boundary evolution reads
\be
\la{5.4}
\wh D_a=A_++A_-+a\,q^{\wh N},
\qquad
T_a=-E_0\,\frac{\wh D_a}{2},
\qquad
e^{-\beta T_a}=e^{\frac{\wt\beta}{2}\wh D_a},
\ee
in the conventions of \eqref{2.4}.
We parametrize the brane parameter $a$ by
\be
\la{5.5}
a=q^{\kappa+\frac12}=e^{-\l(\kappa+\frac12)},
\ee
where $\kappa$ is the Laplace transform variable, \cf \eqref{5.3}.
The expansion in $a$ used below is therefore an expansion around infinite tension.

The brane amplitude without a matter insertion is the thermal trace
$\Tr_{\mc H}[e^{\frac{\wt\beta}{2}\wh D_a}]$. Its only divergence is in the $a$-independent term, the trace of $e^{-\beta T}$, which diverges as shown at the end of Sec.~\ref{sec:decorated}.
Subtracting the $a=0$ trace leaves a finite
amplitude with the ordinary-trumpet expansion
\cite{Okuyama:2023byh}
\ba
\la{5.6}
Z_{\rm brane}^{\rm sub}(a,\beta)
&\equiv\Tr_{\mc H}\!\left[e^{\frac{\wt\beta}{2}\wh D_a}
-e^{\frac{\wt\beta}{2}\wh D_0}\right]
=\sum_{b\geq1}\frac{a^b}{1-q^b}\,I_b(\wt\beta).
\ea
Using \eqref{5.5} and setting $\mathfrak b=\l b$, we have
\be
\la{5.7}
\frac{a^b}{1-q^b}
=\frac{e^{-\kappa\mathfrak b}}{2\sinh(\mathfrak b/2)}.
\ee
The coefficient $a^b/(1-q^b)$ of $I_b$ in \eqref{5.6} is therefore exactly
the Laplace weight times the inverse orbit factor $1/\mc J(\mathfrak b)$ of the
JT neck transform
\eqref{5.3}, with the neck integral replaced by a
discrete sum. In a sense, extracting Taylor coefficients at $a=0$ is the discrete
counterpart of inverting that Laplace transform.

\paragraph{Morse limit.}
The Morse Hamiltonian arises as a continuum limit of $\wh D_a$, the brane JT
limit of \cite{Okuyama:2023byh}. With the slice coordinate
$\varphi_n=-\l n-\log\l$, \ie $q^n=\l e^{\varphi_n}$, and with $\varphi_n$ and
$\kappa$ held fixed as $\l\to0$, the action of $\wh D_a$ on smooth profiles
$f(\varphi_n)$ gives
\be
\la{5.8}
\frac{2-\wh D_a}{\l^2}\ \longrightarrow\
H_\kappa,
\qquad
\frac{q^{\wh N}}{\l}\ \longrightarrow\ e^\varphi,
\ee
 where $H_\kappa$ was defined in \eqref{5.3}. Since
$\widetilde\beta\lambda^2/2\to\tau$ by \eqref{4.2}, the first limit gives
$e^{-\wt\beta}e^{\wt\beta\wh D_a/2}\to e^{-\tau H_\kappa}$.\footnote{Since $2-\wh D_a=2(T_a+E_0)/E_0$, this is the origin of the factor $e^{-\wt\beta}$ in the JT limit \eqref{8.1} of the amplitudes, as in \eqref{4.3}.}

Having identified the formal JT limit of the brane transfer
operator (used here only as motivation),
we now include the matter insertion. The matter line, running from its boundary endpoint to the brane, crosses the open chords of the slice where it is inserted, with $q^\Delta$ per crossing, as discussed in Sec.~\ref{sec:decorated}. This gives the insertion $q^{\Delta\wh N}$ in the brane trace. By \eqref{5.8}, $q^{\Delta\wh N}/\l^\Delta$ tends to the insertion $e^{\Delta\varphi}$ of \eqref{5.3}.

\subsection{Definition of the amplitudes}
\la{sec:fq-definition}

As the discrete inversion of the neck transform suggests, we define the amplitudes by expanding the inserted brane trace around $a=0$,\footnote{\la{fn:trace-class}%
The trace \eqref{5.9} is finite and entire in the brane parameter $a$.
For $0<q<1$ the operators $A_\pm$ and $q^{\wh N}$ have norm at most one,
so $e^{\wt\beta\wh D_a/2}$ is bounded and entire in $a$ and $\wt\beta$, while
$t^{\wh N}$ is trace class.}
\be
\la{5.9}
F_\Delta(a,\beta)
\equiv\Tr_{\mc H}\!\left[q^{\Delta\wh N}\,
e^{\frac{\wt\beta}{2}\wh D_a}\right].
\ee
Namely,
\be
\la{5.10}
\text{definition of $Z_{\Delta}$:}\qquad
F_\Delta(a,\beta)
=\sum_{m\geq0}\frac{a^m}{1-q^{m+\Delta}}\;
Z_\Delta(m,\beta).
\ee
 We call
$Z_\Delta(m,\beta)$ the \emph{matter-decorated trumpet amplitudes}, or
simply the amplitudes. Here ``trumpet'' and ``neck'' are names for
features of chord diagrams: the neck is where the $m$ Hamiltonian chords
counted by $a^m$ end on the brane and, in the wormhole, where the chords
joining the two boundaries pass from one to the other
(Sec.~\ref{sec:fq-sectors}).

Since
$\wh D_0=\wh D$ by \eqref{5.4}, the trace at $a=0$ is the wormhole \eqref{2.12},
$F_\Delta(0,\beta)=\mc W_\Delta(\beta)$, and the $m=0$ (zero-label) amplitude is
\be
\la{5.11}
Z_\Delta(0,\beta)=(1-q^\Delta)\,\mc W_\Delta(\beta).
\ee
In the gluing sum, this term contributes $(1-q^\Delta)\,\mc W_\Delta(\beta_1)\,\mc W_\Delta(\beta_2)$, the diagrams in which only the matter line joins the two boundaries (Sec.~\ref{sec:fq-zero-label}). Each factor $\mc W_\Delta(\beta_i)$ is the wormhole with the other boundary shrunk to zero length, whose diagrams consist of chords with both ends on boundary $i$ and of closed loops.

By \eqref{5.7} at $b=m+\Delta$, the coefficient $a^m/(1-q^{m+\Delta})$ in \eqref{5.10} is $a^{-\Delta}$ times the JT neck kernel of \eqref{5.3} at $\mathfrak b=\mathfrak b_m$, where
\be
\la{5.12}
\mathfrak b_m=\l\,(m+\Delta).
\ee
We call $\mathfrak b_m$ the \emph{shifted label} and use it throughout the
rest of the paper.

\subsection{Normalization and closed loops}
\la{sec:fq-normalization}

The normalization of the amplitudes $Z_\Delta$ in the expansion \eqref{5.10} is set by the denominator $1-q^{m+\Delta}$. Gluing does not fix it, as discussed in Sec.~\ref{sec:general}, and neither does the JT limit \eqref{8.1}, which some rescalings $Z_\Delta(m,\beta)\mapsto s_mZ_\Delta(m,\beta)$, $w_\Delta(m)\mapsto s_m^{-2}w_\Delta(m)$ preserve.\footnote{\la{fn:denominator}For example, replacing the denominator by
$1-q^{m+c}$ with $c>0$, \ie $s_m=(1-q^{m+c})/(1-q^{m+\Delta})$, gives
$s_m\to1$ when $\l m\to\mathfrak b>0$.} We now show that the choice made in \eqref{5.10} corresponds to a criterion from the chord expansion of the brane trace, which separates closed loops around the neck from the one-boundary factor.

\paragraph{Chord expansion of the brane trace.}
In the expansion of the exponential in \eqref{5.9},
$e^{\wt\beta\wh D_a/2}=\sum_{k\geq0}\frac{1}{k!}\big(\frac{\wt\beta}{2}\big)^k\wh D_a^{\,k}$, the power
$\wh D_a^{\,k}$ is the sum over all ordered sequences of $k$
terms, each one of $A_+$, $A_-$ or $a\,q^{\wh N}$. On the chord state
$\ket n$, each term is a \emph{move} of the number $n$ of open
Hamiltonian chords. The trace is unchanged by the diagonal change of
basis $\ket n\to\sqrt{(q;q)_n}\,\ket n$, after which the matrix
elements are
\be
\la{5.13}
\begin{array}{l@{\qquad}l@{\qquad}l}
\text{open:} & n\to n+1, & \text{weight }1,\\[2pt]
\text{close:} & n\to n-1, & \text{weight }1-q^{n},\\[2pt]
\text{neck:} & n\to n, & \text{weight }a\,q^{n}.
\end{array}
\ee
A term of the expansion corresponds to a starting value $n_0$ and a sequence of $k$ moves
returning to $n_0$, as the trace requires. Its weight is the product
of the weights \eqref{5.13} times $q^{\Delta n_0}$ from the matter
insertion, which by cyclicity may be placed at the slice where the
trace starts. Each neck move carries one factor $a$, so the
coefficient of $a^m(\wt\beta/2)^k/k!$ in $F_\Delta(a,\beta)$ is the sum,
over all such terms with $k$ moves of which exactly $m$ are neck moves,
of their weights stripped of the factor $a^m$.

These weights can be read as crossings in the standard way \cite{Ismail:1987qHermite} by picturing the $n$ open chords as
horizontal lines stacked at heights $1,\dots,n$ in the strip between
the asymptotic boundary (top) and the brane (bottom), with boundary
time running from left to right, as in Fig.~\ref{fig:stack-rules}. We then split the
closing weight as
$1-q^n=(1-q)\sum_{p=1}^{n}q^{\,n-p}$ and read the term $p$ as ``close the chord at height $p$''. A diagram
is then equivalently $n_0$ together with the sequence of moves, each closing
labeled by its height $p\in\{1,\dots,n\}$, with weight
$(1-q)q^{n-p}$.

Each such labeled sequence is drawn uniquely by updating the stack with two rules: an opened chord is put on top, and when the chord at height $p$ is closed it leaves the stack and every chord above it moves down by one, so that the order of the chords never changes (Fig.~\ref{fig:stack-rules}). In this drawing the powers of $q$ count crossings. A closing chord at height $p$ leaves upwards towards the boundary and crosses the $n-p$ chords above it. An opening chord comes down from the boundary onto the top of the stack and crosses nothing. A neck move is a vertical line from the boundary to the brane, which we call a \emph{neck line}, and crosses all $n$ chords. The matter insertion is a vertical line at the starting slice, which crosses all $n_0$ chords, with weight $q^\Delta$ per crossing. In addition, each closing, \ie each chord with both ends on the boundary, carries a factor $1-q$, and each neck move a factor $a$. These are the chord-diagram rules of DSSYK, with the brane as the lower edge of the strip.

\begin{figure}[!t]
\centering
\resizebox{0.85\textwidth}{!}{%
\begin{tikzpicture}[x=1cm,y=1cm,line cap=round,line join=round,
  bd/.style={very thick,diagramInk},
  br/.style={thick,diagramInk!55},
  st/.style={thick,diagramInk,rounded corners=4pt},
  qd/.style={circle,fill=diagramInk,inner sep=1.1pt},
  lab/.style={font=\scriptsize},
  hl/.style={font=\scriptsize,gray},
]
\begin{scope}
  \draw[bd] (0,3.2)--(5,3.2); \draw[br] (0,0)--(5,0);
  \node[lab,anchor=south] at (2.5,3.25) {asymptotic boundary};
  \node[lab,anchor=north] at (2.5,-0.05) {brane (neck)};
  \foreach \h/\y in {1/0.6,2/1.2,3/1.8,4/2.4} \node[hl,anchor=east] at (-0.05,\y) {\h};
  \fill[diagramInk!8] (0,0.02) rectangle (5,0.9);
  \draw[st] (0,0.6)--(5,0.6);
  \draw[st] (0,1.2)--(1.6,1.2)--(1.6,3.2);
  \draw[st] (0,1.8)--(2.0,1.8)--(2.45,1.2)--(5,1.2);
  \draw[st] (0,2.4)--(2.0,2.4)--(2.45,1.8)--(5,1.8);
  \draw[st] (3.6,3.2)--(3.6,2.4)--(5,2.4);
  \node[qd] at (1.6,1.8) {}; \node[qd] at (1.6,2.4) {};
  \foreach \n/\y in {a/0.6,b/1.2,c/1.8,d/2.4} \node[lab,anchor=south] at (0.6,\y) {\n};
  \foreach \n/\y in {a/0.6,c/1.2,d/1.8,e/2.4} \node[lab,anchor=south] at (4.6,\y) {\n};
  \node[lab,anchor=west] at (1.65,2.85) {close $p=2$};
  \node[lab,anchor=west] at (3.65,2.85) {open};
  \node[lab,diagramInk!70,anchor=west] at (2.5,2.15) {$-1$};
  \node[lab,diagramInk!70,anchor=west] at (2.5,1.55) {$-1$};
  \node[lab,diagramInk!60,anchor=west] at (2.0,0.32) {below $p$: untouched};
  \node[font=\footnotesize,align=left,anchor=west,text width=6.6cm] at (5.6,1.6) {%
    \textbf{Stack rules}\par\smallskip
    $\bullet$ Closing the chord at height $p$: it crosses the $n-p$ chords above it.\par
    $\bullet$ The chords above $p$ move down by one.\par
    $\bullet$ The chords below $p$ do not move.\par
    $\bullet$ An opened chord goes on top.\par
    $\bullet$ The order of the chords never changes.\par\smallskip
    Hence a chord moves only \emph{down}, and only when a chord \emph{below} it is closed.};
\end{scope}
\end{tikzpicture}}
\caption{Stack rules of the chord expansion of the brane trace. When
the chord at height $p$ is closed (here $p=2$), it crosses the chords
above it, which then move down by one; chords below it are not touched
(gray band). An opened chord always enters on top.}
\la{fig:stack-rules}
\end{figure}

\paragraph{Closed chord loops.}
The trace identifies the two edges of the strip height by height (the
dotted edges of Fig.~\ref{fig:loops}). A \emph{closed loop} around the neck is a chord that is open at the start, is never closed within the period, and
returns to its starting height. Such a loop crosses the $m$ neck lines
and the matter line once each, and nothing else. The weight of a loop is therefore $q^{m+\Delta}$,
independently of the rest of the diagram (Fig.~\ref{fig:loops}).

Since chords move only down, and only when a chord below them is
closed, the loops are the bottom block of the initial stack that no
closing reaches; a never-closed chord above a closed one ends lower than
it started and is not a loop (Fig.~\ref{fig:loops}(c)). Deleting the
lowest loop lowers every other chord by one and leaves the closing
weights $(1-q)q^{n-p}$ unchanged, so the diagrams with $r$ loops are the
diagrams without closed loops multiplied by $(q^{m+\Delta})^r$, and summing over
the number $r$ of closed loops factorizes the inserted trace as
\be
\la{5.14}
F_\Delta(a,\beta)=\sum_{m\geq0}a^m\sum_{r\geq0}
\big(q^{m+\Delta}\big)^{r}\,Z^{\rm lf}_\Delta(m,\beta)
=\sum_{m\geq0}\frac{a^m}{1-q^{m+\Delta}}\,Z^{\rm lf}_\Delta(m,\beta),
\ee
where $Z^{\rm lf}_\Delta(m,\beta)$ is the sum over chord diagrams with
$m$ neck lines which are  \emph{loop-free}, \ie with no closed loop around the neck. Comparison with
\eqref{5.10} gives $Z_\Delta(m,\beta)=Z^{\rm lf}_\Delta(m,\beta)$, \ie
the amplitude is the loop-free part of the trace.

Our normalization in \eqref{5.10} thus means that closed loops, which never reach the asymptotic boundary, are assigned to the neck rather than to the one-boundary factor. This is the \emph{loop criterion} anticipated in the Introduction.%
\footnote{Without matter, the loop criterion is the removal of the vacuum bubbles in
the chord diagrams of the brane trace of \cite[App.~F]{Cao:2025pir} (Sec.~\ref{sec:fq-relation}).}

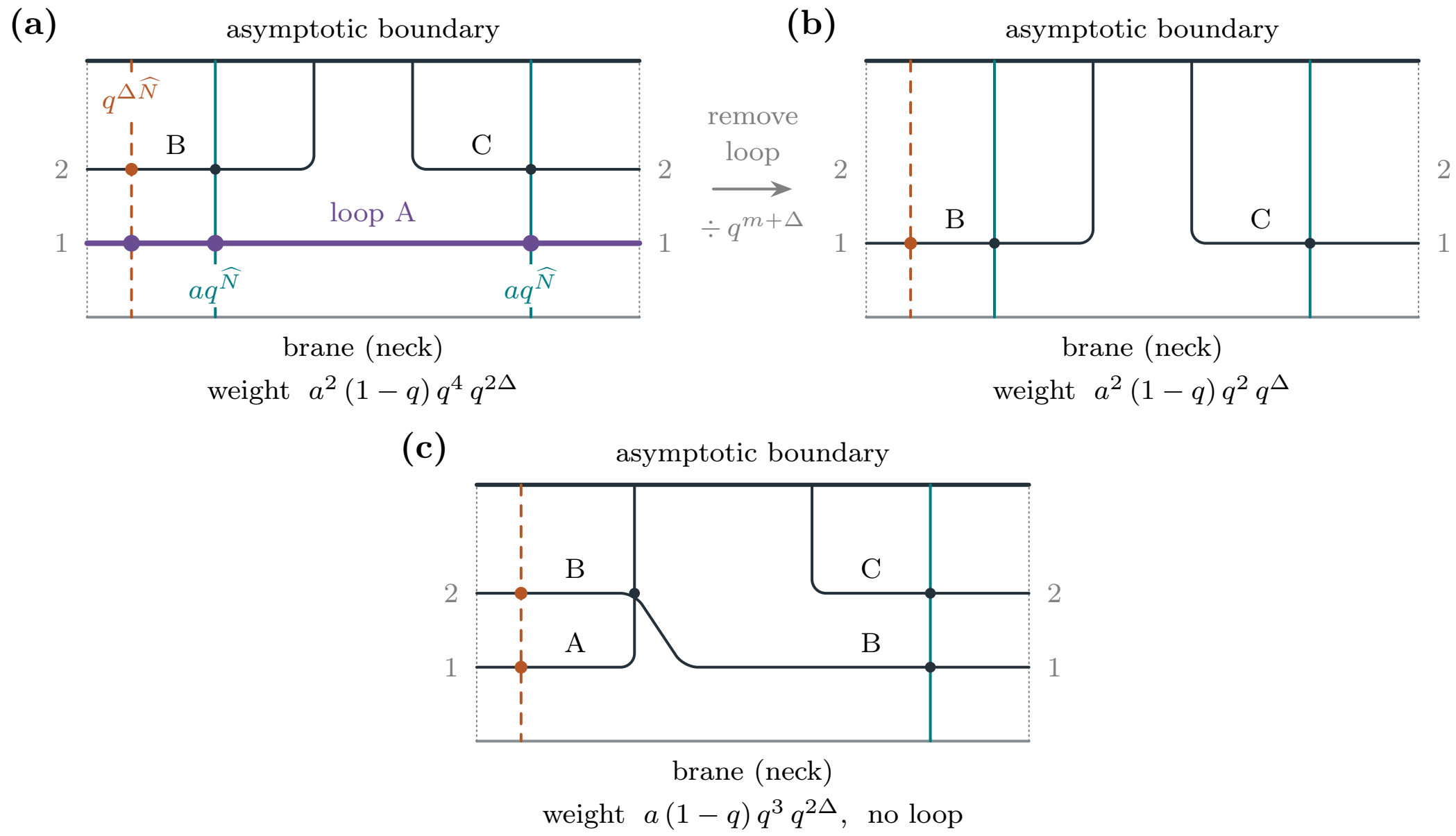
\begin{figure}[!t]
\centering
\resizebox{0.9\textwidth}{!}{%
\begin{tikzpicture}[x=1cm,y=1cm,line cap=round,line join=round,
  bd/.style={very thick,diagramInk},
  br/.style={thick,diagramInk!55},
  st/.style={thick,diagramInk,rounded corners=4pt},
  lp/.style={line width=1.6pt,diagramLoop},
  nk/.style={thick,diagramTeal},
  mt/.style={thick,dashed,diagramMatter},
  qd/.style={circle,fill=diagramInk,inner sep=1.1pt},
  md/.style={circle,fill=diagramMatter,inner sep=1.3pt},
  ld/.style={circle,fill=diagramLoop,inner sep=1.7pt},
  lab/.style={font=\scriptsize},
]
\newcommand{\loopstrip}[3]{%
  \begin{scope}[shift={(#1,#2)}]
    \draw[bd] (0,2.6)--(5.6,2.6);
    \draw[br] (0,0)--(5.6,0);
    \draw[densely dotted,gray] (0,0)--(0,2.6);
    \draw[densely dotted,gray] (5.6,0)--(5.6,2.6);
    \node[lab,anchor=south] at (2.8,2.65) {asymptotic boundary};
    \node[lab,anchor=north] at (2.8,-0.05) {brane (neck)};
    \node[font=\small\bfseries,anchor=south east] at (-0.15,2.65) {#3};
    \foreach \h/\y in {1/0.75,2/1.5}{
      \node[lab,gray,anchor=east] at (-0.05,\y) {\h};
      \node[lab,gray,anchor=west] at (5.65,\y) {\h};}
  \end{scope}}
\loopstrip{0}{0}{(a)}
\begin{scope}
  \draw[lp] (0,0.75)--(5.6,0.75);
  \draw[st] (0,1.5)--(2.3,1.5)--(2.3,2.6);
  \draw[st] (3.3,2.6)--(3.3,1.5)--(5.6,1.5);
  \draw[mt] (0.45,2.6)--(0.45,0);
  \draw[nk] (1.3,2.6)--(1.3,0);
  \draw[nk] (4.5,2.6)--(4.5,0);
  \node[md] at (0.45,1.5) {}; \node[qd] at (1.3,1.5) {}; \node[qd] at (4.5,1.5) {};
  \node[ld] at (0.45,0.75) {}; \node[ld] at (1.3,0.75) {}; \node[ld] at (4.5,0.75) {};
  \node[lab,diagramLoop,anchor=south] at (2.9,0.78) {loop A};
  \node[lab,anchor=south] at (0.9,1.52) {B};
  \node[lab,anchor=south] at (4.0,1.52) {C};
  \node[lab,diagramMatter,fill=white,inner sep=1pt] at (0.45,2.25) {$q^{\Delta\wh N}$};
  \node[lab,diagramTeal,fill=white,inner sep=1pt] at (1.3,0.32) {$aq^{\wh N}$};
  \node[lab,diagramTeal,fill=white,inner sep=1pt] at (4.5,0.32) {$aq^{\wh N}$};
  \node[lab,anchor=north] at (2.8,-0.45) {weight $\;a^2\,(1-q)\,q^{4}\,q^{2\Delta}$};
\end{scope}
\draw[-{Stealth[length=2.2mm]},thick,gray] (6.35,1.3)--(7.15,1.3);
\node[lab,gray,align=center,anchor=south] at (6.75,1.4) {remove\\loop};
\node[lab,gray,anchor=north] at (6.75,1.2) {$\div\, q^{m+\Delta}$};
\loopstrip{7.9}{0}{(b)}
\begin{scope}[shift={(7.9,0)}]
  \draw[st] (0,0.75)--(2.3,0.75)--(2.3,2.6);
  \draw[st] (3.3,2.6)--(3.3,0.75)--(5.6,0.75);
  \draw[mt] (0.45,2.6)--(0.45,0);
  \draw[nk] (1.3,2.6)--(1.3,0);
  \draw[nk] (4.5,2.6)--(4.5,0);
  \node[md] at (0.45,0.75) {}; \node[qd] at (1.3,0.75) {}; \node[qd] at (4.5,0.75) {};
  \node[lab,anchor=south] at (0.9,0.77) {B};
  \node[lab,anchor=south] at (4.0,0.77) {C};
  \node[lab,anchor=north] at (2.8,-0.45) {weight $\;a^2\,(1-q)\,q^{2}\,q^{\Delta}$};
\end{scope}
\loopstrip{3.95}{-4.3}{(c)}
\begin{scope}[shift={(3.95,-4.3)}]
  \draw[st] (0,0.75)--(1.6,0.75)--(1.6,2.6);
  \draw[st] (0,1.5)--(1.6,1.5)--(2.1,0.75)--(5.6,0.75);
  \draw[st] (3.4,2.6)--(3.4,1.5)--(5.6,1.5);
  \draw[mt] (0.45,2.6)--(0.45,0);
  \draw[nk] (4.6,2.6)--(4.6,0);
  \node[md] at (0.45,0.75) {}; \node[md] at (0.45,1.5) {};
  \node[qd] at (1.6,1.5) {};
  \node[qd] at (4.6,0.75) {}; \node[qd] at (4.6,1.5) {};
  \node[lab,anchor=south] at (1.0,0.77) {A};
  \node[lab,anchor=south] at (1.0,1.52) {B};
  \node[lab,anchor=south] at (4.0,0.77) {B};
  \node[lab,anchor=south] at (4.0,1.52) {C};
  \node[lab,anchor=north] at (2.8,-0.45) {weight $\;a\,(1-q)\,q^{3}\,q^{2\Delta}$, \ no loop};
\end{scope}
\end{tikzpicture}}
\caption{
Chord diagrams of the inserted brane trace: asymptotic boundary on top, brane at the bottom, boundary time
from left to right, heights in gray, and dotted edges identified by
the trace. Vertical teal lines are neck lines $aq^{\wh N}$, the dashed
line is the matter line $q^{\Delta\wh N}$, and dots mark crossings,
of weight $q$ (black) or $q^\Delta$ (orange).
(a)~A diagram with $m=2$ neck lines starting with two
open chords. The lowest chord A returns to height $1$ and is
a closed loop around the neck; its crossings (violet) with the two neck lines and the matter
line give the factor $q^{m+\Delta}$. Chord B is closed and
C is opened; after the identification they form a single chord with both ends
on the boundary. (b)~The same sequence of moves with the loop deleted: every other crossing is unchanged, and the weight is smaller by $q^{2+\Delta}$. (c)~A
chord that is never closed but is not a loop: closing A, which crosses B, lets
B drop from height $2$ to height $1$, so that after the identification B continues
as A, and C as B.}
\la{fig:loops}
\end{figure}

\paragraph{Threshold property.}
At lowest order in the boundary length, the loop criterion reduces to a simple condition on the threshold coefficient. At order $a^m$, the diagrams with the minimal number $m$ of
Hamiltonian insertions consist of the $m$ neck lines alone. Any chord open at the start would be a closed loop, since all insertions are neck lines and none of them closes it, so in the loop-free diagram nothing is crossed and the weight is one: only the factor $(\wt\beta/2)^m/m!$ of the $m$-th order term of $e^{\wt\beta\wh D_a/2}$ remains. Hence
\be
\la{5.15}
Z_\Delta(m,\beta)=\frac{1}{m!}\Big(\frac{\wt\beta}{2}\Big)^m+O(\wt\beta^{\,m+2}),
\ee
where the next order requires one opening and one closing, hence two
more insertions. The amplitude and the ordinary trumpet $I_m(\wt\beta)$ thus have the same leading term at small $\wt\beta$, which does not depend on $\Delta$: this is the condition we used in \eqref{3.6}.
By itself, it fixes the normalization, since the rescalings $Z_\Delta(m,\beta)\mapsto s_mZ_\Delta(m,\beta)$ considered at the beginning of this subsection multiply the threshold coefficient by $s_m$, and it is thus equivalent to the loop criterion.

\section{Sector factorization}
\la{sec:fq-chords}

We now ask how the amplitudes \eqref{5.10}, defined by a one-boundary trace, are related to the wormhole itself. In other words, we examine whether the chord expansion of the wormhole, sorted by the number of Hamiltonian chords joining the two boundaries, is reproduced term by term by gluing the one-boundary amplitudes $Z_\Delta(m,\beta)$ with a weight that accounts for what happens at the neck, \ie far from the boundaries.

\subsection{The sectors and the neck weight}
\la{sec:fq-sectors}

The wormhole $\mc W_\Delta(\beta_1+\beta_2)$ has the chord expansion of Sec.~\ref{sec:decorated}. Opened along the matter chord into the strip of Fig.~\ref{fig:annulus-chords}c, whose edges the trace identifies, and not cut at the neck, it has a stack expansion of the same kind as the brane trace of Sec.~\ref{sec:fq-normalization}, with the stack running from boundary~1 (top) to boundary~2 (bottom). Insertions of boundary~1 open and close chords at the top, those of boundary~2 at the bottom.
 In the expansion in powers of $\wt\beta/2$, as in \eqref{5.13}, each term is an initial stack together with a sequence of moves, each closing labeled by the height of the chord it closes, as in Sec.~\ref{sec:fq-normalization}; we call such a term a \emph{history} of the stack. These histories are now the chord diagrams on the annulus of Sec.~\ref{sec:decorated}. They have weight $q$ per crossing of two Hamiltonian chords, $q^\Delta$ per crossing with the matter line, and $1-q$ per Hamiltonian chord with both endpoints on the boundaries.
 For a single boundary this is the correspondence between transfer-matrix histories and chord diagrams of
\cite{Ismail:1987qHermite}, and it extends directly to the annulus (App.~\ref{app:sectors-histories}). In a history the factors come in the order of the moves, while in the diagram they are attached to the crossings, which is what allows us to sort them by region (Fig.~\ref{fig:hist-conf}).

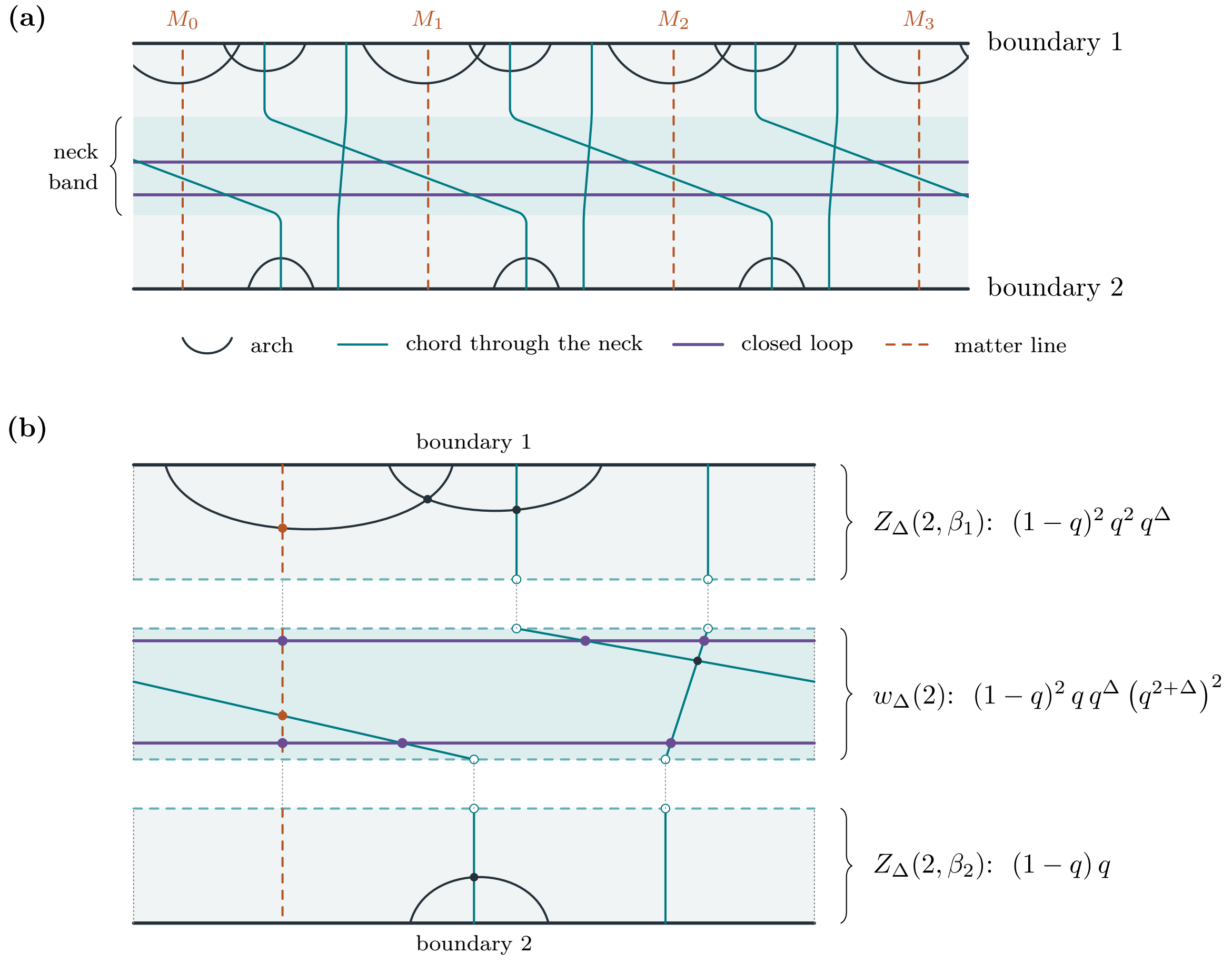
\begin{figure}[!tp]
\centering
\resizebox{\textwidth}{!}{%
\begin{tikzpicture}[x=1.05cm,y=1.05cm,line cap=round,line join=round,font=\small]
\fill[diagramWash] (-0.6,0) rectangle (9.6,3.0);
\fill[diagramTeal!13] (-0.6,0.9) rectangle (9.6,2.1);
\draw[very thick,diagramInk] (-0.6,3.0)--(9.6,3.0);
\draw[very thick,diagramInk] (-0.6,0)--(9.6,0);
\node[anchor=west] at (9.7,3.0) {boundary 1};
\node[anchor=west] at (9.7,0) {boundary 2};
\begin{scope}
\clip (-0.6,-0.01) rectangle (9.6,3.01);
\draw[line width=1.1pt,diagramLoop] (-0.6,1.15)--(9.6,1.15);
\draw[line width=1.1pt,diagramLoop] (-0.6,1.55)--(9.6,1.55);
\foreach \x in {0,3,6,9}{\draw[thick,dashed,diagramMatter] (\x,0)--(\x,3.0);}
\foreach \s in {-3,0,3,6,9}{
  \draw[thick,diagramInk] (\s+0.5,3.0) .. controls (\s+0.65,2.55) and (\s+1.35,2.55) .. (\s+1.5,3.0);
  \draw[thick,diagramInk] (\s+2.2,3.0) .. controls (\s+2.5,2.35) and (\s+3.4,2.35) .. (\s+3.7,3.0);
  \draw[thick,diagramInk] (\s+0.8,0) .. controls (\s+0.95,0.5) and (\s+1.45,0.5) .. (\s+1.6,0);
}
\foreach \s in {-3,0,3,6,9}{
  \draw[thick,diagramTeal,rounded corners=3pt] (\s+1.0,3.0)--(\s+1.0,2.1)--(\s+4.2,0.9)--(\s+4.2,0);
  \draw[thick,diagramTeal,rounded corners=3pt] (\s+2.0,3.0)--(\s+2.0,2.1)--(\s+1.9,0.9)--(\s+1.9,0);
}
\end{scope}
\foreach \x/\l in {0/0,3/1,6/2,9/3}{
  \node[diagramMatter,anchor=south] at (\x,3.05) {\scriptsize $M_{\l}$};}
\draw[decorate,decoration={brace,amplitude=4pt}] (-0.75,0.9)--(-0.75,2.1);
\node[anchor=east,align=right] at (-0.9,1.5) {\scriptsize neck\\[-2pt]\scriptsize band};
\begin{scope}[shift={(0,-0.6)}]
  \draw[thick,diagramInk] (0,0) .. controls (0.1,-0.25) and (0.5,-0.25) .. (0.6,0);
  \node[anchor=west] at (0.7,-0.08) {\scriptsize arch};
  \draw[thick,diagramTeal] (1.9,-0.08)--(2.5,-0.08);
  \node[anchor=west] at (2.6,-0.08) {\scriptsize chord through the neck};
  \draw[line width=1.1pt,diagramLoop] (6.0,-0.08)--(6.6,-0.08);
  \node[anchor=west] at (6.7,-0.08) {\scriptsize closed loop};
  \draw[thick,dashed,diagramMatter] (8.6,-0.08)--(9.2,-0.08);
  \node[anchor=west] at (9.3,-0.08) {\scriptsize matter line};
\end{scope}
\node[font=\small\bfseries] at (-1.9,3.3) {(a)};
\node[font=\small\bfseries] at (-1.9,-1.72) {(b)};
\begin{scope}[shift={(-0.6,-7.75)},xscale=1.3]
\tikzset{cutd/.style={circle,draw=diagramTeal,fill=white,inner sep=1.1pt},
  hd/.style={circle,fill=diagramInk,inner sep=1.1pt},
  mdd/.style={circle,fill=diagramMatter,inner sep=1.2pt},
  ldd/.style={circle,fill=diagramLoop,inner sep=1.3pt}}
\fill[diagramWash] (0,4.2) rectangle (6.4,5.6);
\draw[very thick,diagramInk] (0,5.6)--(6.4,5.6);
\draw[thick,diagramTeal!60,dashed] (0,4.2)--(6.4,4.2);
\draw[densely dotted,gray] (0,4.2)--(0,5.6); \draw[densely dotted,gray] (6.4,4.2)--(6.4,5.6);
\node[anchor=south] at (3.2,5.62) {\scriptsize boundary 1};
\draw[thick,dashed,diagramMatter,name path=Mt] (1.4,5.6)--(1.4,4.2);
\draw[thick,diagramTeal,name path=Nt1] (3.6,5.6)--(3.6,4.2);
\draw[thick,diagramTeal] (5.4,5.6)--(5.4,4.2);
\draw[thick,diagramInk,name path=A1] (0.3,5.6) .. controls (0.5,4.55) and (2.8,4.55) .. (3.0,5.6);
\draw[thick,diagramInk,name path=A2] (2.4,5.6) .. controls (2.6,4.85) and (4.2,4.85) .. (4.4,5.6);
\fill[name intersections={of=A1 and Mt,name=i}] (i-1) circle[radius=0] node[mdd]{};
\fill[name intersections={of=A1 and A2,name=j}] (j-1) circle[radius=0] node[hd]{};
\fill[name intersections={of=A2 and Nt1,name=k}] (k-1) circle[radius=0] node[hd]{};
\foreach \x in {3.6,5.4} \node[cutd] at (\x,4.2) {};
\fill[diagramTeal!13] (0,2.0) rectangle (6.4,3.6);
\draw[thick,diagramTeal!60,dashed] (0,3.6)--(6.4,3.6);
\draw[thick,diagramTeal!60,dashed] (0,2.0)--(6.4,2.0);
\draw[densely dotted,gray] (0,2.0)--(0,3.6); \draw[densely dotted,gray] (6.4,2.0)--(6.4,3.6);
\draw[line width=1.1pt,diagramLoop,name path=L1] (0,3.45)--(6.4,3.45);
\draw[line width=1.1pt,diagramLoop,name path=L2] (0,2.2)--(6.4,2.2);
\draw[thick,dashed,diagramMatter,name path=Mb] (1.4,3.6)--(1.4,2.0);
\draw[thick,diagramTeal,name path=Ca] (3.6,3.6)--(6.4,2.95);
\draw[thick,diagramTeal,name path=Cb] (0,2.95)--(3.2,2.0);
\draw[thick,diagramTeal,name path=C2] (5.4,3.6)--(5.0,2.0);
\fill[name intersections={of=Ca and C2,name=a}] (a-1) circle[radius=0] node[hd]{};
\fill[name intersections={of=Cb and Mb,name=b}] (b-1) circle[radius=0] node[mdd]{};
\foreach \P in {Mb,Ca,C2}{\fill[name intersections={of=L1 and \P,name=c}] (c-1) circle[radius=0] node[ldd]{};}
\foreach \P in {Mb,Cb,C2}{\fill[name intersections={of=L2 and \P,name=d}] (d-1) circle[radius=0] node[ldd]{};}
\foreach \x in {3.6,5.4} \node[cutd] at (\x,3.6) {};
\foreach \x in {3.2,5.0} \node[cutd] at (\x,2.0) {};
\fill[diagramWash] (0,0) rectangle (6.4,1.4);
\draw[very thick,diagramInk] (0,0)--(6.4,0);
\draw[thick,diagramTeal!60,dashed] (0,1.4)--(6.4,1.4);
\draw[densely dotted,gray] (0,0)--(0,1.4); \draw[densely dotted,gray] (6.4,0)--(6.4,1.4);
\node[anchor=north] at (3.2,-0.02) {\scriptsize boundary 2};
\draw[thick,dashed,diagramMatter] (1.4,1.4)--(1.4,0);
\draw[thick,diagramTeal,name path=Nb1] (3.2,1.4)--(3.2,0);
\draw[thick,diagramTeal] (5.0,1.4)--(5.0,0);
\draw[thick,diagramInk,name path=B1] (2.6,0) .. controls (2.75,0.75) and (3.75,0.75) .. (3.9,0);
\fill[name intersections={of=B1 and Nb1,name=e}] (e-1) circle[radius=0] node[hd]{};
\foreach \x in {3.2,5.0} \node[cutd] at (\x,1.4) {};
\foreach \x in {1.4,3.6,5.4} \draw[densely dotted,diagramInk!50] (\x,3.6)--(\x,4.2);
\foreach \x in {1.4,3.2,5.0} \draw[densely dotted,diagramInk!50] (\x,1.4)--(\x,2.0);
\draw[decorate,decoration={brace,amplitude=4pt}] (6.65,5.6)--(6.65,4.2);
\node[anchor=west,align=left] at (6.85,4.9) {$Z_\Delta(2,\beta_1)$:\ \ $(1-q)^2\,q^2\,q^{\Delta}$};
\draw[decorate,decoration={brace,amplitude=4pt}] (6.65,3.6)--(6.65,2.0);
\node[anchor=west,align=left] at (6.85,2.8) {$w_\Delta(2)$:\ \ $(1-q)^2\,q\,q^{\Delta}\,\big(q^{2+\Delta}\big)^2$};
\draw[decorate,decoration={brace,amplitude=4pt}] (6.65,1.4)--(6.65,0);
\node[anchor=west,align=left] at (6.85,0.7) {$Z_\Delta(2,\beta_2)$:\ \ $(1-q)\,q$};
\end{scope}
\end{tikzpicture}}
\caption{In both panels boundary~1 is at the top, boundary~2 at the bottom and the band around the neck in between. (a)~A chord diagram of the decorated wormhole in the universal
cover of the annulus, a strip shown over three periods, with
two chords through the neck per period (teal), two closed loops
(violet), arches on both boundaries (black) and the lifts $M_\ell$ of
the matter line (dashed orange). The chords through the neck are drawn vertical outside the shaded band around the neck, which also contains the loops. One of the
two chords through the neck winds once around the neck relative to the
matter line, so that each of its lifts crosses one $M_\ell$; the other
does not wind. (b)~The same diagram over one turn around the neck (dotted edges identified), cut along the edges of the band. Near each boundary, the chords through the neck become vertical neck lines and give a loop-free diagram of the brane trace, a term of $Z_\Delta(2,\beta_i)$; the band carries the relative twist, the windings and the loops, a term of $w_\Delta(2)$. Filled dots mark the crossings, and the weights shown multiply to $(1-q)^5q^{8+4\Delta}$, the weight of the diagram in (a), as in \eqref{6.2}.}
\label{fig:cover}
\end{figure}

Let $S_m(\beta_1,\beta_2)$ denote the sector of chord diagrams with $m$ chords through the
neck, \ie the sum over diagrams with exactly $m$ chords joining the two
boundaries,
\be
\la{6.1}
S_m(\beta_1,\beta_2)=\sum_{K_1,K_2\geq0}
\frac{(\wt\beta_1/2)^{K_1}(\wt\beta_2/2)^{K_2}}{K_1!\,K_2!}
\sum_{C}\omega(C),
\ee
where $C$ runs over the chord configurations with $K_1$ and $K_2$
insertion points on the two boundaries and $m$ chords joining them (any number of closed loops included) and $\omega(C)$ is the
weight \eqref{D.2} given to $C$ by the chord rules above. Summing
over $m$ gives $\sum_mS_m=\mc W_\Delta(\beta_1+\beta_2)$.
Our main result is that each sector is a product of two one-boundary amplitudes.
\begin{theorem}[Sector factorization]
\la{prop:sectors}
For every $m\geq0$, $0<q<1$, $\Delta>0$ and $\beta_1,\beta_2>0$, the
sector $S_m$ factorizes into the loop-free one-boundary amplitudes of
Sec.~\ref{sec:fq-normalization},
\be
\la{6.2}
S_m(\beta_1,\beta_2)=w_\Delta(m)\,Z_\Delta(m,\beta_1)\,
Z_\Delta(m,\beta_2),
\ee
 with the positive weight, independent of $\beta_1,\beta_2$,
\be
\la{6.3}
w_\Delta(m)=\frac{(1-q)^m}{1-q^{m+\Delta}}
\sum_{\sigma\in\mathsf S_m}\sum_{\bm k\in\mathbb Z^m}
q^{N(\sigma,\bm k)}\,q^{\Delta\sum_i|k_i|}.
\ee
The sum runs over the relative twists, encoded in the permutation $\sigma$, and the windings $\bm k=(k_1,\dots,k_m)$ of the $m$ chords through the neck, whose mutual crossings are counted by
\be
\la{6.4}
N(\sigma,\bm k)=\sum_{i<j}|k_i-k_j+\chi_{ij}|, \ \text{with }
\chi_{ij}=1 \text{ if } \sigma(i)>\sigma(j) \text{ and $0$ otherwise,}
\ee
and the factor $1/(1-q^{m+\Delta})$ is the sum over closed loops.
\end{theorem}
The twists and windings are made precise below.
 The proof of the theorem is given in App.~\ref{app:sectors}. Below we sketch the argument on the universal cover of the annulus, reading the weight of each diagram off a convenient drawing of it (Fig.~\ref{fig:cover}). It uses two properties of the diagrams (App.~\ref{app:sectors-histories}): the weight of a diagram is fixed by how the lifts of its chords interleave, whatever the drawing, and every choice of chords and windings occurs exactly once.

\paragraph{Factorization.}
We draw the diagrams in the universal cover of the annulus (Fig.~\ref{fig:cover}), the strip $\mathbb R_{x}\times[0,1]$ with boundary~1 at height $1$, boundary~2 at height $0$ and
deck translation $x\mapsto x+1$, one turn around the neck. Every curve on the annulus is repeated periodically in the cover, as copies that we call its lifts; those of the matter line are the segments $M_\ell$ at $x=\ell\in\mathbb Z$,
and we call the interval $(\ell,\ell+1)$ a period. The chords with both endpoints on one boundary, which we call arches, stay
near their boundary. The chords through the neck can be drawn vertical
outside a band around the neck, since the weights do not depend on the drawing, so that they cross one another and the
matter line only inside the band, where the closed loops also lie.

Near boundary $i=1,2$, a chord through the neck crosses exactly the arches of that boundary that straddle the endpoint of the chord there. If $n$ arch lifts straddle this endpoint (an arch whose lift spans more than one period may straddle it with several lifts), these crossings therefore cost $q^n$. This is the weight of a neck line crossing $n$ open chords in the loop-free brane trace.
 Likewise the arches straddling the matter endpoint cross the matter line, with weight $q^\Delta$ each, as the open chords cross the insertion $q^{\Delta\wh N}$ in the brane trace. Hence the arches of boundary~$i$ with the $m$ endpoints form a loop-free diagram of the brane trace, in which the band plays the role of the brane. The diagrams on either side thus sum to the loop-free one-boundary factor $Z_\Delta(m,\beta_i)$.
 What happens in the band, the
crossings of the chords through the neck with one another and with the
matter line and the closed loops, is independent of what happens near
the boundaries, since the two can be chosen independently, so each sector is a product (Fig.~\ref{fig:cover}(b)),
which is \eqref{6.2}, with $w_\Delta(m)$ the sum over the band. The neck label $m$ thus counts
the Hamiltonian chords that cross the neck, and the neck lines of the
brane trace are these chords seen from one side.

\paragraph{The neck weight $w_\Delta(m)$.}
The sum over the band depends on the relative order and the windings of the chords through the neck. To write it explicitly, we label the chords through the neck $1,\ldots,m$ by the order of their endpoints along boundary~1, starting from the matter endpoint. We let $\sigma(i)$ be the rank of the endpoint of chord $i$ along boundary~2, again starting from the matter endpoint. Since the matter endpoints fix an origin on both boundaries, the labels are not identified under cyclic shifts, and the permutation $\sigma\in\mathsf S_m$ is the relative twist.

Each chord through the neck has exactly one lift that starts in the
period $(0,1)$ of boundary~1. If the lift of chord $i$ ends in the
period $(k_i,k_i+1)$ of boundary~2, the integer $k_i$ is the winding of
chord $i$ around the neck relative to the matter line, and we collect the windings in $\bm k=(k_1,\dots,k_m)\in\mathbb Z^m$.
This lift crosses $M_\ell$ for $|k_i|$ values of $\ell$ --- the sign of $k_i$ records only the direction of the winding.
We draw the chords so that two lifts cross at most once, \ie without the removable pairs of crossings that a wiggled drawing would add, and count crossings up to common deck translation. The number $N(\sigma,\bm k)$ of crossings of the $m$ chords with one another then depends only on $\sigma$ and $\bm k$: two lifts cross exactly when the order of their endpoints is reversed between the two boundaries, which gives \eqref{6.4}.
In Fig.~\ref{fig:cover}(b), for instance, $\sigma$ is the identity and $\bm k=(1,0)$, so that $N=1$.

Each of the $m$ chords carries the closing weight $1-q$, which the neck lines of the brane trace do not carry. Each closed loop crosses the $m$ chords and the matter line, and so contributes $q^{m+\Delta}$; \eg in Fig.~\ref{fig:cover}(b) the two loops give $q^{2(m+\Delta)}$. Including both, with the sum over the number of loops, the sum over the band is the weight \eqref{6.3}. Its double sum converges, since $q^{N(\sigma,\bm k)}\leq1$ and $\sum_{\bm k\in\mathbb Z^m}q^{\Delta\sum_i|k_i|}=[(1+q^\Delta)/(1-q^\Delta)]^m$. It is evaluated in closed form in Corollary~\ref{cor:spectral}.

Since $\sum_mS_m=\mc W_\Delta(\beta_1+\beta_2)$, summing \eqref{6.2} over $m$ gives the gluing identity, for all
$0<q<1$, $\Delta>0$ and independent $\beta_1,\beta_2>0$,
\be
\la{6.5}
\mc W_\Delta(\beta_1+\beta_2)=\sum_{m\geq0}w_\Delta(m)\,Z_\Delta(m,\beta_1)\,Z_\Delta(m,\beta_2).
\ee

\subsection{Spectral form and closed form of the neck weight}
\la{sec:fq-gluing}

The amplitudes $Z_\Delta(m,\beta)$ and the neck weights \eqref{6.3} satisfy the hypotheses of the uniqueness statement of Sec.~\ref{sec:general}: the amplitudes are real and entire in $\wt\beta$ (footnote~\ref{fn:trace-class}) and have the threshold property \eqref{5.15}, and by Theorem~\ref{prop:sectors} the sectors, which have non-negative Taylor coefficients, add up to $\mc W_\Delta$ term by term in the Taylor expansion. That statement then gives both in closed form, in terms of the orthonormal polynomials of the weight $\mc K_\Delta$. These are the continuous $q$-ultraspherical (Rogers) polynomials
\cite{AskeyIsmail:1983Ultraspherical,Ismail:2009OrthogonalPolynomials},
\be
\la{6.6}
C_m(\cos\theta;t|q)=\sum_{j=0}^m
\frac{(t;q)_j(t;q)_{m-j}}{(q;q)_j(q;q)_{m-j}}\,e^{i(m-2j)\theta}.
\ee
 Their squared norms with the dressed measure $\mc K_\Delta\,d\theta/(2\pi)$ are the quantities $h_m$ in the orthogonality relation \eqref{C.1}.

\begin{corollary}[Spectral form and neck weight]
\la{cor:spectral}
For every $m\geq0$, $0<q<1$ and $\Delta>0$, the amplitudes are
\be
\la{6.7}
Z_\Delta(m,\beta)
=(1-q^{m+\Delta})\,
\frac{(q^\Delta;q)_m}{(q^{2\Delta};q)_m}
\int_0^\pi\frac{d\theta}{2\pi}\,\mc K_\Delta(\theta)\,
C_m(\cos\theta;q^\Delta|q)\,e^{\wt\beta\cos\theta},
\ee
 where $C_m$ are the Rogers polynomials \eqref{6.6}, and the neck weight \eqref{6.3} is
\be
\la{6.8}
w_\Delta(m)=
\frac{(q;q)_m\,(q^{2\Delta};q)_m}
{(1-q^{m+\Delta})\,(q^\Delta;q)_m^2}.
\ee
\end{corollary}
\begin{proof}
By the uniqueness statement of Sec.~\ref{sec:general} with $w=w_\Delta$ and $Z=Z_\Delta$, the product $\sqrt{w_\Delta(m)}\,Z_\Delta(m,\beta)$ is given by \eqref{3.4}, with $p_m=C_m/\sqrt{h_m}$. By \eqref{6.6}, the leading coefficient of $C_m(x;t|q)$ in $x=\cos\theta$ is $2^m(t;q)_m/(q;q)_m$, so that the leading coefficient of $p_m$ is $\zeta_m=2^m(t;q)_m/[(q;q)_m\sqrt{h_m}]$ and the right-hand side of \eqref{3.4} is $\pm\wt\beta^{\,m}/(\zeta_m\,m!)$ up to higher orders in $\wt\beta$. Comparison with the threshold expansion \eqref{5.15} gives the plus sign and, as in \eqref{3.7}, $\sqrt{w_\Delta(m)}=2^m/\zeta_m=1/(N_m\sqrt{h_m})$, where $N_m$ is the prefactor of the integral in \eqref{6.7}, so that $N_mh_m=(t;q)_m/(q;q)_m$. This is \eqref{6.8}, and \eqref{3.4} then becomes \eqref{6.7}.
\end{proof}
By the proof, the product $\sqrt{w_\Delta(m)}\,Z_\Delta(m,\beta)$ is the transform of the orthonormal Rogers polynomial $C_m/\sqrt{h_m}$, and the gluing identity \eqref{6.5} is Parseval's identity for this basis. Comparing the closed form \eqref{6.8} with the chord sum \eqref{6.3} at $t=q^\Delta$, for arbitrary $\Delta>0$, gives the following identity.

\begin{proposition}[Twist--winding identity]
\la{prop:shi}
Let $0<q<1$, $0<t<1$ and $m\geq0$. With $N(\sigma,\bm k)$ as in \eqref{6.4},
\be
\la{6.9}
\sum_{\sigma\in\mathsf S_m}\sum_{\bm k\in\mathbb Z^m}
q^{N(\sigma,\bm k)}\,t^{\sum_i|k_i|}
=\frac{(q;q)_m}{(1-q)^m}\,\frac{(t^2;q)_m}{(t;q)_m^2}.
\ee
 At $m=0$ both sides are equal to one.
\end{proposition}

A combinatorial proof, independent of the uniqueness statement and of the orthogonal polynomials, is given in App.~\ref{app:sectors-shi}. The identity \eqref{6.9} has an interesting group-theoretic reading. The pairs $(\sigma,\bm k)$ label the elements $\hat\sigma$ of the extended affine symmetric group (the bijections of $\mathbb Z$ with $\hat\sigma(i+m)=\hat\sigma(i)+m$), with $\hat\sigma(i)=\sigma(i)+m\,k_i$, and $N(\sigma,\bm k)$ is their length \cite{Shi:1986KL,Bjorner:1996AffinePermutations}. The sum $\sum_i|k_i|$ counts the crossings of the chords with the lifts of the matter line, \ie how many windows away from the fundamental window $1,\dots,m$ the entries $\hat\sigma(1),\dots,\hat\sigma(m)$ lie.
 The identity thus expresses the norms of the Rogers polynomials as a sum over the extended affine symmetric group.\footnote{At $t=1$, with the common winding quotiented, the sum is $m$ times Bott's formula \cite{Bott:1956Morse} for the length generating function of the affine symmetric group (App.~\ref{app:sectors-free}). For $q$ a prime power, we checked that the identity \eqref{6.9} can also be derived from the Iwahori--Matsumoto formula for the double cosets of $GL_m$ over a local field \cite[Prop.~3.2]{IwahoriMatsumoto:1965}, \cite[Ch.~V, Eq.~(2.9)]{Macdonald:1995HallPolynomials}, and the classical count of sublattices \cite{Tamagawa:1963}, \cite[Ch.~V, Sec.~4]{Macdonald:1995HallPolynomials}, and then holds for every $q$, since at each order in $t$ both sides are polynomials in $q$. Our two proofs, from the orthogonal polynomials above and combinatorial in App.~\ref{app:sectors-shi}, are self-contained and do not require the $p$-adic and Hecke-algebra machinery of these counts.}

\paragraph{The identity at $m=1$ and $m=2$.}
The double sum in \eqref{6.3} can be evaluated directly in the first two sectors, which shows how the twists and windings combine. At $m=1$ there is no twist and a chord through the neck
never crosses its own translates, so $N=0$ and the double sum is
$\sum_{k\in\mathbb Z}t^{|k|}=(1+t)/(1-t)$, in agreement with \eqref{6.9}. At $m=2$, the two chords cross $|k_1-k_2|$ times for the
identity twist and $|k_1-k_2+1|$ times for the transposition, which
exchanges their order of arrival on boundary~2.
Summing first over the
windings at fixed relative winding $d=k_1-k_2$,
$\sum_{k_1-k_2=d}t^{|k_1|+|k_2|}=t^{|d|}\big(\tfrac{1+t^2}{1-t^2}+|d|\big)$,
and then over $d$,
\be
\la{6.10}
\sum_{d\in\mathbb Z}t^{|d|}\Big(\frac{1+t^2}{1-t^2}+|d|\Big)
\big(q^{|d|}+q^{|d+1|}\big)
=\frac{(1+q)(1-t^2)(1-qt^2)}{(1-t)^2(1-qt)^2},
\ee
which agrees with \eqref{6.9}.

The factorization \eqref{6.2}, with $w_\Delta(m)$ in the closed form
\eqref{6.8}, was also checked by exhaustive enumeration of chord
diagrams, exact in $q$ and to third order in $q^\Delta$, for up to eight Hamiltonian insertions on the two boundaries (see App.~\ref{app:sectors} for details).

\paragraph{Direct derivation from the brane trace.}
The spectral form \eqref{6.7} can also be derived directly from the brane trace \eqref{5.9}, without Theorem~\ref{prop:sectors}; we sketch the argument. The eigenfunctions of the brane transfer operator are the continuous big $q$-Hermite polynomials \cite{Okuyama:2023byh}. In their Poisson kernel \cite[Eq.~(1.12)]{Chen:2007RogersSzego}, take coincident arguments, equal brane parameters and $q^\Delta$ as the expansion variable. Multiplied by their spectral measure, the result is the product of $\mc K_\Delta$ and the generating function $\sum_n\frac{(q^\Delta;q)_n}{(q^{2\Delta};q)_n}\,C_n(x;q^\Delta|q)\,a^n$ of the Rogers polynomials \cite[Eq.~(24)]{Cohl:2022Generating}. By \eqref{5.10}, this gives \eqref{6.7} and then, with the orthogonality relation \eqref{C.1}, the gluing identity \eqref{6.5} with the weight \eqref{6.8}. On its own, it does not identify the weight with the chord sum \eqref{6.3} over twists and windings, which requires Theorem~\ref{prop:sectors}, and so does not give the identity \eqref{6.9}.

\subsection{Relation to other constructions}
\la{sec:fq-relation}

We compare our chord construction with three related ones.

\paragraph{Single-sided black hole trumpets of \cite{Cao:2025pir}.}
Without matter, the structure of Sec.~\ref{sec:fq-normalization} was found in \cite{Cao:2025pir}: the chords not
connected to the asymptotic boundary, called vacuum bubbles, form concentric loops crossing only the chords that end on the
brane, and one of the regularizations of the divergent brane trace considered there removes them diagram by diagram. This is the
matter-free form of our loop criterion and, according to \cite{Cao:2025pir}, it is what justifies gluing two trumpets into a
double cone. The matter trumpet of \cite{Cao:2025pir} is a different observable from ours. Its brane reflects matter, so that no matter line crosses the neck and vacuum bubbles can be of either kind.

\paragraph{Late-time ramp trumpets of \cite{Raz:2025wjw}.}
A trumpet and a gluing factor for chords were obtained in \cite{Raz:2025wjw} for a different observable: the part of the connected double-trace correlator $\langle\tr e^{-\beta_1H}\,\tr e^{-\beta_2H}\rangle_J$ of DSSYK without matter that survives at late times and gives the ramp. Ours reduces at $\Delta=0$ to the matrix-model amplitude \eqref{2.9} (Sec.~\ref{sec:fq-zero-label}). The two agree in the Gaussian random-matrix limit $q\to0$, where the result of \cite[Eq.~(4.14)]{Raz:2025wjw} is, up to normalization, $\sum_{k\geq1}k\,I_k(2\beta_1)\,I_k(2\beta_2)$, with $k$ the number of chords joining the two traces: this is \eqref{2.9} at $q=0$, with the matter-free weight $w_0(m)=m$ of \eqref{7.7} and the loop-free factors $I_m(\wt\beta)$. For $q>0$ they are different functions of $\beta_1,\beta_2$, with the same late-time ramp.

\paragraph{Heavy-string wormhole of \cite{Bettaque:2026syk}.}
\la{sec:sd-bettaque}
The observable of \cite{Bettaque:2026syk} is the average over the couplings of the product of two thermal one-point
functions~\cite{Bettaque:2026magic} of a \emph{fixed} Majorana string of size $p\Delta$, with $\Delta$ of order $1/\l$ in the
Schwarzian regime. Since the string is fixed, a chord diagram contributes only if the index sets of the $K$ Hamiltonian chords
joining the two traces (their $K$ is our $m$) reproduce that of the string \cite[Eq.~(3.4)]{Bettaque:2026syk}, which in the
double-scaling limit localizes the amplitude on $K=\Delta$ for integer $\Delta$. In the decorated wormhole \eqref{2.12} the matter operator has random couplings,
so that no such condition arises and all sectors contribute. The two constructions also glue along different surfaces. The heavy-string wormhole is cut \emph{along}
the open slice through the operator insertion: its two halves are disk wavefunctions with $K$ open chords. They are glued by a single projector, $K=\Delta$, with the $\Delta!$ permutations of the chords as its weight. This gluing is derived from the SYK model in the double-scaling limit.
 Ours cuts \emph{across} the matter line and along the closed neck, the cut along which trumpets are glued in JT gravity.

\section{Further exact properties of the amplitudes}
\la{sec:fq-spectral}

\subsection{The case \texorpdfstring{$\Delta=1$}{Delta=1}}
\la{sec:fq-D1}

At $\Delta=1$ the
amplitudes can be computed directly from the matter-free brane expansion
\eqref{5.6}.
Since $q^{\widehat N}=\partial_a\widehat D_a$ and the trace is cyclic, we obtain\footnote{Duhamel's
formula reads
$\partial_a e^{\widetilde\beta\widehat D_a/2}
=(\widetilde\beta/2)\int_0^1dv\,
e^{(1-v)\widetilde\beta\widehat D_a/2}
q^{\widehat N}e^{v\widetilde\beta\widehat D_a/2}$.
Cyclicity makes the trace of the integrand independent of $v$, and
the $a=0$ subtraction in \eqref{7.1} does not depend on $a$, giving
the displayed identity.}
\be
\la{7.1}
F_1(a,\beta)=\frac{2}{\widetilde\beta}\,\partial_a
\Tr_{\mathcal H}\!\left[e^{\widetilde\beta\widehat D_a/2}
-e^{\widetilde\beta\widehat D_0/2}\right] = \frac{2}{\widetilde\beta}\,\partial_a Z_{\rm brane}^{\rm sub}(a,\beta).
\ee
Differentiating the expansion \eqref{5.6} and
comparing with \eqref{5.10} therefore gives
\be
\la{7.2}
Z_1(m,\beta)=\frac{2b}{\widetilde\beta}I_b(\widetilde\beta)
=I_m(\widetilde\beta)-I_{m+2}(\widetilde\beta),
\qquad b=m+1.
\ee
At $\Delta=1$ the matter line crosses each chord with weight $q$, as a
Hamiltonian chord does, so the neck carries $b=m+1$ lines, which gives the factor $b$ in \eqref{7.2}. Accordingly, the
denominator $1-q^{m+1}$ in \eqref{5.10} is that of the term $b$ in \eqref{5.6}, and the shifted
label \eqref{5.12} is $\l b$, an ordinary-trumpet label. This case is thus degenerate: as for its JT counterpart \eqref{4.14}, the result tests the normalization and the counting of labels, not the dynamics of the matter line.

The gluing weights are $w_1(m)=1/(1-q)$, independent of $m$, and the
dressed weight \eqref{2.15} is $\mc K_1(\theta)=4\sin^2\theta/(1-q)$, so that
$\mc W_1(\beta)=2I_1(\wt\beta)/[(1-q)\wt\beta]$, and
\eqref{6.5} reduces to the Bessel identity
$\sum_{b\geq1}b^2I_b(x)I_b(y)=\frac{xy}{2(x+y)}I_1(x+y)$,
with $x=\widetilde\beta_1$ and $y=\widetilde\beta_2$, a consequence of the addition theorem
\cite[Sec.~10.44(ii)]{DLMF}.

\subsection{Expansion in ordinary trumpets}
\la{sec:fq-trumpet-form}

The amplitudes $Z_\Delta(m,\beta)$ have an expansion in ordinary trumpets (which truncates to a finite sum for integer $\Delta$) with closed-form coefficients,\footnote{At positive integer $\Delta=d$ the sum in
\eqref{7.3} stops at $j=d$, since $(q^{-d};q)_j=0$ for $j>d$: the amplitude is
a superposition of the $d+1$ ordinary trumpets with labels
$m,m+2,\ldots,m+2d$. At $\Delta=1$, $c^{(1)}_0=1$, $c^{(1)}_1=-1$, $c^{(1)}_j=0$
for $j\geq2$ and $\gamma^{(1)}_{m,1}=1$, so \eqref{7.3} truncates to
\eqref{7.2}.}
\be
\la{7.3}
\begin{gathered}
Z_\Delta(m,\beta)
=\sum_{j\geq0}c_j^{(\Delta)}\,\gamma^{(\Delta)}_{m,j}\,
I_{m+2j}(\wt\beta), \qquad c_j^{(\Delta)}=q^{\Delta j}(q^{-\Delta};q)_j/(q;q)_j,
\\[3pt]
\gamma^{(\Delta)}_{m,0}=1,\quad
\gamma^{(\Delta)}_{m,j}
=\frac{(1-q^{m+2j})\,(q^{m+1};q)_{j-1}}{(q^{m+\Delta+1};q)_j},\qquad j\geq1.
\end{gathered}
\ee
The structure of \eqref{7.3} is that of the Fourier expansion in $\theta$ of the factor multiplying the Boltzmann weight in \eqref{6.7}, since $I_b(\wt\beta)$ is the transform \eqref{2.8} of $2\cos(b\theta)$. The Fourier coefficients below frequency $m$ vanish because $C_m$ is orthogonal to the polynomials of lower degree, which is why the sum starts at the ordinary trumpet $I_m$. The $c^{(\Delta)}_j$ are the coefficients
 of the $q$-binomial expansion
$(z;q)_\infty/(q^\Delta z;q)_\infty=\sum_{j\geq0}c^{(\Delta)}_jz^j$
\cite{gasper2011basic} of each of the two factors of
$\mc K_\Delta$ in \eqref{2.15}.
The closed form of the coefficients is derived in
App.~\ref{app:fq-spectral}.

\paragraph{Trumpet-basis form of the gluing.}
Expanding $e^{\wt\beta\cos\theta}=\sum_{b\geq0}\epsilon_bI_b(\wt\beta)\cos(b\theta)$,
with $\epsilon_0=1$ and $\epsilon_b=2$ for $b\geq1$, in \eqref{2.14} gives
\be
\la{7.4}
\mc W_\Delta(\beta_1+\beta_2)=\sum_{b,b'\geq0}\mathsf K_{bb'}\,I_b(\wt\beta_1)I_{b'}(\wt\beta_2),\qquad
\mathsf K_{bb'}=\epsilon_b\epsilon_{b'}\int_0^\pi\frac{d\theta}{2\pi}\,\mc K_\Delta(\theta)\cos(b\theta)\cos(b'\theta),
\ee
a positive-definite matrix, nonzero only for $b\equiv b'$ mod $2$. Unlike the undecorated measure $b\,\delta_{bb'}$ of \eqref{2.9}, it is not diagonal: in the basis of ordinary trumpets the matter line mixes the labels.
By \eqref{7.3}, the gluing identity is the factorization
$\mathsf K=C^{\mathsf T}\,{\rm diag}(w_\Delta)\,C$, with $C_{m,m+2j}=c^{(\Delta)}_j\gamma^{(\Delta)}_{m,j}$
unit upper-triangular: the amplitudes are the unique
combinations of ordinary trumpets, each starting with $I_m$ at unit
coefficient, that diagonalize the kernel, and the loop criterion is the
unit normalization of this leading coefficient.
This unit-triangular
(Cholesky-type) factorization is unique; the uniqueness statement of Sec.~\ref{sec:general} extends this to one-boundary factors that are not assumed to be
combinations of ordinary trumpets.

\subsection{Zero label and matter-free limit}
\la{sec:fq-zero-label}

\paragraph{The zero-label contribution.}

Since the Rogers polynomial of degree zero is $C_0(\cos\theta;t|q)=1$, the spectral form \eqref{6.7} gives
$Z_\Delta(0,\beta)
=(1-q^\Delta)\mc W_\Delta(\beta)$, consistently with \eqref{5.11}.
This $m=0$ term is required in the gluing sum, with positive weight
$w_\Delta(0)=(1-q^\Delta)^{-1}$; it collects the chord diagrams in
which no Hamiltonian chord joins the two boundaries
(Sec.~\ref{sec:fq-sectors}). Unlike the label $b=0$, absent from the matter-free wormhole
\eqref{2.9}, it contributes here.
Explicitly, the sector $S_0$ of \eqref{6.2} is
$w_\Delta(0)Z_\Delta(0,\beta_1)Z_\Delta(0,\beta_2)
=(1-q^\Delta)\,\mc W_\Delta(\beta_1)\,\mc W_\Delta(\beta_2)$, as noted after \eqref{5.11}. Each of the two factors $\mc W_\Delta(\beta_i)$ contains the loop sum $1/(1-q^\Delta)$, in which every loop crosses only the matter line. The factor $1-q^\Delta$ removes one of the two loop sums, and the other is the weight $w_\Delta(0)$, since the closed loops belong to the neck and are counted once.

\paragraph{The matter-free limit.}

The amplitudes reduce to ordinary trumpets as
matter is removed. Indeed, in \eqref{7.3},
$c_j^{(\Delta)}\to\delta_{j0}$ as $\Delta\to0$ at fixed $q$, giving
$Z_\Delta(m,\beta)\to I_m(\wt\beta)$, the ordinary trumpet \eqref{2.8},
for all $m\geq0$.
Only the labels $m\geq1$ have a counterpart in the matter-free wormhole \eqref{2.9}.

As $\Delta\to0$, the $m=0$ term $F_\Delta(0,\beta)$ of \eqref{5.10} diverges through its closed-loop factor $(1-q^\Delta)^{-1}$, as the trace without the insertion diverges through its $a$-independent term (Sec.~\ref{sec:fq-motivation}). We therefore compare with the subtracted amplitude \eqref{5.6} and do the same $a=0$ subtraction at
finite $\Delta$ before taking the matter-free limit.
By \eqref{5.10} and the amplitude limit above, we obtain
\bea
\la{7.5}
\lim_{\Delta\to0}
\big[F_\Delta(a,\beta)-F_\Delta(0,\beta)\big]
&=\sum_{m\geq1}\frac{a^m}{1-q^m}\,I_m(\wt\beta)
=Z_{\rm brane}^{\rm sub}(a,\beta),
\eea
where the last equality is the ordinary-trumpet
expansion \eqref{5.6}, and the limit may be taken inside the sum over $m$.\footnote{
The coefficient of $a^m$ in $F_\Delta$ is a sum of chord diagrams with
non-negative weights that depend on $\Delta$ only through the factor
$q^{\Delta n}$ of the matter line, with $n$ the number of chords it crosses
(Sec.~\ref{sec:fq-normalization}). It therefore increases as $\Delta$
decreases, up to its $\Delta=0$ value $I_m(\wt\beta)/(1-q^m)$ for
$m\geq1$, and the convergent series \eqref{5.6} dominates the sum over $m$
for $|a|<1$.}

The gluing weights, instead, do not have a finite limit: as $\Delta\to0$,
at fixed $q$ and fixed $m\geq1$,
\be
w_\Delta(m)=\frac{2}{1-q^\Delta}-\coth\frac{\l m}{2}+O(\Delta),
\qquad w_\Delta(0)=\frac{1}{1-q^\Delta}.
\ee
Since $1-q^\Delta\sim\lambda\Delta$, the gluing
weights (and thus also the wormhole amplitude) diverge proportionally
to $1/(\lambda\Delta)$; in the trace \eqref{2.12} this is the loss of suppression of states with many open Hamiltonian chords. At leading order the weights are thus equal for all $m\geq1$, not the undecorated measure $b$ of
\eqref{2.9}.
The reason for this apparent discrepancy is the common winding $\bm k\to\bm k+c(1,\ldots,1)$ of the $m$
chords through the neck, which leaves $N(\sigma,\bm k)$ in \eqref{6.4}
unchanged. With matter, the windings are measured relative to the matter
line and weighted by $t^{\sum_i|k_i|}$, $t=q^\Delta$, and the sum over
$c$ gives, as $t\to1$, the divergent factor
$\sum_{c\in\mathbb Z}t^{\sum_i|k_i+c|}\simeq2/[m(1-t)]$. Without matter
there is no line to measure the common winding against, and we quotient it, as the twist is quotiented in JT
(footnote~\ref{fn:unfolding}). At finite $q$ this is a prescription, as is the subtraction of the divergent $m=0$ term above. Quotienting the common winding in \eqref{6.3} at $\Delta=0$,
with closed loops that cross only the $m$ chords (each of weight $q^m$),
gives exactly the undecorated measure,
\be
\la{7.7}
w_0(m)\equiv\frac{(1-q)^m}{1-q^m}\sum_{\sigma\in\mathsf S_m}\;
\sum_{\bm k\in\mathbb Z^m/\mathbb Z(1,\ldots,1)}q^{N(\sigma,\bm k)}=m,
\qquad m\geq1,
\ee
as shown in App.~\ref{app:sectors-free}. Similarly, in JT, the sum over the windings $\ell$ of the matter geodesic is $\sum_{\ell\in\mathbb Z}e^{-\Delta|\ell|\mathfrak b}\simeq2/(\Delta\mathfrak b)$ (App.~\ref{app:neck-unfold}), which is the factor $2/[m(1-t)]$ above at $\mathfrak b=\l m$, $t=q^\Delta\simeq1-\l\Delta$. Dividing the flat measure $\frac{\Gamma(\Delta)^2}{\Gamma(2\Delta)}\,d\mathfrak b\simeq\frac2\Delta\,d\mathfrak b$ by it restores $\mathfrak b\,d\mathfrak b$ at leading order as $\Delta\to0$.

Without matter, the loop-free one-boundary factors are the ordinary trumpets $I_m(\wt\beta)$, by \eqref{5.6}. Each chord sector of the undecorated wormhole \eqref{2.9} with $m\geq1$ is therefore its term $b=m$. The sector $m=0$ has a divergent loop sum and is absent from \eqref{2.9}.

An interesting related regime is $q\to0$ at fixed crossing weight $t=q^\Delta$ (that is, $\Delta\to0$ at fixed $\l\Delta$). This is the Gaussian random-matrix limit, where Hamiltonian chords do not cross one another. There the amplitudes and the weights are elementary:
 by \eqref{7.3} and \eqref{6.8},
\be
\la{7.8}
Z(m,\beta)=I_m(\wt\beta)-(1-t)\sum_{j\geq1}t^{\,j-1}I_{m+2j}(\wt\beta),
\qquad
w(0)=\frac{1}{1-t},\qquad w(m)=\frac{1+t}{1-t}\quad(m\geq1).
\ee
 Only the terms with $N(\sigma,\bm k)=0$ of the chord sum \eqref{6.3} survive,
and the weight does not depend on $m\geq1$, as the flat measure of
\eqref{4.13} does not depend on the neck circumference. In this limit the
zero label is not suppressed: $w(0)$ and $w(m)$ are of the same order.

\subsection{Recurrence coefficients}
\la{sec:fq-local-evolution}

For the weight \eqref{2.15}, the local evolution \eqref{3.8} of the symmetrized amplitudes
$\widetilde Z_m(\beta)=\sqrt{w_\Delta(m)}\,Z_\Delta(m,\beta)$ has the coefficients
\be
\la{7.9}
B_m=\left[\frac{(1-q^{m})(1-q^{m+2\Delta-1})}
{(1-q^{m+\Delta-1})(1-q^{m+\Delta})}\right]^{1/2},
\qquad m\geq1,
\ee
which follow from  \eqref{6.8} through $B_m^2=w_\Delta(m)/w_\Delta(m-1)$, as in Sec.~\ref{sec:general}. They are the entries of the Jacobi matrix $\mathsf J_\Delta$ introduced there. By the sector factorization, the Krylov index $m$ of Sec.~\ref{sec:general} is the number of chords through the
neck.\footnote{The $B_m$ are the Lanczos coefficients of the measure $(1-q^\Delta)\mc K_\Delta\,d\theta/(2\pi)$ \cite{Nandy:2024zcd}; for chord numbers as Krylov indices in DSSYK see \cite{Rabinovici:2023yex,Ambrosini:2024sre}; the same Jacobi operator appears in \cite{Aguilar-Gutierrez:2026nmd}, with the earlier Rogers representation in \cite{Berkooz:2025ydg}.}

In the JT limit, at fixed $\mathfrak b_{m+1/2}=\l(m+\Delta+\frac12)$,
\be
\la{7.10}
B_{m+1}=1-\frac{\l^2\,\Delta(\Delta-1)}{8\sinh^2(\mathfrak b_{m+1/2}/2)}+O(\l^4),
\ee
so that $(2-\mathsf J_\Delta)/\l^2$ tends to the P\"oschl--Teller operator $H^{\rm PT}_\Delta$ of \eqref{4.16} on sequences $f(\mathfrak b_m)$ with $f$ smooth on $\mathfrak b>0$; this is the operator of the neck heat equation of the fixed-neck amplitudes (Sec.~\ref{sec:neck-generic}). This operator limit does not select the solution of \eqref{4.16} at $\mathfrak b=0$, where both solutions are square-integrable for $0<\Delta<3/2$; the behavior $\mathfrak b^\Delta$ of \eqref{4.17} follows from the explicit limit of the amplitudes (App.~\ref{app:fq-JT}).

\section{\texorpdfstring{JT limit of the finite-$q$ construction}{JT limit of the finite-q construction}}
\la{sec:fq-JT-fixed}

We compare the amplitudes $Z_\Delta(m,\beta)$ and the weights
$w_\Delta(m)$ with three results in the JT limit of Sec.~\ref{sec:neck-fixed}: the gluing formula \eqref{4.13}, with its flat
measure and the fixed-neck amplitudes \eqref{4.15}; the exact value
\eqref{4.14} of the insertion at $\Delta=1$; and the small-neck behavior
\eqref{4.17}.

The two gluing formulas, \eqref{6.5} at finite $q$ and \eqref{4.13} in JT, are expansions in orthogonal functions, of the wormhole $\mc W_\Delta$ and of its JT limit \eqref{4.3}--\eqref{4.6}. Let us examine their similar structures. On the JT side, the fixed-neck amplitudes \eqref{4.15} are thermal superpositions of the eigenfunctions $\chi_{\Delta,k}$ of the P\"oschl--Teller operator \eqref{4.16}, whose orthogonality \eqref{A.11} turns the gluing integral \eqref{4.13} into \eqref{4.6}. On the DSSYK side, by the spectral form of the amplitudes (Corollary~\ref{cor:spectral}) and the remark following it, the products $\sqrt{w_\Delta(m)}\,Z_\Delta(m,\beta)$ are transforms of the orthonormal Rogers polynomials of the dressed weight $\mc K_\Delta$, and the gluing identity is Parseval's identity for them. Moreover, Rogers polynomials tend to Jacobi functions, which are the profiles $\chi_{\Delta,k}$ up to $k$- and $\mathfrak b$-dependent factors, as shown in \cite{Koornwinder:1990Jacobi,Koelink:1995Identities} for $\Delta\geq1$ and in App.~\ref{app:fq-JT} for every $\Delta>0$. Finally, the dressed weight $\mc K_\Delta$ tends to the weight $C^{\rm Sch}_\Delta|h^{\rm Sch}_\Delta|^2$ of \eqref{4.6}, by \eqref{B.7}.
 These remarks, altogether, indicate that the decomposition of the wormhole into sectors, the sum over $m$ in \eqref{6.5}, goes over in the JT limit into its decomposition by neck circumference, the integral over $\mathfrak b$ in \eqref{4.13}.

This structure, however, fixes only the products $\sqrt{w_\Delta(m)}\,Z_\Delta(m,\beta)$, as discussed in Sec.~\ref{sec:general}. The separate limits of the amplitudes and of the weights depend on their normalization.

For our amplitudes, normalized by the loop criterion of Sec.~\ref{sec:fq-normalization}, we can get a more precise result: $\l\,w_\Delta(m)$ tends to a constant, independent of $\mathfrak b$, and a single $m$-independent rescaling takes the amplitudes to the fixed-neck JT amplitudes as normalized by the Schwarzian path integral,
\be
\la{8.1}
\l\,w_\Delta(m)\ \longrightarrow\
\frac{\Gamma(\Delta)^2}{\Gamma(2\Delta)},
\qquad
\l^{-(\Delta+1)}e^{-\wt\beta}\,Z_\Delta(m,\beta)
\ \longrightarrow\ Z^{\rm fix}_\Delta(\mathfrak b,\tau) ,
\ee
for every $\Delta,\tau,\mathfrak b>0$, in the JT scaling \eqref{4.1} at fixed $\mathfrak b=\l m$.

The rescaling factors of the amplitudes in \eqref{8.1} have a simple origin: $e^{-\wt\beta}$ removes the spectral-edge energy, and $\l^{-\Delta}$ converts the insertion $q^{\Delta\wh N}$ into its JT counterpart $e^{\Delta\varphi}$ (\cf \eqref{5.8}). The remaining $\l^{-1}$ accounts for the spacing $\l$ of the neck circumferences, so that $\l^2\sum_m w_\Delta(m)\to\frac{\Gamma(\Delta)^2}{\Gamma(2\Delta)}\int_0^\infty d\mathfrak b$, the flat measure of \eqref{4.13}.

The limit of the weight is elementary. With
$(q^a;q)_m=(1-q)^m\Gamma_q(a+m)/\Gamma_q(a)$, \eqref{6.8} reads
\be
\la{8.2}
w_\Delta(m)=\frac{\Gamma_q(\Delta)^2}{\Gamma_q(2\Delta)}\,
\frac{\Gamma_q(m+1)\,\Gamma_q(m+2\Delta)}{\Gamma_q(m+\Delta)^2\,(1-q^{m+\Delta})}.
\ee
The first factor tends to $\Gamma(\Delta)^2/\Gamma(2\Delta)$, as in
footnote~\ref{f:1}. In the second, we set $q^m\to e^{-\mathfrak b}$. As $q\to1$,
$(q^{m+a};q)_\infty/(q^m;q)_\infty\to(1-e^{-\mathfrak b})^{-a}$, so that
$\Gamma_q(m+a)\propto(1-q)^{-a}(1-e^{-\mathfrak b})^a$ up to $a$-independent factors.
The powers of $1-e^{-\mathfrak b}$ cancel the denominator
$1-q^{m+\Delta}\to1-e^{-\mathfrak b}$, and the second factor behaves as $(1-q)^{-1}$. Since $\l/(1-q)\to1$, this gives the limit of $\l\,w_\Delta(m)$ in \eqref{8.1}.

The proofs of the limit of the amplitudes in \eqref{8.1} and of the convergence of the gluing sum are instead given in App.~\ref{app:fq-JT}, where the full result is stated as Proposition~\ref{prop:JT}. The proof holds for every $\Delta>0$, which is needed to take $\Delta$ arbitrarily small, and shows that the gluing sum converges to the gluing integral, with no additional contribution from small or large necks, the zero label included. In particular, the limit reproduces the exact value \eqref{4.14} at $\Delta=1$ and the small-neck behavior \eqref{4.17}.

The chord form \eqref{6.3} of the weight shows how the flat measure arises. Its two factors have separate limits as $\l m\to\mathfrak b$. The sum over twists and
windings, with the factor $(1-q)^m$, tends to $\frac{\Gamma(\Delta)^2}{\Gamma(2\Delta)}\,e^{-\mathfrak b/2}\mc J(\mathfrak b)/\l$, where $\mc J$ is the orbit factor \eqref{4.11}. The closed-loop
factor $1/(1-q^{m+\Delta})$ tends to $e^{\mathfrak b/2}/\mc J(\mathfrak b)$. In JT, the flat measure comes from the twist integral and the winding sum alone, \eqref{A.1}--\eqref{A.3}. At finite $q$, it is the limit of the whole weight: the closed loops cancel the orbit factor carried by the sum over twists and windings. In the brane trace, the same closed loops are removed to define the amplitudes, and there they play the role of the inverse orbit factor, the kernel $1/\mc J$ of \eqref{5.3}.

Table~\ref{tab:dictionary} summarizes the correspondence between the
fixed-neck JT construction and its finite-$q$ counterpart.

\begin{table}[!htb]
\centering
\renewcommand{\arraystretch}{1.3}
\begin{tabular}{>{\raggedright\arraybackslash}p{0.44\textwidth}>{\raggedright\arraybackslash}p{0.48\textwidth}}
\hline
\textbf{JT gravity at fixed neck} & \textbf{DSSYK at finite $q$}\\
\hline
neck circumference $\mathfrak b>0$ & shifted label $\mathfrak b_m=\l(m+\Delta)$, $m\geq0$\\
Schwarzian momentum $k$ & spectral angle $\theta=\l k$\\
Laplace transform in $\mathfrak b$, kernel $e^{-\kappa\mathfrak b}$ & expansion in the brane parameter $a=q^{\kappa+\frac12}$\\
inverse orbit factor $1/[2\sinh(\mathfrak b/2)]$ & closed-loop sum $1/(1-q^{m+\Delta})$ in the brane trace\\
Morse Hamiltonian $H_\kappa$ & brane transfer operator $T_a$\\
endpoint insertion $e^{\Delta\varphi(u_0)}$ & insertion $q^{\Delta\wh N}$\\
twist integral and winding sum & sum over chord twists $\sigma\in\mathsf S_m$, windings $\bm k\in\mathbb Z^m$ and closed loops\\
flat measure $\frac{\Gamma(\Delta)^2}{\Gamma(2\Delta)}\,d\mathfrak b$ & gluing weight $w_\Delta(m)$\\
neck profile $\chi_{\Delta,k}(\mathfrak b)$ & Rogers polynomial $C_m(\cos\theta;q^\Delta|q)$\\
P\"oschl--Teller operator $H^{\rm PT}_\Delta$ & Jacobi operator $\mathsf J_\Delta$ (Sec.~\ref{sec:fq-local-evolution})\\
$\Delta=1$: $Z^{\rm fix}_1=(\mathfrak b/\tau)\,Z^{\rm JT}_{\rm trumpet}$ & $\Delta=1$: $Z_1(m,\beta)=\frac{2(m+1)}{\wt\beta}\,I_{m+1}(\wt\beta)$ (Sec.~\ref{sec:fq-D1})\\
\hline
\end{tabular}
\caption{Correspondence between the fixed-neck JT construction and its finite-$q$ counterpart. The JT entries are recovered from the finite-$q$ ones in the JT limit at fixed $\mathfrak b_m\to\mathfrak b$.}
\label{tab:dictionary}
\end{table}

In the JT limit the zero-label amplitude (Sec.~\ref{sec:fq-zero-label}) is of order $\l^\Delta$ after the rescaling of \eqref{8.1}, the same power as the small-neck behavior $\mathfrak b^\Delta$ of \eqref{4.17} at $\mathfrak b_0=\l\Delta$.

\section{Classical limit and a sine-dilaton interpretation}
\la{sec:sinedilaton}

In Sec.~\ref{sec:fq-JT-fixed} the amplitudes were studied in the JT limit, at fixed Schwarzian boundary length $\tau$ and neck circumference $\mathfrak b$. The classical limit of sine-dilaton gravity holds instead the renormalized boundary length fixed. The two regimes are different but overlap in a corner of the parameter space. Here we derive the asymptotics of the amplitudes in the classical regime and compare them with classical sine-dilaton trumpets.

\subsection{Classical asymptotics}
\la{sec:sd-limit}

We call \emph{classical regime} the limit
\be
\la{9.1}
\l\to0,\qquad
\beta_s=\l\wt\beta\ \ \text{fixed},\qquad
\mathfrak b=\l m\ \ \text{fixed},\qquad
\Delta\ \ \text{fixed}.
\ee
The JT regime also holds $\mathfrak b$ fixed as $\l\to0$, but differs
in which ratio of boundary length and neck label is held fixed. Since
$\tau\simeq\l\beta_s/2$ by \eqref{4.2},
\be
\la{9.2}
\frac{\mathfrak b}{\beta_s}\times\frac{\tau}{\mathfrak b^2}
\simeq\frac{\l}{2\mathfrak b}\ \longrightarrow\ 0 ,
\ee
so the classical regime, at fixed $\mathfrak b/\beta_s$, has
$\tau/\mathfrak b^2\to0$, and the JT regime, at fixed $\tau/\mathfrak b^2$, has
$\mathfrak b/\beta_s\to0$. The two overlap where both ratios are small, and we
name this region the \emph{JT corner}. In JT units it is the small-$\tau$ limit
considered in Sec.~\ref{sec:neck-cut}, where the fixed-neck Schwarzian boundary is
classical (App.~\ref{app:JT-classical}).

For $b>0$, let $\eta$ be defined by
\be
\la{9.3}
\sinh\eta=\frac{b}{\wt\beta}=\frac{\l b}{\beta_s}.
\ee
In the classical regime, at $b=m$, $\sinh\eta=\mathfrak b/\beta_s$. The uniform Debye asymptotics of the Bessel functions at large order \cite[Eq.~(10.41.3)]{DLMF}
give $I_{m+s}(\wt\beta)/I_m(\wt\beta)\to e^{-s\eta}$ at fixed $s$, and the amplitudes obey\footnote{In \eqref{9.4} and \eqref{9.6}, $A\sim B$ means $A/B\to1$ in the classical regime \eqref{9.1}. The relation \eqref{9.4} follows from the expansion \eqref{7.3}, using
$c^{(\Delta)}_j\to(-1)^j\binom{\Delta}{j}$,
$\gamma^{(\Delta)}_{m,j}\to1$ and
$I_{m+2j}/I_m\to e^{-2j\eta}$ term by term, the sum over $j$ converging
geometrically at fixed $\eta>0$.}
\be
\la{9.4}
Z_\Delta(m,\beta)\ \sim\ (1-e^{-2\eta})^\Delta\,I_m(\wt\beta)=\Big(\frac{2m}{\wt\beta}\Big)^{\!\Delta}e^{-\Delta\eta}\,I_m(\wt\beta).
\ee
It is convenient to introduce
\be
\la{9.5}
\overline Z_\Delta(b,\beta)\equiv\Big(\frac{2b}{\wt\beta}\Big)^{\!\Delta}
I_b(\wt\beta),\qquad b>0 .
\ee
Since $I_{m+\Delta}/I_m\to e^{-\Delta\eta}$, \eqref{9.4} gives
\be
\la{9.6}
Z_\Delta(m,\beta)\ \sim\ \overline Z_\Delta(m+\Delta,\beta)\ \sim\ e^{-\Delta\eta}\,\overline Z_\Delta(m,\beta).
\ee
At $\Delta=1$ the first relation in \eqref{9.6} is an identity
for every $q$: by \eqref{7.2},
\be
\la{9.7}
Z_1(m,\beta)=\overline Z_1(m+1,\beta).
\ee
The asymptotics \eqref{9.6} are a property of our amplitudes, derived on the DSSYK side with no bulk input. We now attempt a bulk interpretation, necessarily heuristic in the absence of a fully-fledged closed-channel quantization of sine-dilaton gravity with matter.

\subsection{\texorpdfstring{$\overline Z_\Delta$ as a sine-dilaton trumpet dressed with a matter probe}{Z as a sine-dilaton trumpet dressed with a matter probe}}
\la{sec:sd-bulk}

In sine-dilaton gravity \cite{Blommaert:2023opb,Blommaert:2024ymv,Bossi:2024ffa}, with the Euclidean action \eqref{F.1} recalled in App.~\ref{app:sd-classical}, the bulk quantum parameter is $\hbar=\l$ \cite{Blommaert:2024ymv,Cui:2025sgy}.\footnote{Complex Liouville string theory provides an independent worldsheet definition of sine-dilaton gravity on higher-genus surfaces \cite{Collier:2024kwt,Collier:2024lys,Collier:2025pbm}, in a quantization whose disk amplitude differs from the DSSYK-related one used here \cite{Blommaert:2025avl,Blommaert:2025eps}; we do not rely on it.}
Without matter, the closed channel, where spatial slices are closed curves surrounding the neck, has been quantized in \cite{Blommaert:2025avl}. The length of the closed geodesic $\gamma$ at the neck is quantized in units of $\hbar$ \cite{Blommaert:2024whf,Blommaert:2025avl},\footnote{The length is measured in the auxiliary Weyl-rescaled metric $ds^2_{\rm AdS}=e^{-i\Phi}ds^2$, with $\Phi$ the dilaton and $ds^2$ the line element of the action \eqref{F.1}, which is locally AdS$_2$ on classical solutions. The momentum conjugate to this length is $2\pi$-periodic, and the quantization follows from treating the periodicity as a gauge redundancy.}
\be
\la{9.8}
\oint_{\gamma}ds\,e^{-i\Phi/2}=\hbar\, b=\lambda\, b,
\ee
with integer $b\geq1$ for physical states. The amplitude of the trumpet of label $b$ with an asymptotic boundary of length $\beta$ is the Bessel function $I_b(\wt\beta)$ of \eqref{2.8}, and the physical inner product gives
\be
\la{9.9}
\braket{b}{b'}=\frac{1}{b}\delta_{bb'},
\qquad
\mathbb I_{\rm phys}=\sum_{b\geq1}b\,\ket b\bra b,
\qquad b,b'\geq1,
\ee
so that gluing two trumpets with the measure $b$ gives the matter-free wormhole \eqref{2.9}. With the matter line, the decorated wormhole has been obtained in the bulk only in the open channel, by cutting \emph{along} the matter geodesic \cite{Cui:2025sgy}.\footnote{The amplitude is a trace over the Hilbert space of the interval joining the asymptotic boundaries; with the identification of its length states with the chord states \cite{Lin:2022rbf,Blommaert:2024ymv,Cui:2025sgy}, it agrees with the chord trace \eqref{2.12}. The two-sided chord Hilbert spaces with matter of \cite{Lin:2023trc,Okuyama:2024yya} are likewise open-channel spaces, defined on slices joining the two boundaries.} The counterpart of \eqref{9.9} for a matter line of fixed dimension $\Delta$, states labelled by the closed geodesic around the neck and their measure, is not available.

In the classical regime \eqref{9.1}, $\hbar=\l$ is small, and the matter line, with $\Delta$ fixed, is a
probe: it multiplies the matter-free wormhole by the factor $e^{-\Delta L}$, with $L$ its renormalized length on the classical wormhole geometry. This is a statement about the two-boundary amplitude, while its splitting into one-boundary factors is a separate step. To compute $L$, we glue two classical trumpets of neck length $\l b$, the hyperbolic metrics \eqref{F.3} of App.~\ref{app:sd-classical}, with $\eta_i$ given by \eqref{9.3} at $\wt\beta_i$, into a hyperbolic cylinder. In its universal cover, the geodesic joining boundary points at relative position $x_{12}$ along the neck obeys
\be
\la{9.10}
e^{-L}=\big(2\sinh\eta_1\big)\big(2\sinh\eta_2\big)
\Big(2\cosh\frac{x_{12}}{2}\Big)^{-2}.
\ee
This is the JT relation \eqref{A.2}, with $x'(u_i)$ replaced by $2\sinh\eta_i=2b/\wt\beta_i$. It is derived in App.~\ref{app:sd-classical}, together with the subtraction at the endpoints that renormalizes $L$; the JT corner fixes this subtraction at leading order in $\l$.

The quantity $\overline Z_\Delta$ is thus the amplitude of a classical trumpet
of label $b$ times $(2\sinh\eta)^\Delta=(2b/\wt\beta)^\Delta$, the factor of $e^{-\Delta L}$ in \eqref{9.10} that depends only on the boundary of this trumpet. A natural ansatz is that $\overline Z_\Delta$ is the would-be matter-decorated trumpet of sine-dilaton gravity evaluated in the classical regime: the probe leaves the matter-free trumpet amplitude unchanged, and each trumpet carries the factor of $e^{-\Delta L}$ on its side, as the endpoint factors $x_i'(u_i)^\Delta$ do in JT at fixed neck (App.~\ref{app:neck-unfold}). The remaining factor $(2\cosh\frac{x_{12}}2)^{-2\Delta}$ is left to the neck: its integral over the twist, with the sum over windings, gives $\Gamma(\Delta)^2/\Gamma(2\Delta)$ as in \eqref{A.3}, the limit of $\l w_\Delta(m)$ in \eqref{8.1}, which holds in the classical regime as well since the weights do not depend on $\beta$.

In this interpretation, it would be tempting to read from the first relation in \eqref{9.6} that the matter line crossing the neck shifts its length from $\l m$ to $\l(m+\Delta)$. At $\Delta=1$ this reading is exact, by \eqref{9.7}, since the matter line acts as one of the $m+1$ chords through the neck (Sec.~\ref{sec:fq-D1}). This reading is not forced for general $\Delta$. By the second relation in \eqref{9.6}, the shift of the label by $\Delta$ cannot be distinguished from the factor $e^{-\Delta\eta}$, which amounts to subtracting $\eta$ from $L$, a subtraction that depends on the boundary length. Moreover, at integer $\Delta$ the amplitude is, by \eqref{7.3}, a combination of
ordinary trumpets of labels $m,m+2,\ldots,m+2\Delta$, while still obeying \eqref{9.6} at leading order, so that no sharp label is needed. Whether the threaded neck has length $\l m$
or $\l(m+\Delta)$, or no sharp length, is an open question.

An intermediate step towards a closed-channel quantization of sine-dilaton gravity with matter is the heavy-matter regime, with $\Delta\sim1/\l$ so that $q^\Delta$ stays fixed as $\l\to0$. Then $\l\Delta$ is of order one, so that a shift of the label by $\Delta$ changes the neck length at the classical level, and the matter line backreacts on the geometry, as in \cite{Bettaque:2026syk}. This is beyond the scope of this paper: the classical saddle must then include the backreacting matter line, and the asymptotics \eqref{9.6}, derived at fixed $\Delta$, do not apply.

\section{Outlook}
\la{sec:discussion}

The duality between DSSYK and sine-dilaton gravity is established at the level of the disk, and only specific amplitudes of higher topology have been derived in the bulk. On the DSSYK side, amplitudes beyond the disk with matter may be defined by the ETH two-matrix model of the Hamiltonian and the matter operator, used as a completion of the disk to all topologies \cite{Jafferis:2022uhu,Okuyama:2023kdo}, and the matter-decorated wormhole is its genus-zero two-boundary amplitude with one Gaussian matter line \cite{Okuyama:2023yat}. In this framework, we have obtained the gluing of the wormhole at finite $q$, \ie beyond the JT limit, into matter-decorated trumpets, labelled by the number of chords through the neck. It extends the JT gluing at fixed neck circumference and reduces to it in the JT limit. In the classical regime of sine-dilaton gravity, the trumpets take the form expected for a probe matter line on a pure-gravity trumpet. The gluing is exact and derived entirely from chord diagrams. Trumpets and gluing weight are obtained in closed form, the latter by summing over the twists and windings of the chords through the neck and over the closed loops. The gluing is unique, up to normalization, among the gluings whose one-boundary factors have the threshold property, and the normalization is fixed by assigning the closed loops to the gluing weight.

The most direct open problem is a bulk derivation. Our trumpets and gluing weight are defined by chord diagrams. In sine-dilaton gravity they should arise from the quantization of closed slices crossed by a matter line of fixed dimension, with $w_\Delta(m)$ as the measure on the states labelled by $m$. The closed-channel quantization without matter \cite{Blommaert:2025avl}, the quantization with an observer of dynamical mass \cite{Blommaert:2025rgw} and the Hilbert space with matter on slices joining the two boundaries \cite{Cui:2025sgy} are natural starting points. A closed slice crossed by a matter line appears in the chord description of the baby universe of two coupled DSSYK models \cite{Sontag:2026iiu}: the slice through the handle is crossed by the worldline of a heavy operator and by $b\geq0$ Hamiltonian chords, and the states labelled by $b$ span the Hilbert space of the baby universe (see also \cite{Aguilar-Gutierrez:2026aym}). There each chord through the handle carries a fixed suppression factor and, in the ribbon phase of that model, the state with $b$ chords has the norm $[b]_q!=\sum_{\sigma\in\mathsf S_b}q^{N(\sigma,\bm 0)}$ of the chord inner product, which is the sum over twists in \eqref{6.3} without windings. In $w_\Delta(m)$ the same sum is extended to the windings around the matter line and to the closed loops. Whether the sectors $m$, with this measure, define a Hilbert space of the same kind is left open.

It would be interesting to see whether our construction extends to wormholes threaded by several matter
lines, crossing or not. For two non-crossing matter lines threading the same neck, each boundary is divided into two arcs. The gluing weight should be a matrix, indexed by the way the chords through the neck are distributed between the two arcs. Another extension is to the decorated wormholes of deformed DSSYK
models, with Askey--Wilson spectral weights
\cite{Aguilar-Gutierrez:2026nmd} or finite-cutoff branes
\cite{Aguilar-Gutierrez:2026ogo,Aguilar-Gutierrez:2026fcg}. Whenever the
wormhole has a spectral representation \eqref{1.1} with a positive density of compact support that is an even function of the energy, the uniqueness argument of Sec.~\ref{sec:general} applies
unchanged, whereas the loop criterion and the shifted label depend
on how the deformation enters the chord expansion.

Beyond the wormhole, the one-boundary amplitudes and weights discussed in this paper may be used to sew higher-genus amplitudes with Gaussian matter. Without matter, the correlators of the DSSYK matrix
model are sewn from the trumpets $I_b(\wt\beta)$, the gluing factor $b$ and
discrete analogues of the Weil--Petersson volumes
\cite{Mirzakhani:2006fta,Mirzakhani:2006eta,Eynard:2007kz,Okuyama:2023kdo,Raz:2025wjw,Giacchetto:2025nnb,DoNorbury:2025qmirz,DoNorbury:2025dssyk,Ahmed:2026qWP}.
With matter, the decorated trumpets and the weight $w_\Delta(m)$ would
replace trumpet and gluing factor on the threaded neck, and the missing
ingredient is the decorated analogue of the volumes, the counterpart of
the sewing with matter geodesics of JT gravity
\cite{Saad:2019pqd,Jafferis:2022wez,Stanford:2022fdt,Iliesiu:2024cnh}.
In such a sewing, a threaded neck would be summed over $m\geq0$ with the
weight $w_\Delta(m)$, the zero label included, and an empty neck over $b\geq1$
with the measure $b$. With the prescription of Sec.~\ref{sec:fq-zero-label} for the matter-free limit, the threaded-neck weight reduces to the empty-neck measure as $\Delta\to0$, for $m\geq1$. The two rules differ by whether a matter line is present to measure the common winding of the chords through the neck: the winding is summed over for the threaded neck and quotiented for the empty one. Whether the two rules combine consistently in a single sewing is left as an interesting open problem.
The genus-one correction to $\mc W_\Delta$ in the ETH two-matrix model of
DSSYK \cite{Jafferis:2022uhu,Okuyama:2023aup,Alfinito:2026pst,Okuyama:2024eyf}
is a concrete target where such a sewing can be examined, and the
supersymmetric chord Hilbert spaces
\cite{Berkooz:2020SUSYDS,Boruch:2023N2wormholes,Ahmed:2026N2volumes} offer
a parallel setting.

\acknowledgments

This work is supported by the INFN grant GAST.
The use of the AI model Claude Fable 5.2 (Anthropic) is acknowledged for assistance with
computations, streamlining of some derivations, explicit enumeration checks, and the TikZ code of the figures. The author takes full responsibility for the accuracy
and content of the manuscript.

\appendix

\section{Fixed-neck JT amplitudes: classical gluing and neck profiles}
\la{app:JT-dictionary}

This appendix collects various fixed-neck JT computations used in the main text.

\subsection{Fixed-neck gluing}
\la{app:neck-unfold}

The matter line on the JT double trumpet represents propagation of a bulk scalar dual to the
boundary operator $\mc O$, with mass squared
$m_{\rm scalar}^2=\Delta(\Delta-1)$ in units of the AdS$_2$ radius.
On the AdS$_2$ covering space, $SL(2,\mathbb R)$ covariance fixes its
boundary-to-boundary propagator, up to a position-independent
normalization, to $e^{-\Delta L}$, where $L$ is the renormalized
geodesic length between the endpoints
\cite{Mertens:2017mtv,Mertens:2022irh}. On the double trumpet, the
propagator is obtained by summing these contributions over all integer
winding sectors \cite{Saad:2019pqd,Jafferis:2022wez}.

We now express the matter propagator in terms of the boundary
profiles, before averaging over their fluctuations.
For fixed boundary curves $x_1(u)$ and $x_2(u)$ on the
two sides of the double trumpet, we denote the relative
position of the marked endpoints on the covering space by
$x_{12}=x_0+\alpha+\ell\mathfrak b$, where
$x_0=x_1(u_1)-x_2(u_2)$.
Here $\alpha\in[0,\mathfrak b)$ is the relative twist
and $\ell\in\mathbb Z$ labels the matter-geodesic winding sector. Then, we use the unfolding identity\footnote{\label{fn:unfolding}%
For the undecorated double trumpet there is no sum over marked matter paths, and the twist integral alone gives the factor $\mathfrak b$ in the usual gluing measure.}
\be
\la{A.1}
\sum_{\ell\in\mathbb Z}\int_0^{\mathfrak b}d\alpha\,
F(x_0+\alpha+\ell\mathfrak b)
=\int_{-\infty}^{\infty}dx_{12}\,F(x_{12}),
\ee
applied to $F=e^{-\Delta L}$, where the renormalized length $L$ of
the geodesic joining these endpoints obeys\footnote{%
This is the $\Delta=1$ form of the standard Schwarzian
transformation law for the renormalized two-point function
$e^{-\Delta L}$ (the bilocal operator)
\cite{Blommaert:2018oro}: each endpoint contributes one factor of the boundary Jacobian
$x'(u)$, while the dependence on the relative hyperbolic coordinate remains
$(2\cosh(x_{12}/2))^{-2}$. For the classical trajectories
$x_i'=\mathfrak b/\tau_i$, this reduces to the
classical length relation used in
App.~\ref{app:JT-classical}.}
\be
\la{A.2}
e^{-L}=x_1'(u_1)\,x_2'(u_2)\,
\Big(2\cosh\frac{x_{12}}{2}\Big)^{-2}.
\ee
The resulting
relative-position integral in \eqref{A.1} is
\be
\la{A.3}
\int_{-\infty}^{\infty}dx_{12}\,
\Big(2\cosh\frac{x_{12}}{2}\Big)^{-2\Delta}
=\frac{\Gamma(\Delta)^2}{\Gamma(2\Delta)}.
\ee
The factor \eqref{A.3} is independent of both the fluctuating curves
and the neck circumference.
The matter insertion therefore leaves only the separate
endpoint factors $[x_1'(u_1)]^\Delta$ and
$[x_2'(u_2)]^\Delta$, and the two boundary path integrals
decouple at fixed neck.

\subsection{Classical JT gluing and the small-\texorpdfstring{$\tau$}{tau} limit}
\la{app:JT-classical}

At small boundary lengths $\tau_i$ and fixed neck circumference $\mathfrak b$, the
Schwarzian boundaries are classical, $x_i(u)=\mathfrak b u/\tau_i$.
The
renormalized length $L$ of the geodesic joining the matter endpoints on them,
at relative position $x_{12}$ on the covering space as in \eqref{A.1},
is given by \eqref{A.2} at $x_i'=\mathfrak b/\tau_i$:\footnote{\label{fn:geodesic}In
the double-trumpet metric $d\sigma^2+\cosh^2\sigma\,dx^2$ of
Sec.~\ref{sec:neck-cut}, with cutoff $\epsilon$, place the regulated
boundaries at $\sigma=R_1$ and $\sigma=-R_2$.
The hyperbolic distance $D$ satisfies
$\cosh D=\cosh R_1\cosh R_2\cosh x_{12}+\sinh R_1\sinh R_2$, so for $R_{i}\to\infty$,
$D=R_1+R_2+2\log\cosh(x_{12}/2)+O(e^{-2R_i})$.
 The induced boundary metric fixes $e^{R_i}\simeq2/(\epsilon x_i')$, \ie $2\tau_i/(\epsilon\mathfrak b)$
 for $x_i=\mathfrak b u/\tau_i$.
 Subtracting the standard $-2\log\epsilon$ divergence gives \eqref{A.4}.}
\be
\la{A.4}
e^{-L}=\frac{\mathfrak b^2}{\tau_1\tau_2}\,\Big(2\cosh\frac{x_{12}}{2}\Big)^{-2}.
\ee
In the scalar propagator $e^{-\Delta L}$, as in \eqref{A.1}, the winding sum and the twist integral over $[0,\mathfrak b)$
unfold $x_{12}$ onto the real line, and the $x_{12}$ integral of $(2\cosh\frac{x_{12}}2)^{-2\Delta}$ gives the factor
\eqref{A.3}.
Thus,  we obtain the classical JT gluing formula
\be
\la{A.5}
\mc W^{\rm cl}_\Delta(\tau_{1}, \tau_{2})
=
\frac{\Gamma(\Delta)^2}{\Gamma(2\Delta)}
\int_0^\infty d\mathfrak b\;
\Big[Z_{\rm trumpet}^{\rm JT}(\mathfrak b,\tau_1)\Big(\frac{\mathfrak b}{\tau_1}\Big)^{\!\Delta}\Big]
\Big[Z_{\rm trumpet}^{\rm JT}(\mathfrak b,\tau_2)\Big(\frac{\mathfrak b}{\tau_2}\Big)^{\!\Delta}\Big].
\ee
The gluing measure $d\mk b$ is flat and the integrand factorizes: each bracket is
$Z^{\rm fix}_\Delta(\mathfrak b,\tau_i)$ of \eqref{4.13} with
$\langle e^{\Delta\varphi(u_0)}\rangle_{\mathfrak b,\tau_i}$ replaced by its probe-geodesic value $(\mathfrak b/\tau_i)^\Delta$.
Using \eqref{4.7}, the Gaussian neck integral gives the leading term
of \eqref{4.6} as $S=\tau_1+\tau_2\to0$,
$\mc W^{\rm cl}_\Delta=\Gamma(\Delta)S^{-\Delta-\frac12}/(2\sqrt\pi)$.

\subsection{Neck profiles}
\la{app:neck-bp}

The explicit neck profiles in \eqref{4.15} are obtained from
\cite{Jafferis:2022wez} by splitting off the factor $|h_\Delta^{\rm Sch}(k)|$, so that they have unit oscillatory
amplitude at large $\mathfrak b$, as follows from the large-argument asymptotics of ${}_2F_1$. For $k>0$, in terms of the Gauss
hypergeometric function ${}_2F_1$,
\bea
\la{A.6}
\chi_{\Delta,k}(\mathfrak b)
&=\frac{\sqrt\pi}{\Gamma(\Delta+\frac12)}
|h_\Delta^{\rm Sch}(k)|\,\mathfrak a_\Delta(\mathfrak b)
\
{}_2F_1\!\left(\frac12+2ik,\frac12-2ik;
\Delta+\frac12;-\sinh^2\frac{\mathfrak b}{4}\right),
\\
\mathfrak a_\Delta(\mathfrak b)
&=\left(\sinh\frac{\mathfrak b}{4}\right)^\Delta
\left(\cosh\frac{\mathfrak b}{4}\right)^{1-\Delta}.
\eea
At $\Delta=1$, where $|h^{\rm Sch}_1(k)|=2k$ and the hypergeometric function reduces to $\sin(k\mathfrak b)/(4k\sinh(\mathfrak b/4))$, one finds $\chi_{1,k}=\sin(k\mathfrak b)$.

\paragraph{Proof of the Morse relation \eqref{5.3}.}

The profile \eqref{A.6} has the integral representation
\be
\la{A.7}
|h_\Delta^{\rm Sch}(k)|\,|\Gamma(2ik)|^2\,\chi_{\Delta,k}(\mathfrak b)
=\int_0^\infty dx\,x^{\Delta-1}\,
e^{-x\coth\frac{\mathfrak b}{2}}\,K_{2ik}\Big(\frac{x}{\sinh\frac{\mathfrak b}{2}}\Big).
\ee
The two forms are related by the Laplace transform of the Macdonald function
\cite[6.621.3]{Gradshteyn:2014}. For $\kappa>-\frac12$ the operator $H_\kappa$ of \eqref{5.3} has no bound states, and its
eigenfunctions of energy $k^2$, $k>0$, are $\psi_k(\varphi)=\mathcal N_k\,e^{-\varphi/2}\,W_{-\kappa,ik}(e^\varphi)$, with $W$ the
Whittaker function. Their behavior at $\varphi\to-\infty$, a standing wave of amplitude $2|\Gamma(2ik)|/|\Gamma(\frac12+\kappa+ik)|$,
fixes the normalization to $\delta(k-k')$,
$\mathcal N_k^2=|\Gamma(\frac12+\kappa+ik)|^2/(2\pi|\Gamma(2ik)|^2)$, so that
\be
\la{A.8}
\Tr\!\left(e^{-\tau H_\kappa}e^{\Delta\varphi}\right)
=\int_0^\infty dk\,e^{-\tau k^2}\,\mathcal N_k^2\int_0^\infty dx\,x^{\Delta-2}\,W_{-\kappa,ik}(x)^2.
\ee
The square of the Whittaker function has the integral representation \cite[6.669.4]{Gradshteyn:2014}
\be
\la{A.9}
W_{-\kappa,ik}(x)^2=\frac{2x}{|\Gamma(\frac12+\kappa+ik)|^2}\int_0^\infty dt\;
e^{-x\cosh t}\,\tanh^{2\kappa}\frac t2\;K_{2ik}(x\sinh t).
\ee
With $\tanh\frac t2=e^{-\mathfrak b/2}$ one has $\cosh t=\coth\frac{\mathfrak b}{2}$, $\sinh t=1/\sinh\frac{\mathfrak b}{2}$ and
$dt=d\mathfrak b/(2\sinh\frac{\mathfrak b}{2})$, up to orientation, so that the measure $\tanh^{2\kappa}\frac t2\,dt$ is the kernel
$e^{-\kappa\mathfrak b}\,d\mathfrak b/(2\sinh\frac{\mathfrak b}{2})$ of \eqref{5.3}. Inserting \eqref{A.9} in \eqref{A.8},
the factors $|\Gamma(\frac12+\kappa+ik)|^2$ cancel, the integral over $x$ is \eqref{A.7}, and
\be
\mathcal N_k^2\int_0^\infty dx\,x^{\Delta-2}\,W_{-\kappa,ik}(x)^2
=\frac1\pi\,|h_\Delta^{\rm Sch}(k)|\int_0^\infty d\mathfrak b\,
\frac{e^{-\kappa\mathfrak b}}{2\sinh\frac{\mathfrak b}{2}}\,\chi_{\Delta,k}(\mathfrak b).
\ee
By \eqref{4.15}, the integral over $k$ with the factor $e^{-\tau k^2}$ is \eqref{5.3}. The relation thus holds for every
$\Delta>0$ and $\kappa>-\frac12$, momentum by momentum.

The functions $\chi_{\Delta,k}$ satisfy the orthogonality relation \cite{Jafferis:2022wez},
which, converted to the normalization \eqref{A.6}, reads
\be
\la{A.11}
\int_0^\infty d\mathfrak b\,
\chi_{\Delta,k}(\mathfrak b)\chi_{\Delta,k'}(\mathfrak b)
=\frac\pi2\delta(k-k').
\ee
Substituting \eqref{4.15} in the geometric gluing formula \eqref{4.13} and
using this relation yields, for $S=\tau_1+\tau_2$,
\bea
\la{A.12}
&\frac{\Gamma(\Delta)^2}{\Gamma(2\Delta)}
\int_0^\infty d\mathfrak b\,
Z^{\rm fix}_\Delta(\mathfrak b,\tau_1)Z^{\rm fix}_\Delta(\mathfrak b,\tau_2)
=C_\Delta^{\rm Sch}\int_0^\infty dk\,
|h_\Delta^{\rm Sch}(k)|^2e^{-Sk^2}
=\mc W^{\rm JT}_\Delta(S),
\eea
so that \eqref{4.13} reproduces the spectral wormhole amplitude
\eqref{4.6}, including its overall normalization.
Near $\mathfrak b=0$, the two independent solutions of the
P\"oschl--Teller equation \eqref{4.16} behave as $\mathfrak b^\Delta$
and $\mathfrak b^{1-\Delta}$ (at $\Delta=1/2$ the exponents coincide and
the second solution contains $\mathfrak b^{1/2}\log\mathfrak b$); the
expression \eqref{A.6}
selects the solution behaving as $\mathfrak b^\Delta$, normalized as in
\eqref{A.11}, since
$\mathfrak a_\Delta\simeq(\mathfrak b/4)^\Delta$ and ${}_2F_1\to1$ as
$\mathfrak b\to0$.

\section{Proof of the JT limit}
\la{app:fq-JT}

The limits \eqref{8.1} of Sec.~\ref{sec:fq-JT-fixed} and the convergence of the gluing sum are proved in the following proposition, with precise conditions under which they hold,
\begin{proposition}[JT limit]
\la{prop:JT}
Let $\Delta>0$ be real and $\tau,\mathfrak b>0$. In the JT scaling
\eqref{4.1}, along any sequence $(\l,m)$ with $m\to\infty$ and $\l m\to\mathfrak b$ (equivalently
$\mathfrak b_m\to\mathfrak b$, since $\mathfrak b_m-\l m=\l\Delta\to0$), the limits \eqref{8.1} hold, \ie
\be
\l\,w_\Delta(m)\ \longrightarrow\
\frac{\Gamma(\Delta)^2}{\Gamma(2\Delta)},
\qquad
\l^{-(\Delta+1)}e^{-\wt\beta}\,Z_\Delta(m,\beta)
\ \longrightarrow\ Z^{\rm fix}_\Delta(\mathfrak b,\tau).
\nonumber\ee
Moreover, for $\tau_1,\tau_2>0$ and fixed $0<\mathfrak b'<\mathfrak b''$, the terms of the gluing sum \eqref{6.5} with $\mathfrak b'\leq\mathfrak b_m\leq\mathfrak b''$, multiplied by $\l^{-2\Delta}e^{-\wt\beta_1-\wt\beta_2}$, sum to the JT gluing integral \eqref{4.13} restricted to $[\mathfrak b',\mathfrak b'']$ as $\l\to0$, and the remaining terms vanish as $\l\to0$ and then $\mathfrak b'\to0$, $\mathfrak b''\to\infty$.
\end{proposition}
\begin{proof}
The weight limit is derived in Sec.~\ref{sec:fq-JT-fixed}; the rest of this appendix proves the amplitude limit, for every real $\Delta>0$, and the statement on the gluing sum.

The ingredients are standard: for $\Delta\geq1$, the continuum limit of the Rogers
polynomial follows from the convergence of $q$-ultraspherical polynomials to Jacobi functions
\cite{Koornwinder:1990Jacobi,Koelink:1995Identities}, and its
identification with the neck profiles $\chi_{\Delta,k}$ of
\cite{Jafferis:2022wez} follows from the hypergeometric representation of a Legendre
function \cite{DLMF}. The new ingredients here are the control of the two
ends of the Fourier sum for $0<\Delta<1$, where the limiting integrand
is unbounded, the tail bound that allows the limit to be taken under
the thermal integral, the selection of the small-neck branch, and the
convergence of the gluing sum itself. Although they rely on simple estimates, their derivation is given in full for completeness.

\paragraph{Rogers polynomials at fixed neck.}
Fix $\tau,\mathfrak b>0$, take the scaling \eqref{4.1} along any
sequence with $\mathfrak b_m=\l(m+\Delta)\to\mathfrak b$, and set
$\theta=\l k$ in \eqref{6.7}.
We write $\varrho_j=(q^\Delta;q)_j/(q;q)_j$. The Fourier expansion of the Rogers polynomial
\eqref{6.6} reads
\be
\la{B.1}
\l^{2\Delta-1}C_m(\cos\l k;q^\Delta|q)
=\l\sum_{j=0}^m
(\l^{\Delta-1}\varrho_j)
(\l^{\Delta-1}\varrho_{m-j})e^{i(m-2j)\l k}.
\ee
The product representation of $\varrho_{j}$ and its limit are
\bea
\la{B.2}
\varrho_j
&=\frac{(q^\Delta;q)_\infty}{(q;q)_\infty}
\frac{(q^{j+1};q)_\infty}{(q^{j+\Delta};q)_\infty},
\qquad
\l^{\Delta-1}\varrho_j
\longrightarrow\frac{(1-e^{-y})^{\Delta-1}}{\Gamma(\Delta)},
\qquad \l j\to y>0.
\eea
The convergence is uniform on compact positive $y$ intervals, by
the $q$-gamma limit \cite[Thm.~2.2]{Zhang:2008qg} and the product-ratio
limit used in Sec.~\ref{sec:fq-JT-fixed}. Near the two ends of the sum in \eqref{B.1},
$\l j\to0$ and $\l(m-j)\to0$, where convergence is not uniform, we use the bound
\be
\la{B.3}
\varrho_j\leq c(j+1)^{\Delta-1},\qquad 0\leq j\leq m,
\ee
with $c$ independent of small $\l$.\footnote{
By definition, $\varrho_j=\prod_{i=0}^{j-1}(1-q^{\Delta+i})/(1-q^{1+i})$, and each factor is at
most $e^{\l}(\Delta+i)/(1+i)$. For $\Delta\geq1$ this holds because $(1-q^x)/x$ is decreasing in $x$; for $\Delta<1$ it follows from $1-q^x=q^x(q^{-x}-1)$, because $(q^{-x}-1)/x$ is increasing. Hence $\varrho_j\leq e^{\l j}\,\Gamma(j+\Delta)/[\Gamma(\Delta)\,j!]$. Since $j\leq m$ and $\l m\to\mathfrak b$ in the scaling considered, the exponential $e^{\l j}\leq e^{\l m}$ is bounded uniformly in small $\l$. Finally, the quotient $\Gamma(j+\Delta)/[j!\,(j+1)^{\Delta-1}]$ is finite for every $j\geq0$ and tends to $1$ as $j\to\infty$, so it is bounded in $j$, which gives \eqref{B.3}.}

Let $0<\epsilon<\mathfrak b/2$. For $\l j\leq\epsilon$, $\l(m-j)$ is in a compact positive interval, so that $\l^{\Delta-1}\varrho_{m-j}$ is bounded, by \eqref{B.2}. With \eqref{B.3}, the absolute contribution of this range to the rescaled sum is therefore at most
\[
C\l^\Delta\sum_{j\leq\epsilon/\l}(j+1)^{\Delta-1}
\leq C'(\epsilon+\l)^\Delta,
\]
where the sum is bounded by an integral of the monotonic function $x^{\Delta-1}$. The other end is
identical by $j\leftrightarrow m-j$. Taking $\l\to0$ at fixed $\epsilon$, and then
$\epsilon\to0$, proves Riemann-sum convergence for all $\Delta>0$,
uniformly on compact $k$ intervals.
By \eqref{B.2}, the limit of the sum \eqref{B.1} is the integral over $y\in(0,\mathfrak b)$, with $\l j\to y$, of $[(1-e^{-y})(1-e^{-(\mathfrak b-y)})]^{\Delta-1}e^{i(\mathfrak b-2y)k}/\Gamma(\Delta)^2$. To evaluate it, set $y=\mathfrak b/2-v$ and use
\[
(1-e^{-y})(1-e^{-(\mathfrak b-y)})
=e^{-\mathfrak b/2}\left(2\cosh\frac{\mathfrak b}{2}-2\cosh v\right);
\]
the phase is $e^{2ikv}$ and, since the rest of the integrand is even in $v$, it may be replaced by $\cos(2kv)$. Hence
\be
\la{B.4}
\l^{2\Delta-1}C_m(\cos\l k;q^\Delta|q)
\longrightarrow
\frac{e^{(1-\Delta)\mathfrak b/2}}{\Gamma(\Delta)^2}
\mc P_\Delta(\mathfrak b,k),
\ee
where
\be
\la{B.5}
\mc P_\Delta(\mathfrak b,k)
=\int_{-\mathfrak b/2}^{\mathfrak b/2}dv\,
\left(2\cosh\frac{\mathfrak b}{2}-2\cosh v\right)^{\Delta-1}
\cos(2kv).
\ee

\paragraph{Thermal integral and identification.}
By \eqref{6.7}, the amplitude is a prefactor times the integral of the dressed weight, the Rogers polynomial and the Boltzmann factor. The limit of the polynomial is \eqref{B.4}. For the prefactor, the same product asymptotics give
\be
\la{B.6}
\l^{-\Delta}(1-q^{m+\Delta})
\frac{(q^\Delta;q)_m}{(q^{2\Delta};q)_m}
\longrightarrow
\frac{\Gamma(2\Delta)}{\Gamma(\Delta)}
e^{(\Delta-1)\mathfrak b/2}
\left(2\sinh\frac{\mathfrak b}{2}\right)^{1-\Delta}.
\ee
For the dressed weight with the Boltzmann factor, the spectral-edge limit of Sec.~\ref{sec:spectral-ingredients} is,
uniformly on compact $k$ intervals \cite[App.~B,
Thm.~B.2]{Koornwinder:1990Jacobi},
\bea
\la{B.7}
\frac{\l^{1-2\Delta}}{2\pi}\,
\mc K_\Delta(\l k)e^{\wt\beta(\cos\l k-1)}
&\longrightarrow
C_\Delta^{\rm Sch}|h_\Delta^{\rm Sch}(k)|^2e^{-\tau k^2},
\eea
with $C_\Delta^{\rm Sch}$ and $h_\Delta^{\rm Sch}$ of \eqref{4.6}.
The convergence in \eqref{B.4} and \eqref{B.7} holds on compact $k$ intervals,
while after $\theta=\l k$ the integral \eqref{6.7} extends to
$k=\pi/\l$.
To bound the contribution from $k>K$ uniformly in small $\l$,
note that positivity of the Fourier coefficients gives
$|C_m(\cos\theta;q^\Delta|q)|\leq C_m(1;q^\Delta|q)$, so that \eqref{B.4} at $k=0$ bounds the rescaled polynomial uniformly
in $k$. We write the shifted Boltzmann factor in \eqref{B.7} as $e^{\wt\beta(\cos\theta-1)}=\big(e^{\wt\beta(\cos\theta-1)/2}\big)^2$. One factor is at most
$e^{-2\tau k^2/\pi^2}$, since $1-\cos\theta\geq2\theta^2/\pi^2$ and
$\wt\beta\geq2\tau/\l^2$. The other, integrated against the rescaled
spectral weight, is
$\l^{-2\Delta}e^{-\wt\beta/2}\mc W_\Delta(\beta/2)$, which has a
finite limit by \eqref{4.3}--\eqref{4.6}. With the bounded prefactor \eqref{B.6}, the contribution from $k>K$ is
$O(e^{-2\tau K^2/\pi^2})$.
Compact convergence and this tail bound justify combining the three limits
\eqref{B.4}, \eqref{B.6} and \eqref{B.7} under the integral;
the exponential factors in $\mathfrak b$ cancel, and the powers of
$\l$, with the Jacobian $d\theta=\l\,dk$, combine to $\l^{\Delta+1}$:
\be
\la{B.8}
\l^{-(\Delta+1)}e^{-\wt\beta}Z_\Delta(m,\beta)
\longrightarrow
\frac{(2\sinh\frac{\mathfrak b}{2})^{1-\Delta}}
{2\pi\Gamma(\Delta)}
\int_0^\infty dk\,|h_\Delta^{\rm Sch}(k)|^2e^{-\tau k^2}
\mc P_\Delta(\mathfrak b,k).
\ee

The Mehler--Dirichlet representation
\cite[Sec.~14.12(ii)]{DLMF}, with
$\xi=\mathfrak b/2$, $\mu=\Delta-\tfrac12$ and
$\nu=-\tfrac12+2ik$, evaluates \eqref{B.5} as an associated
Legendre function,
\be
\la{B.9}
\mc P_\Delta(\mathfrak b,k)
=\sqrt{2\pi}\,2^{\Delta-1}\Gamma(\Delta)
\left(\sinh\frac{\mathfrak b}{2}\right)^{\Delta-\frac12}
P^{\frac12-\Delta}_{-\frac12+2ik}
\!\left(\cosh\frac{\mathfrak b}{2}\right).
\ee
The hypergeometric form of $P^\mu_\nu$
\cite[Eq.~(14.3.6)]{DLMF} matches \eqref{A.6}, including the
prefactor, and gives
\be
\la{B.10}
\frac{(2\sinh\frac{\mathfrak b}{2})^{1-\Delta}}
{2\Gamma(\Delta)}\,\mc P_\Delta(\mathfrak b,k)
=\frac{\chi_{\Delta,k}(\mathfrak b)}{|h_\Delta^{\rm Sch}(k)|},
\qquad k>0.
\ee
The integral representation requires $\mu>-1/2$, precisely
$\Delta>0$.
Equation \eqref{B.8} is consequently \eqref{4.15}, proving the
amplitude limit \eqref{8.1}.

\paragraph{Small-neck branch.}

The branch at $\mathfrak b=0$ can also be read directly from
\eqref{B.5}: at $k=0$, rescaling $v=\mathfrak b s/2$ gives a beta integral, and at fixed $k$
\be
\la{B.11}
\mc P_\Delta(\mathfrak b,k)
=\mathfrak b^{2\Delta-1}
\frac{\Gamma(\Delta)^2}{\Gamma(2\Delta)}
\left[1+O(\mathfrak b^2)\right],
\ee
which gives $Z^{\rm fix}_\Delta(\mathfrak b,\tau)\simeq\mathfrak b^\Delta\mc W^{\rm JT}_\Delta(\tau)/\Gamma(\Delta)$ as in \eqref{4.17}, including at
$\Delta=1/2$, where the indicial exponents $\Delta$ and $1-\Delta$ of \eqref{4.16} coincide (App.~\ref{app:neck-bp}).

\paragraph{Convergence of the gluing sum.}
With
$\hat Z_m(\tau)\equiv\l^{-(\Delta+1)}e^{-\wt\beta}
Z_\Delta(m,\beta)$, we write the rescaled gluing sum as
\[
G_\l\equiv\l\sum_{m\geq0}[\l w_\Delta(m)]\,\hat Z_m(\tau_1)\hat Z_m(\tau_2),
\]
which by \eqref{6.5} equals $\l^{-2\Delta}e^{-\wt\beta_1-\wt\beta_2}\mc W_\Delta(\beta_1+\beta_2)$.
The amplitude limit holds along every sequence
$\mathfrak b_m\to\mathfrak b>0$ and is continuous in $\mathfrak b$, so it is uniform on compact positive neck intervals, as is
the weight limit. The part of $G_\l$ with
$\mathfrak b'\leq\mathfrak b_m\leq\mathfrak b''$ therefore converges to the JT integral \eqref{4.13} restricted to $[\mathfrak b',\mathfrak b'']$.
By \eqref{A.12}, the full JT integral equals $\mc W^{\rm JT}_\Delta(\tau_1+\tau_2)$, the limit of $G_\l$ by \eqref{4.3}, so the complementary part of $G_\l$ tends to the JT integral over the complement, which vanishes as $\mathfrak b'\to0$ and $\mathfrak b''\to\infty$. All summands are nonnegative
(Sec.~\ref{sec:general}), so both tails,
including $m=0$, are controlled without cancellations.
\end{proof}

\section{Properties of the Rogers polynomials}
\la{app:fq-spectral}

This appendix collects the properties of the Rogers
polynomials used in Secs.~\ref{sec:fq-chords} and~\ref{sec:fq-spectral}.
Using the expression \eqref{2.15} of $\mc K_\Delta$, one finds that they satisfy
the orthogonality relation
\cite[Sec.~13.2]{Ismail:2009OrthogonalPolynomials}
\bea
\la{C.1}
&\int_0^\pi\frac{d\theta}{2\pi}\,\mc K_\Delta(\theta)
C_m(\cos\theta;t|q)C_n(\cos\theta;t|q)
=h_m\,\delta_{mn},\qquad h_m\equiv\frac{(t^2;q)_m}{(1-tq^m)(q;q)_m},\qquad t=q^\Delta.
\eea

\paragraph{Trumpet form.}
The expansion \eqref{7.3} follows, for every $\Delta>0$, from the connection formula, which expresses the Chebyshev polynomials of the second
kind, $U_n(\cos\theta)=\sin((n+1)\theta)/\sin\theta$, in Rogers polynomials
\cite{AskeyIsmail:1983Ultraspherical,Ismail:2009OrthogonalPolynomials},
\be
\la{C.2}
U_n(x)=\sum_{k=0}^{\lfloor n/2\rfloor}
\frac{t^k\,(q/t;q)_k\,(q;q)_{n-k}}{(q;q)_k\,(qt;q)_{n-k}}\,
\frac{1-tq^{n-2k}}{1-t}\;C_{n-2k}(x;t|q).
\ee
 By the orthogonality relation \eqref{C.1}, the integral of
$\mc K_\Delta\,C_m\,U_n$ vanishes unless $n=m+2k$ with $k\geq0$, when it is
$h_m$ times the coefficient of $C_m$ in $U_{m+2k}$. With $N_m$ the prefactor of
\eqref{6.7}, so that $N_mh_m=(t;q)_m/(q;q)_m$, this gives
\be
\la{C.3}
N_m\int_0^\pi\frac{d\theta}{2\pi}\,\mc K_\Delta(\theta)\,
C_m(\cos\theta;t|q)\,U_{m+2k}(\cos\theta)
=b_k\equiv t^k\,\frac{(q/t;q)_k\,(q^{m+1};q)_k}{(q;q)_k\,(tq^{m+1};q)_k}.
\ee
 Since $e^{\wt\beta\cos\theta}=I_0(\wt\beta)+2\sum_{n\geq1}I_n(\wt\beta)\cos n\theta$ by \eqref{2.8}, and $2\cos n\theta=U_n-U_{n-2}$ with $U_{-1}=0$, the Boltzmann factor reads
$e^{\wt\beta\cos\theta}=\sum_{n\geq0}\big[I_n(\wt\beta)-I_{n+2}(\wt\beta)\big]U_n(\cos\theta)$,
and \eqref{6.7} becomes
\be
\la{C.4}
Z_\Delta(m,\beta)=\sum_{k\geq0}b_k\,\big[I_{m+2k}(\wt\beta)-I_{m+2k+2}(\wt\beta)\big]
=\sum_{j\geq0}(b_j-b_{j-1})\,I_{m+2j}(\wt\beta),\qquad b_{-1}=0.
\ee
 It remains to show that $b_j-b_{j-1}=c^{(\Delta)}_j\gamma^{(\Delta)}_{m,j}$, with $t=q^\Delta$. At $j=0$ both sides equal one. For $j\geq1$, the ratio $b_j/b_{j-1}=(t-q^j)(1-q^{m+j})/[(1-q^j)(1-tq^{m+j})]$ and the identity $(t-q^j)(1-q^{m+j})-(1-q^j)(1-tq^{m+j})=(t-1)(1-q^{m+2j})$ give
\[
b_j-b_{j-1}=b_{j-1}\Big(\frac{b_j}{b_{j-1}}-1\Big)
=\frac{(t-1)(1-q^{m+2j})}{(1-q^j)(1-tq^{m+j})}\,b_{j-1}.
\]
Inserting $b_{j-1}$ from \eqref{C.3} and using $(t-1)\,t^{j-1}(q/t;q)_{j-1}=t^j(1/t;q)_j$, $(1-q^j)(q;q)_{j-1}=(q;q)_j$ and $(1-tq^{m+j})(tq^{m+1};q)_{j-1}=(tq^{m+1};q)_j$, one finds $c^{(\Delta)}_j\gamma^{(\Delta)}_{m,j}$, and \eqref{C.4} becomes \eqref{7.3}.
At positive integer $\Delta=d$, $(q/t;q)_k=(q^{1-d};q)_k$ vanishes for
$k\geq d$, and the sum stops at $j=d$.

\section{Proof of the sector factorization theorem}
\la{app:sectors}

We prove Theorem~\ref{prop:sectors}, \ie the factorization \eqref{6.2} with the weight $w_\Delta(m)$ of
\eqref{6.3}. The expansion of the wormhole trace is a sum over histories of the stack of open chords: sequences of openings and closings in boundary time. This description is sequential, and the weight of each closing depends on the whole stack, so that neither the number $m$ of chords joining the two boundaries nor their twists and windings can be read off directly. We therefore rewrite the sum as a sum over chord configurations, pairings of the insertion points drawn in the universal cover of the annulus and weighted by their crossings (Claim~1 of Sec.~\ref{app:sectors-histories}). This is the standard correspondence of the disk, carried over to the annulus. In the new description the weight is a product over pairs of crossing curves, and the crossings near each boundary and those around the neck can be separated (Claim~2 of Sec.~\ref{app:sectors-cut}), which gives the factorization. Fig.~\ref{fig:hist-conf}(b) shows the two descriptions of one diagram.

\subsection{From transfer-matrix histories to chord configurations}
\la{app:sectors-histories}

We start from the wormhole trace
\eqref{2.12}, with the evolution operators of the two boundaries kept
separate and $t=q^\Delta$,
\be
\la{D.1}
\mc W_\Delta(\beta_1+\beta_2)
=\Tr_{\mc H}\!\left[t^{\wh N}
e^{\wt\beta_1\wh D/2}e^{\wt\beta_2\wh D/2}\right],
\ee
and expand both exponentials. The
coefficient of $(\wt\beta_1/2)^{K_1}(\wt\beta_2/2)^{K_2}/(K_1!\,K_2!)$
is a sum over sequences of $K_1$ moves of $\wh D$ (actions of $A_+$ or $A_-$, which open or close a Hamiltonian chord) from boundary~1 and
$K_2$ moves from boundary~2.

We use weights corresponding to the rescaled basis of Sec.~\ref{sec:fq-normalization}, together with its picture of the $n$ open chords as a stack of horizontal lines: an opening
has weight $1$, and a closing across $c$ open chords has weight
$(1-q)q^c$. The $K_1$ boundary-1 moves act at the top of the stack,
the $K_2$ boundary-2 moves at the bottom. At the top or at the bottom of the stack, with $n$ open chords, the closing
weights sum to $\sum_{c=0}^{n-1}(1-q)q^c=1-q^n$, so moves at either end reproduce the matrix elements of $\wh D$ in the rescaled basis, $1$ for an opening and $1-q^n$ for a closing, and the two exponentials in \eqref{D.1} can act at opposite ends of the same stack.

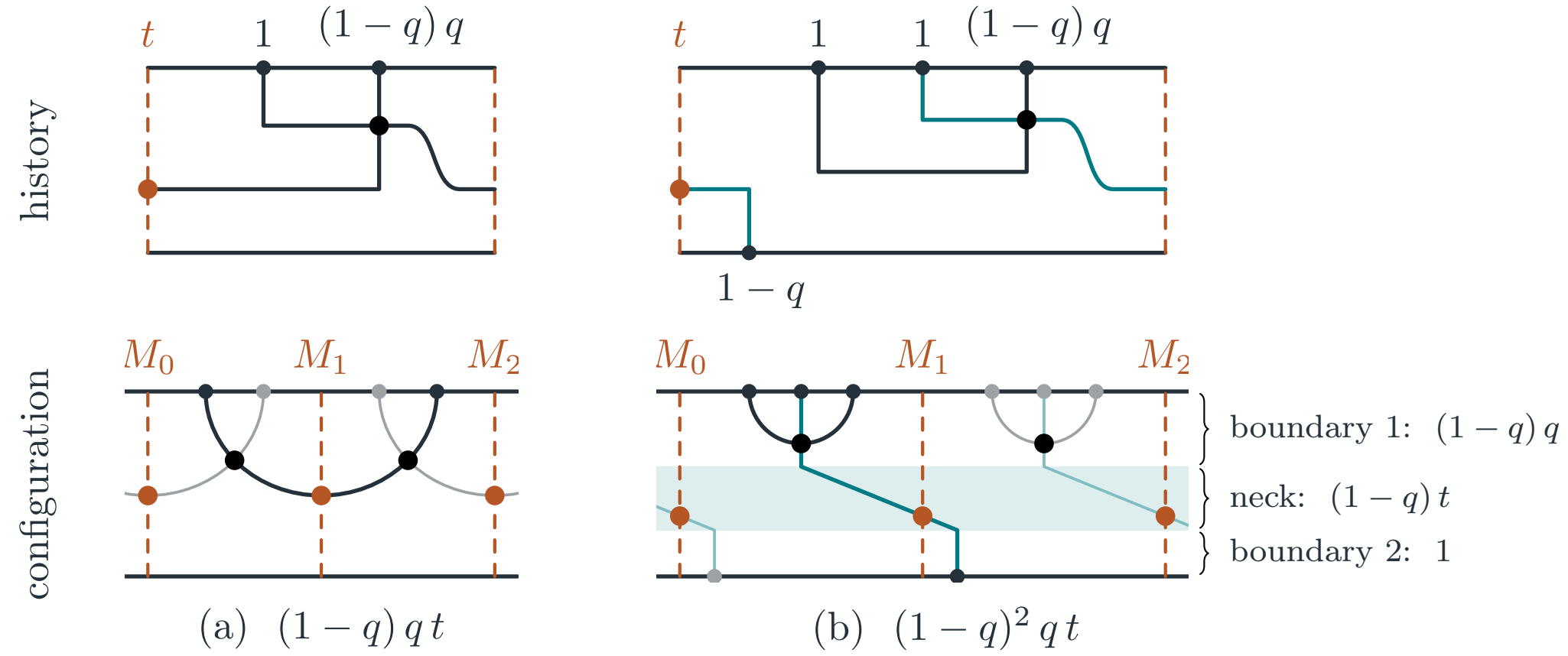
\begin{figure}[!t]
\centering
\begin{tikzpicture}[x=1cm,y=1cm,line cap=round,line join=round,font=\small,text=diagramInk]
\tikzset{bdy/.style={diagramInk,line width=1pt},
  mat/.style={diagramMatter,dashed,line width=0.8pt},
  ch/.style={diagramInk,line width=1pt},
  tr/.style={diagramInk!45,line width=0.7pt},
  sd/.style={diagramTeal,line width=1pt},
  sdt/.style={diagramTeal!50,line width=0.7pt},
  ins/.style={circle,fill=diagramInk,inner sep=1.3pt},
  xq/.style={circle,fill=black,inner sep=1.7pt},
  xt/.style={circle,fill=diagramMatter,inner sep=1.7pt}}
\node[rotate=90] at (-0.75,3.6) {history};
\node[rotate=90] at (-0.75,0.8) {configuration};
\begin{scope}[shift={(0,0)}]
 \begin{scope}[shift={(0.2,2.8)}]
  \draw[bdy] (0,1.6)--(3,1.6); \draw[bdy] (0,0)--(3,0);
  \draw[mat] (0,0)--(0,1.6); \draw[mat] (3,0)--(3,1.6);
  \draw[ch] (0,0.55)--(2,0.55)--(2,1.6);
  \draw[ch] (1,1.6)--(1,1.1)--(2.25,1.1) .. controls (2.5,1.1) and (2.45,0.55) .. (2.7,0.55)--(3,0.55);
  \node[ins] at (1,1.6) {}; \node[ins] at (2,1.6) {};
  \node[xq] at (2,1.1) {};
  \node[xt] at (0,0.55) {};
  \node[anchor=south] at (1,1.65) {$1$}; \node[anchor=south] at (2.1,1.65) {$(1-q)\,q$};
  \node[anchor=south,text=diagramMatter] at (0,1.65) {$t$};
 \end{scope}
 \begin{scope}[shift={(0.2,0)}]
  \clip (-0.2,-0.05) rectangle (3.2,2.15);
  \draw[bdy] (-0.2,1.6)--(3.2,1.6); \draw[bdy] (-0.2,0)--(3.2,0);
  \foreach \x/\l in {0/0,1.5/1,3/2} {\draw[mat] (\x,0)--(\x,1.6); \node[anchor=south,text=diagramMatter] at (\x,1.62) {$M_\l$};}
  \draw[tr] (-1.0,1.6) arc[start angle=180,end angle=360,x radius=1.0,y radius=0.9];
  \draw[ch] (0.5,1.6) arc[start angle=180,end angle=360,x radius=1.0,y radius=0.9];
  \draw[tr] (2.0,1.6) arc[start angle=180,end angle=360,x radius=1.0,y radius=0.9];
  \foreach \x in {0.5,2.5} \node[ins] at (\x,1.6) {};
  \foreach \x in {1.0,2.0} \node[ins,fill=diagramInk!45] at (\x,1.6) {};
  \foreach \x in {0,1.5,3} \node[xt] at (\x,0.7) {};
  \foreach \x in {0.75,2.25} \node[xq] at (\x,1.0047) {};
 \end{scope}
 \node at (1.7,-0.45) {(a)\ \ $(1-q)\,q\,t$};
\end{scope}
\begin{scope}[shift={(4.6,0)}]
 \begin{scope}[shift={(0.2,2.8)}]
  \draw[bdy] (0,1.6)--(4.2,1.6); \draw[bdy] (0,0)--(4.2,0);
  \draw[mat] (0,0)--(0,1.6); \draw[mat] (4.2,0)--(4.2,1.6);
  \draw[sd] (0,0.55)--(0.6,0.55)--(0.6,0);
  \draw[ch] (1.2,1.6)--(1.2,0.7)--(3.0,0.7)--(3.0,1.6);
  \draw[sd] (2.1,1.6)--(2.1,1.15)--(3.3,1.15) .. controls (3.55,1.15) and (3.5,0.55) .. (3.75,0.55)--(4.2,0.55);
  \node[ins] at (1.2,1.6) {}; \node[ins] at (2.1,1.6) {}; \node[ins] at (3.0,1.6) {}; \node[ins] at (0.6,0) {};
  \node[xq] at (3.0,1.15) {};
  \node[xt] at (0,0.55) {};
  \node[anchor=south] at (1.2,1.65) {$1$}; \node[anchor=south] at (2.1,1.65) {$1$}; \node[anchor=south] at (3.1,1.65) {$(1-q)\,q$};
  \node[anchor=north] at (0.7,-0.05) {$1-q$};
  \node[anchor=south,text=diagramMatter] at (0,1.65) {$t$};
 \end{scope}
 \begin{scope}[shift={(0.2,0)}]
  \begin{scope}
  \clip (-0.2,-0.05) rectangle (4.4,2.15);
  \fill[diagramTeal!13] (-0.2,0.4) rectangle (4.4,0.95);
  \draw[bdy] (-0.2,1.6)--(4.4,1.6); \draw[bdy] (-0.2,0)--(4.4,0);
  \foreach \x/\l in {0/0,2.1/1,4.2/2} {\draw[mat] (\x,0)--(\x,1.6); \node[anchor=south,text=diagramMatter] at (\x,1.62) {$M_\l$};}
  \draw[sdt] (-1.05,1.6)--(-1.05,0.95)--(0.3,0.4)--(0.3,0);
  \draw[sd]  (1.05,1.6)--(1.05,0.95)--(2.4,0.4)--(2.4,0);
  \draw[sdt] (3.15,1.6)--(3.15,0.95)--(4.5,0.4)--(4.5,0);
  \draw[ch] (0.6,1.6) arc[start angle=180,end angle=360,x radius=0.45,y radius=0.45];
  \draw[tr] (2.7,1.6) arc[start angle=180,end angle=360,x radius=0.45,y radius=0.45];
  \foreach \x in {0.6,1.05,1.5} \node[ins] at (\x,1.6) {};
  \foreach \x in {2.7,3.15,3.6} \node[ins,fill=diagramInk!45] at (\x,1.6) {};
  \node[ins] at (2.4,0) {}; \node[ins,fill=diagramInk!45] at (0.3,0) {};
  \foreach \x in {1.05,3.15} \node[xq] at (\x,1.15) {};
  \foreach \x in {0,2.1,4.2} \node[xt] at (\x,0.522) {};
  \end{scope}
  \draw[decorate,decoration={brace,amplitude=2pt}] (4.5,1.58)--(4.5,0.97);
  \node[anchor=west] at (4.62,1.275) {\scriptsize boundary 1:\ \ $(1-q)\,q$};
  \draw[decorate,decoration={brace,amplitude=2pt}] (4.5,0.93)--(4.5,0.42);
  \node[anchor=west] at (4.62,0.675) {\scriptsize neck:\ \ $(1-q)\,t$};
  \draw[decorate,decoration={brace,amplitude=2pt}] (4.5,0.38)--(4.5,0.02);
  \node[anchor=west] at (4.62,0.2) {\scriptsize boundary 2:\ \ $1$};
 \end{scope}
 \node at (2.5,-0.45) {(b)\ \ $(1-q)^2\,q\,t$};
\end{scope}
\end{tikzpicture}
\caption{Histories (top) and their chord configurations in the universal cover (bottom). The upper line is boundary~1 and the lower line is boundary~2; the dashed lines are the matter line and its lifts $M_\ell$, in the top row the edges identified by the trace. A history has the weights of its moves and a factor $t$ for each chord open at the start. Black dots are crossings of two Hamiltonian chords, orange dots crossings with the matter line; translates are drawn in lighter colors. (a)~Boundary~1 opens a chord and closes the chord open at the start across it. The trace joins the two into one arch winding once around the neck, and the factor $q$ of the closing is the crossing of the arch with its own translate. (b)~Boundary~2 closes the chord open at the start; boundary~1 opens an arch, opens a second chord and closes the arch across it. The history gives the factors in the order of the moves, and does not show that the two teal chords are one chord joining the two boundaries, winding once. The configuration sorts the same factors by region: the arch crossing this chord near boundary~1, the chord crossing the matter line in the band around the neck (shaded), and nothing near boundary~2; the factors $1-q$ of the arch and of the chord go to boundary~1 and to the neck.}
\label{fig:hist-conf}
\end{figure}

We call \emph{history} a single term of this expansion (after splitting each closing weight $1-q^n$ into its terms $(1-q)q^c$): an initial number $n_0$ of open chords and a sequence of $K_1$ top and $K_2$ bottom moves, each an opening or the closing of a specified chord, that returns to a stack of $n_0$ chords. Its weight is the product of the weights of its moves times the factor $t^{n_0}$ from the trace, which identifies final and initial stack positions in order.

Chord configurations are defined in the universal cover $\mathbb R_x\times[0,1]$ of the annulus, with boundary~1
at the top, boundary~2 at the bottom, and deck translation $x\mapsto
x+1$. The matter line lifts to the periodically repeated lines $M_\ell=\{\ell\}\times[0,1]$,
$\ell\in\mathbb Z$. We place the $K_i$ moves of boundary~$i$ at insertion points on that boundary, in the fundamental domain $0<x<1$, preserving
their order on each boundary, with all $K_1+K_2$ points at distinct $x$.
This fixes a relative order of the top and bottom moves, and any such order computes \eqref{D.1}.

A \emph{configuration} consists of a pairing of the insertion points, lifted pair by pair modulo common deck translation, and a number
$r\geq0$ of closed loops. Lifted pairs on the same boundary are
\emph{arches}; those on opposite boundaries are \emph{strands}.
In the cover, each loop simply lifts to a horizontal line $\mathbb R_x\times\{y\}$. Two lifts, of pairs or of the matter line, \emph{interleave} if their
endpoints alternate along the boundary circle of the strip, compactified to a disk by adding the points at $x=\pm\infty$.\footnote{For curves in a disk that cross at most once, this is the condition for crossing.} In particular, an arch on boundary~$i$ crosses a strand precisely when the endpoint of the strand on boundary~$i$ lies between the two endpoints of the arch; top and bottom
arches do not cross. A loop crosses every strand and every matter
line, but no arch or other loop.

For $\nu=(K_1+K_2)/2$ paired chords (an integer, since all insertion points are paired), the configuration weight is
\be
\la{D.2}
\omega(C)=(1-q)^\nu q^{X_{\rm HH}}t^{X_{\rm HM}}.
\ee
Here $X_{\rm HM}$ counts interleavings with matter-line lifts, plus the number $r$ of loops, each crossing the matter line once,
and $X_{\rm HH}$ counts interleavings between pairs of Hamiltonian chords modulo simultaneous
deck translation. Explicitly, choosing one lift $\wt c_a$ per arch or strand and denoting by $\wt c_a+j$ its translate by $j$ periods,
\be
\la{D.3}
\begin{aligned}
X_{\rm HH}
={}&\sum_{a<b}\#\{j\in\mathbb Z:\wt c_a,\wt c_b+j\text{ interleave}\}\\
&+\sum_a\#\{j\geq1:\wt c_a,\wt c_a+j\text{ interleave}\}
+r\,\#\{\text{strands}\}.
\end{aligned}
\ee
The second term includes self-crossings of winding arches. Likewise,
\be
\la{D.4}
X_{\rm HM}=\sum_a\#\{\ell:\wt c_a,M_\ell\text{ interleave}\}+r.
\ee
These objects and their crossings can be read from Fig.~\ref{fig:cover}(a) in Sec.~\ref{sec:fq-sectors}, where the strands are called chords through the neck. In each period, the diagram there has two top arches, one bottom arch, two strands, one of them winding, and $r=2$ loops, so $K_1=2\cdot2+2=6$, $K_2=2+2=4$ and $\nu=5$. Its Hamiltonian crossings are one between the two top arches, one between a top arch and a strand, one between the bottom arch and a strand, one between the strands, and four between loops and strands, giving $X_{\rm HH}=8$; its crossings with the matter line are those of a top arch, of the winding strand and of the two loops, giving $X_{\rm HM}=4$. Hence $\omega=(1-q)^5q^8t^4$, the product of the weights in panel (b).

On the disk, the correspondence between transfer-matrix histories and chord diagrams weighted by their crossings is standard \cite{Ismail:1987qHermite}. It extends directly to the annulus, where a history is a configuration read along one period, as in Fig.~\ref{fig:hist-conf}.

\paragraph{Claim 1 (histories and chord configurations).}
\begin{itshape}
\begin{enumerate}
\item[(a)] Histories with $K_1$ top moves and $K_2$ bottom moves are
in weight-preserving one-to-one correspondence with these configurations, with
$n_0=X_{\rm HM}$.
\item[(b)] Histories of the brane trace \eqref{5.9} with $K$ moves, $m$ of them neck
moves, correspond similarly to configurations with $(K-m)/2$ arches on the single asymptotic boundary, drawn at the top, $m$ vertical neck lines ending on the brane,
and $r$ loops. After stripping $a^m$, the weight is \eqref{D.2},
with a factor $1-q$ for each arch only. The neck lines cross neither
one another nor the matter line; loop-free histories have $r=0$.
\end{enumerate}
\end{itshape}

\begin{proof}
(a) A history gives a configuration by periodic repetition: join the chords across the identified edges position by position, and read off the pairs of insertion points with their lifts; the chords that are never closed are the loops. Conversely, following the lifts of a configuration along $x$ gives the openings and closings, and the position of each closed chord is fixed by the order of the stack: from top to bottom, the chords opened by boundary~1, the more recently opened above; then the loops; then the chords opened by boundary~2, the more recently opened below. This order holds in every history and is invariant under deck translations, so that the two maps are inverse to each other.

For the weights, each paired chord closes once per period, which gives $(1-q)^\nu$, and a closing across $c$ chords crosses them in the drawing of the repeated history (Fig.~\ref{fig:hist-conf}). Two curves of this drawing cross at most once, since they can cross only where one of them is closed and a closed chord is crossed by no later closing; they therefore cross exactly when their endpoints interleave, and the factors $q^c$ of one period give $q^{X_{\rm HH}}$. The $n_0$ chords open at the start are those that cross $M_0$, so that $n_0=X_{\rm HM}$.

(b) A neck move is a vertical line crossing the $n$ open chords, with weight $aq^n$ and no factor $1-q$, which leaves the stack unchanged. The argument of (a) applies with openings and closings only at the top and with the neck lines in place of strands, and the loops are those of Sec.~\ref{sec:fq-normalization}. By \eqref{5.14}, $Z_\Delta(m,\beta)$ is the sum over the
configurations with $r=0$.
\end{proof}

Claim~1 was also checked by exhaustive enumeration, history by history, exact in $q$ and to order $t^3$, for up to six Hamiltonian insertions on the two boundaries (16, 132 and 1440 configurations for $K_1+K_2=2$, $4$ and $6$) and for up to five insertions in the brane trace.

\subsection{Separating boundary and neck data}
\la{app:sectors-cut}

Claim~1 identifies the wormhole trace \eqref{D.1} with the sum of the weights \eqref{D.2} over configurations.
Sorting them by the number $m$ of strands gives the sectors $S_m(\beta_1,\beta_2)$ of \eqref{6.1}, with $\sum_mS_m=\mc W_\Delta(\beta_1+\beta_2)$. We now prove the following claim, which is the statement of Theorem~\ref{prop:sectors}.

\paragraph{Claim 2 (factorization).}
\textit{For every $m\geq0$, the sector $S_m(\beta_1,\beta_2)$
factorizes as in \eqref{6.2}, with $w_\Delta(m)$ given by \eqref{6.3}.}

\begin{proof}
\emph{Data.} A configuration with $m$ strands is the same as a choice of top data, bottom data, neck data and a number $r$ of loops, made independently (Fig.~\ref{fig:cover}(b)). The top data are $m$ points $P_1$ marked among the $K_1$ insertion
points of boundary~1, the top endpoints of the strands, and the arches on the remaining points, with their
lifts; the bottom data are defined in the same way on boundary~2. Write the marked points
in one period as $u_1<\cdots<u_m$ at the top and
$v_1<\cdots<v_m$ at the bottom, all in $(0,1)$. The neck data are a permutation $\sigma\in\mathsf S_m$ and integers
$\bm k=(k_1,\dots,k_m)$, defined uniquely by the condition that strand $i$ joins $(u_i,1)$ to
$(v_{\sigma(i)}+k_i,0)$. There is no condition relating
different lifts, since by Claim~1 every configuration is a history, and no common rotation is counted, since the matter line fixes the origin of the cyclic ordering.

\emph{Weights.} Each crossing involves two curves. A top arch crosses neither bottom arches nor loops, so in a crossing that involves a top arch the other curve is a top arch, a strand or a matter line; the same holds for bottom arches. The crossings thus fall into three disjoint groups: those that involve a top arch, those that involve a bottom arch, and the others, which involve only strands, matter lines and loops. A top arch
crosses a strand according only to the top endpoint of the strand, or of its translates if the arch spans several periods. The crossings of the first group therefore depend only on the top data, and they are unchanged if the strands are replaced
by vertical neck lines.
By Claim~1(b), the top data, with a factor $1-q$ per arch, are then
exactly a loop-free brane configuration with its weight. The same holds for the bottom data.

In the third group, strand $i$ crosses $|k_i|$ matter lines.
Its translate by $\ell\geq1$ periods, with endpoints $(u_i+\ell,1)$ and $(v_{\sigma(i)}+k_i+\ell,0)$, has both endpoints to the right of those of strand~$i$ (negative $\ell$ gives the same pairs),
so a strand never crosses its own translates. Two strands cross exactly when their top endpoints and their bottom endpoints are in opposite order.
For $i<j$, the top endpoint of the translate of strand $j$ by $\ell$ periods lies to the
right of $u_i$ exactly when $\ell\geq0$. Its bottom endpoint differs
from that of strand $i$ by
$d_{ij}+\ell$, where
$d_{ij}=v_{\sigma(j)}-v_{\sigma(i)}+k_j-k_i$ is noninteger.
The orders are therefore opposite for
$|\lfloor d_{ij}\rfloor|$ values of $\ell$, and
\be
\la{D.5}
|\lfloor d_{ij}\rfloor|
=|k_i-k_j+\chi_{ij}|,
\qquad \text{with $\chi_{ij}$ as in \eqref{6.4}}.
\ee
The crossings among strands and matter lines thus have weight
$q^{N(\sigma,\bm k)}t^{\sum_i|k_i|}$, with $N$ as in \eqref{6.4},
independent of the endpoint positions. Each loop crosses the $m$
strands and the matter line once, giving $(q^m t)^r$. With the closing factors $(1-q)^m$ of the $m$ strands, the neck factor is
\be
\la{D.6}
(1-q)^m q^{N(\sigma,\bm k)}t^{\sum_i|k_i|}(q^m t)^r,
\ee
and the weight \eqref{D.2} of a configuration is the product of the two boundary weights and of \eqref{D.6}.

\emph{Sum.} Since all weights are nonnegative, we may sum the four sets of data independently, coefficient by
coefficient in the two boundary expansions. The sum over the top data gives the coefficient of
$(\wt\beta_1/2)^{K_1}/K_1!$ in $Z_\Delta(m,\beta_1)$,
by \eqref{5.14}, and the sum over the bottom data the analogous coefficient
in $Z_\Delta(m,\beta_2)$. The sums of \eqref{D.6} over $r$, $\sigma$ and $\bm k$ then give
\be
\la{D.7}
S_m(\beta_1,\beta_2)
=Z_\Delta(m,\beta_1)Z_\Delta(m,\beta_2)
\frac{(1-q)^m}{1-q^m t}
\sum_{\sigma\in\mathsf S_m}\sum_{\bm k\in\mathbb Z^m}
q^{N(\sigma,\bm k)}t^{\sum_i|k_i|}.
\ee
This is \eqref{6.2} with formula \eqref{6.3} for the weight $w_\Delta(m)$; in particular, it does not depend on the relative order
chosen for the insertion points of the two boundaries. For $m=0$, the permutation and winding
sums each reduce to a single term equal to one.
\end{proof}

As a further check, the factorization \eqref{6.2}, with $w_\Delta(m)$ in the closed form \eqref{6.8}, was verified by exhaustive enumeration, exact in $q$ and truncated in $t$, coefficient by coefficient in $\wt\beta_1$ and $\wt\beta_2$ for all $(K_1,K_2)$ with $K_1+K_2\leq8$, to order $t^3$ ($t^4$ for $K_1+K_2\leq6$).\footnote{The script \texttt{sector\_enumeration.py}, available at \url{https://github.com/matteobeccaria/Sector-decomposition}, repeats this check to order $t^3$ and reproduces the counts of configurations quoted above, in exact rational arithmetic at the fixed values $q=\frac13,\frac25,\frac37$ instead of symbolic $q$.}

\section{Proof of the twist--winding identity}
\la{app:shi}

We give the combinatorial proof of Proposition~\ref{prop:shi} and the matter-free limit \eqref{7.7} of the neck weight.

\subsection{The sum over twists and windings}
\la{app:sectors-shi}

To prove Proposition~\ref{prop:shi}, we need to show that
\be
\la{E.1}
\sum_{\sigma\in\mathsf S_m}\sum_{\bm k\in\mathbb Z^m}
q^{N(\sigma,\bm k)}\,t^{\sum_i|k_i|}=\frac{(q;q)_m}{(1-q)^m}\,\frac{(t^2;q)_m}{(t;q)_m^2},
\ee
which implies the closed form \eqref{6.8}. The proof in this section is purely combinatorial and independent of the derivation from the uniqueness statement in Sec.~\ref{sec:fq-gluing}. In the proof, we denote by ${r\brack j}_q=(q;q)_r/\big((q;q)_j(q;q)_{r-j}\big)$ the Gaussian binomial coefficients.

\paragraph{Sum over the twist.} At fixed $\bm k=(k_1,\dots,k_m)$, order the indices
$1,\ldots,m$ by decreasing $k_i$, keeping their original order when two
$k_i$ coincide, and let $I(\sigma,\bm k)$ be the number of inversions of the
sequence of the $\sigma(i)$ read in this order. Then
\be
\la{E.2}
N(\sigma,\bm k)=d(\bm k)-A(\bm k)+I(\sigma,\bm k),
\ee
with $d(\bm k)=\sum_{i<j}|k_i-k_j|$ and $A(\bm k)=\#\{i<j:k_i<k_j\}$.
Indeed, for $i<j$ with $k_i\geq k_j$ the index $i$ comes first,
$|k_i-k_j+\chi_{ij}|=|k_i-k_j|+\chi_{ij}$, and $\chi_{ij}$ is the
inversion indicator of the pair, \ie it equals $1$ exactly when the pair is an inversion of the sequence, $\sigma(i)>\sigma(j)$. For $k_i<k_j$ the index $j$ comes
first, $|k_i-k_j+\chi_{ij}|=|k_i-k_j|-\chi_{ij}$ since $k_j-k_i\geq1$,
and the inversion indicator is $1-\chi_{ij}$. At fixed $\bm k$ the reading
order is fixed, so the sequence runs over all permutations as $\sigma$
runs over $\mathsf S_m$, and $\sum_\sigma q^{I(\sigma,\bm k)}=(q;q)_m/(1-q)^m$, the generating function of permutations by number of inversions \cite[Cor.~1.3.13]{Stanley:EC1}.\footnote{It follows by induction on $m$: inserting $m$ into a permutation of $1,\dots,m-1$ at each of its $m$ positions adds $0,1,\dots,m-1$ inversions, which gives the factor $1+q+\dots+q^{m-1}=(1-q^m)/(1-q)$.}
Since $A(\bm k)\leq m(m-1)/2$, all the sums below converge absolutely for $0<t<1$, and \eqref{E.1} is equivalent to
\be
\la{E.3}
\Lambda_m=\frac{(t^2;q)_m}{(t;q)_m^2},\qquad \Lambda_m\equiv\sum_{\bm k\in\mathbb Z^m}q^{d(\bm k)-A(\bm k)}\,t^{\sum_i|k_i|}.
\ee
This convergence, with $|\lambda_r|<1$ for $r\geq1$ in \eqref{E.5}, is all that the proof uses.

\paragraph{Level decomposition.} For $c\in\mathbb Z+\frac12$, which we call a \emph{level}, let $r_c$
be the number of $k_i$ beyond $c$, \ie with $k_i>c$ if $c>0$ and $k_i<c$
if $c<0$. The level $c$ separates $r_c(m-r_c)$ pairs, and $r_c$ of the
$m$ intervals between $0$ and the $k_i$ cross it. Taking the product over the levels,
\be
\la{E.4}
q^{d(\bm k)}\,t^{\sum_i|k_i|}=\prod_{c}\lambda_{r_c},\qquad \lambda_r=t^r q^{r(m-r)}.
\ee
Here and below $m$ is fixed, so we do not indicate it in $\lambda_r$.

\paragraph{Recursion.} For $0\leq r\leq m$, let $F_r$ be the sum
over $\bm k=(k_1,\ldots,k_r)\in\mathbb Z_{\geq0}^r$, with $A$ and $r_c$ defined as above for $r$ entries, of
$q^{-A(\bm k)}\prod_{c>0}\lambda_{r_c}$. If $j$ of the
$k_i$ are positive (the others being zero), the level $c=\frac12$ gives $\lambda_j$. The increasing
pairs between the zero entries and the positive ones are those with the zero entry
first, and summing over the positions of the positive entries gives
${r\brack j}_{q^{-1}}=q^{-j(r-j)}{r\brack j}_q$, where ${r\brack j}_q$ is the generating function of binary words with $j$ ones and $r-j$ zeros by number of inversions \cite[Prop.~1.7.1]{Stanley:EC1}. Lowering the positive
entries by one leaves their increasing pairs unchanged and maps the
remaining levels to those of $F_j$. Hence
\be
\la{E.5}
F_0=1,\qquad F_r=\sum_{j=0}^r q^{-j(r-j)}{r\brack j}_q\,\lambda_j\,F_j,
\ee
which determines $F_r$ uniquely.\footnote{The term $j=r$ on the right-hand side of \eqref{E.5} is $\lambda_rF_r$, so \eqref{E.5} gives $(1-\lambda_r)F_r$ in terms of the $F_j$ with $j<r$, and $1-\lambda_r\neq0$, since $|\lambda_r|<1$, for $r\geq1$.} The
solution is
\be
\la{E.6}
F_r=\frac{1}{(tq^{m-r};q)_r}.
\ee
Indeed, with $a=tq^{m-r}$, substituting into \eqref{E.5} and multiplying by $(a;q)_r$\footnote{One has $q^{-j(r-j)}\lambda_j=a^j$ and $(a;q)_rF_j=(a;q)_{r-j}$, so that \eqref{E.5} multiplied by $(a;q)_r$ becomes $\sum_j{r\brack j}_q a^j(a;q)_{r-j}=1$.} reduces it to the case $b=0$ of the identity
\be
\la{E.7}
(ab;q)_r=\sum_{j=0}^r{r\brack j}_q\,a^j\,(b;q)_j\,(a;q)_{r-j},
\ee
which is also used below.\footnote{\eqref{E.7} follows by induction on $r$: multiplying the $j$-th term by $1-abq^r=(1-aq^{r-j})+aq^{r-j}(1-bq^j)$ and using ${r\brack j}_q+q^{r+1-j}{r\brack j-1}_q={r+1\brack j}_q$ gives \eqref{E.7} at $r+1$.}

\paragraph{Back to $\Lambda_m$.} In the sum defining $\Lambda_m$, let $p$ of the $k_i$ be positive.
Summing over their positions gives $q^{-p(m-p)}{m\brack p}_q$, as in the recursion with the nonpositive entries in place of the zero ones. In the product \eqref{E.4} over the levels, the level $c=\frac12$, which separates the positive entries from the others, gives $\lambda_p$, the levels $c>\frac12$ involve only the positive entries, which, lowered by one,
give $F_p$, and the levels $c<0$ involve only the nonpositive ones. Changing the sign of the nonpositive entries and reversing
their order preserves their increasing pairs and maps their levels to
positive ones, so they give $F_{m-p}$. With
$1/(tq^{m-p};q)_p=(t;q)_{m-p}/(t;q)_m$ and
$1/(tq^p;q)_{m-p}=(t;q)_p/(t;q)_m$, we get
\be
\la{E.8}
\begin{aligned}
\Lambda_m&=\sum_{p=0}^m q^{-p(m-p)}{m\brack p}_q\,\lambda_p\,F_pF_{m-p}\\
&=\frac{1}{(t;q)_m^2}\sum_{p=0}^m{m\brack p}_q\,t^p\,(t;q)_p\,(t;q)_{m-p}
=\frac{(t^2;q)_m}{(t;q)_m^2},
\end{aligned}
\ee
the last step being again \eqref{E.7}, now with $a=b=t$. This proves \eqref{E.3},
and hence \eqref{E.1}.

\subsection{Matter-free limit of the neck weight}
\la{app:sectors-free}

This subsection proves \eqref{7.7}. Without matter, two chord diagrams
that differ by a common winding $\bm k\to\bm k+c(1,\ldots,1)$ of the $m$
chords through the neck are related by a Dehn twist around the neck,
which fixes both boundaries; with no line to measure the winding against,
they are identified. The counting of Apps.~\ref{app:sectors-histories} and~\ref{app:sectors-cut} goes through at $t=1$, with the matter line replaced by the starting points of the two boundaries, joined by a reference curve through the neck defined up to Dehn twists. The neck weight is then the sum \eqref{7.7} over $\bm k$ modulo the common winding, with closed loops that cross only the $m$ chords.

By \eqref{E.2} the sum over $\sigma$ gives $(q;q)_m/(1-q)^m$ times
$q^{d(\bm k)-A(\bm k)}$, with $d$ and $A$ invariant under the common
shift. Splitting off the common winding in $\Lambda_m$ of \eqref{E.3}
gives
$(1-t)\Lambda_m\to(2/m)\sum_{\bm k\in\mathbb Z^m/\mathbb Z(1,\ldots,1)}q^{d(\bm k)-A(\bm k)}$
as $t\to1$, by dominated convergence, since
$(1-t)\sum_ct^{\sum_i|k_i+c|}\leq d(\bm k)+3$, $d(\bm k)\geq\max_{i,j}|k_i-k_j|$
and $A(\bm k)\leq m(m-1)/2$. The closed form \eqref{E.3} gives
$(1-t)\Lambda_m=(1+t)(t^2q;q)_{m-1}/(tq;q)_{m-1}^2\to2/(q;q)_{m-1}$, hence
$\sum_{\bm k\in\mathbb Z^m/\mathbb Z(1,\ldots,1)}q^{d(\bm k)-A(\bm k)}=m/(q;q)_{m-1}$,
and \eqref{7.7} follows. The factor $m$ counts the $m$ elementary rotations of one boundary relative to the other, whose $m$-th power is the quotiented common winding; they are the discrete counterpart of the twist integral $\int_0^{\mathfrak b}d\alpha$, which gives the factor $\mathfrak b$ in JT (footnote~\ref{fn:unfolding}).

\section{Classical sine-dilaton trumpet and probe geodesic}
\la{app:sd-classical}

This appendix derives the trumpet saddle \eqref{F.3} and the geodesic length \eqref{9.10} used in Sec.~\ref{sec:sd-bulk}.
In the conventions of
\cite{Blommaert:2024ymv,Blommaert:2025avl}, the Euclidean action of
sine-dilaton gravity is
\be
\la{F.1}
I=\frac{1}{\hbar}\bigg[
\frac12\int_{\mc M}d^2x\,\sqrt g\,
\big(\Phi R+2\sin\Phi\big)
+\int_{\partial\mc M}du\,\sqrt h\,
\big(\Phi K-i e^{-i\Phi/2}\big)
\bigg],
\ee
 where $u$ is a boundary coordinate, $\Phi$ is the dilaton, $R$ and $K$ are the
bulk Ricci scalar and boundary extrinsic curvature, and $g$ and $h$ are the
determinants of the bulk and induced boundary metrics on $\mc M$ and
$\partial\mc M$, and $\hbar=\l$ (Sec.~\ref{sec:sd-bulk}). The solutions of its classical equations of motion with
non-constant dilaton form a one-parameter family, labeled
by $\theta$; locally, in the gauge $\Phi=r$, they read
\cite{Grumiller:2002nm,Blommaert:2024ymv,Blommaert:2025avl}
\be
\la{F.2}
ds^2=F(r)\,dt_E^2+\frac{dr^2}{F(r)},\qquad\Phi=r,\qquad
F(r)=2\cos\theta-2\cos r ,
\ee
where $t_E$ is the Euclidean time coordinate. The ADM energy is proportional to $-\cos\theta$, as $E(\theta)$ in \eqref{2.5} \cite{Blommaert:2024ymv}.
Lengths are measured in the Weyl-rescaled metric $ds^2_{\rm AdS}=e^{-i\Phi}ds^2$.  The trumpet saddle has $\theta=i\eta$, with real energy $E(i\eta)=-E_0\cosh\eta$ below the spectral edge.
In terms of the coordinate $\rho$ defined by $e^{-i\Phi}=\cosh\eta-i\rho$, the metric is the hyperbolic trumpet
\be
\la{F.3}
ds^2_{\rm AdS}=\big(\rho^2+\sinh^2\eta\big)\,dt_E^2
+\frac{d\rho^2}{\rho^2+\sinh^2\eta},
\qquad t_E\sim t_E+\l\wt\beta=t_E+\beta_s.
\ee
Its asymptotic region $\rho\to\infty$ is $\Phi\to\frac\pi2+i\infty$,
the sine-dilaton boundary \cite{Blommaert:2024ymv}, and the period of
$t_E$ is the renormalized boundary length \cite{Blommaert:2025avl}.
The neck $\rho=0$ is a closed geodesic of length
$\l\wt\beta\sinh\eta$. Setting it equal to $\l b$, the length \eqref{9.8} at real $b$,
gives \eqref{9.3}.
In the coordinates $(\sigma,x)$ defined by $\rho=\sinh\eta\,\sinh\sigma$,
$x=t_E\sinh\eta$, the metric \eqref{F.3} becomes
$d\sigma^2+\cosh^2\!\sigma\,dx^2$ with $x\sim x+\l b$, which is the
double-trumpet metric of
Sec.~\ref{sec:neck-cut} restricted to $\sigma>0$, with the neck at $\sigma=0$.

Two trumpets with boundaries $\beta_1,\beta_2$ and
the same neck have $\sinh\eta_i=b/\wt\beta_i$, that is, $\eta_i$ is
given by \eqref{9.3} at each $\wt\beta_i$, and gluing
them along $\sigma=0$ gives a single hyperbolic cylinder. In these coordinates a
radial cutoff $\rho=\rho_c$, with $\rho_c\to\infty$ regulating the asymptotic boundary, lies at $e^{|\sigma|}\simeq2\rho_c/\sinh\eta_i$
on side $i$, and the geodesic distance between boundary points at
relative position $x_{12}$ along the neck in the universal cover
follows as in footnote~\ref{fn:geodesic} of App.~\ref{app:JT-classical}:
$e^{-D}=\sinh\eta_1\sinh\eta_2/[4\rho_c^2\cosh^2(x_{12}/2)]$ up to
$O(\rho_c^{-2})$ corrections.
Subtracting $\log(2\rho_c)$ per endpoint gives $e^{-L}=\sinh\eta_1\sinh\eta_2/\cosh^2(x_{12}/2)$, which is \eqref{9.10}, the relation \eqref{A.2} with $x'(u_i)$ replaced by $2\sinh\eta_i=2b/\wt\beta_i=2\l b/\beta_{s,i}$.

The subtraction is fixed by the JT corner. Since $\beta_{s,i}\simeq2\tau_i/\l$ in the JT regime, the factor $2\sinh\eta_i$ is there $\l\times(\l b/\tau_i)$, the classical JT value of $x'(u_i)$ for neck length $\l b$ times one factor $\l$. In $e^{-\Delta L}$ each endpoint therefore carries $\l^\Delta$, which is removed by the factor $\l^{-\Delta}$ in the rescaling of \eqref{8.1}, and the probe factor reduces to its JT value $(\mathfrak b/\tau)^\Delta$. The subtraction also agrees with the prescription of \cite{Blommaert:2024ymv} for the effective-AdS$_2$ disk correlator.

\bibliography{MB-Biblio}
\bibliographystyle{JHEP}
\end{document}